\documentclass[12pt,a4paper]{article}
\usepackage{ifpdf}
\usepackage{times}
\usepackage[scale={0.8,0.9},centering,includeheadfoot]{geometry}
\usepackage{enumitem}
\usepackage{subcaption}
\usepackage{multicol}
\usepackage{lscape}
\usepackage{listings}
\usepackage{float}

\usepackage{pgf}

\usepackage{amsmath,amssymb,xspace,amsthm,bm}
\usepackage{hyperref}

\usepackage[colon,longnamesfirst]{natbib}

\usepackage{dcolumn}\newcolumntype{d}[1]{D{.}{.}{#1}}

\newtheorem{definition}{Definition}

\newtheorem{result}{Result}

\def\qedbox{\ifvmode\else\unskip\fi~\penalty10000
    \hfill{\large$\blacksquare$}}

\def\miscinfo#1#2{{\footnotesize\indent\textsc{#1: }\ignorespaces #2}}   

\def\RINLA{\texttt{R-INLA}\xspace}

\begin{document}
\bibliographystyle{apalike}

\title{\vspace{-15mm}\fontsize{24pt}{10pt}\selectfont\textbf{Variance partitioning in spatio-temporal disease mapping models}}

\author{
    Maria Franco-Villoria \\
    Department of Economics\\
    University of Modena and Reggio Emilia\\
    \and
    Massimo Ventrucci \\
	Department of Statistical Sciences\\    
    University of Bologna\\
    \and
    H{\aa}vard Rue\\
    CEMSE Division, King Abdullah University of Science and Technology, \\
    Thuwal, Saudi Arabia
}
    
\maketitle

\begin{abstract}
Bayesian disease mapping, yet if undeniably useful to describe variation in risk over time and space, comes with the hurdle of prior elicitation on hard-to-interpret random effect precision parameters. We introduce a reparametrized version of the popular spatio-temporal interaction models, based on Kronecker product intrinsic Gaussian Markov Random Fields, that we name the variance partitioning (VP) model. The VP model includes a mixing parameter that balances the contribution of the main and interaction effects to the total (generalized) variance and enhances interpretability. The use of a penalized complexity prior on the mixing parameter aids in coding prior information in a intuitive way. We illustrate the advantages of the VP model using two case studies.
\end{abstract}

\miscinfo{Keywords}{intrinsic Gaussian Markov Random Fields; intrinsic CAR; Kronecker product GMRF; penalized complexity prior; spatio-temporal smoothing}\\
\miscinfo{Address for Correspondence}{Maria Franco-Villoria, Department of Economics, University of Modena and Reggio Emilia. Email: maria.francovilloria@unimore.it}

\section{Introduction}

The Covid-19 pandemic has put the world at stake. In Italy, the first two cases were confirmed on 31$^{st}$ January 2020 and on 9$^{th}$ March 2020 a national lockdown was put in place by the authorities to control and reduce the expansion of the virus. Data on newly infected people have been routinely collected since then to monitor the evolution of the disease. The study of the pandemic evolution can be tackled using disease mapping. Knowledge of how the
infection has spread can help to evaluate the performance of containment measures. In particular, the quantification of the space-time interaction, which describes how the spatial patterns change over time, has been proposed as a way to deepen our understanding on the evolution of the disease \citep{Diggle:1995}.

Disease mapping models \citep{beneito-book, lawson-book, macnab2022, Wakefield2007} aim to describe the variation in risk of a particular disease over space and time. Data are usually available in the form of aggregrated counts at some spatial level, such as counties, municipalities, etc. Additive time and space models have been long used to model disease rates \citep{held:1998}. More recently, the availability of complex data has made it possible to consider more complex models that include an interaction term to appropriately capture space-time relationships in the data (see for example \cite{abellan:2008}, \cite{held:2000}, \cite{waller:1997}, \cite{bernardinelli:1995} to cite a few). Understanding the spatial distribution of disease risk or how it has evolved over time might be useful for public health authorities in planning resource allocation and identification of areas to be prioritized.  In particular, the space-time interaction may reveal important information regarding the nature of the disease, for example suggesting whether a new disease is possibly infectious \citep{Aldstadt:2007} or the existence of additional causes in non-infectious cases \citep{Robertson:2010}. Thus a model that is able to quantify the importance of this term is desirable from a practical point of view; in this paper we introduce a model parametrization that partitions the total variance into main and interaction effects so that the contribution of each of those can be quantified.

Crude disease rates are unrealiable due to sampling variability so smoothing is used to borrow information across neighbouring areas and time points. For this reason, disease mapping has been developed mainly in a Bayesian hierarchical model formulation where the building blocks of a smooth in one and more dimensions can be modelled using intrinsic Gaussian Markov Random Fields (IGMRF) such as the first and second order random walk \citep{lindgren-2008} or the ICAR \citep{besag} models. For modelling interactions statisticians have used tensor products smoothers, where, in a Bayesian framework, the penalty can be seen as a special type of GMRF called Kronecker product GMRFs \citep{rue-2005}. 

In Bayesian spatio-temporal disease mapping the precision parameter of the IGMRFs plays a role in controlling the degree of smoothing applied over time and space. A number of issues related to prior elicitation need to be addressed when dealing with intrinsic models. Firstly, the precision matrix is singular, which means that the total variance that we aim to partition is not finite. In order to define priors on the variance components we can rely upon the concept of generalized variance of an IGMRF; this has been defined by \cite{sorbye-2013} as the geometric mean of the diagonal elements of the generalized inverse of the precision matrix of the IGMRF, and can only be computed upon linear constraints.

 A second issue to bear in mind is that the generalized variance of an IGMRF depends on the structure matrix, and hence it changes depending on things like the temporal and spatial resolution or the size of the dataset at hand. This means that interpretation of the precision parameter becomes case-dependent, making prior elicitation and paremeter interpretation difficult. To avoid this problem, \cite{sorbye-2013} advise scaling the structure matrix so that the generalized variance is equal to 1; this way the precision parameter is automatically rescaled and the prior has the same meaning regardless of the graph structure \citep{Freni-Sterrantino:2018}. Scaling becomes particularly relevant in the context of space-time models, as otherwise differences in the structure matrices of the spatial, temporal and spatio-temporal terms would have an impact on the priors for the corresponding precision parameters that we cannnot control. By scaling the structure matrix of the temporal and spatial random effects, the structure matrix of the interaction, defined as a Kronecker IGMRF, is automatically scaled.

Further to the issues mentioned above, the choice of priors for variance parameters has received much attention in the literature \citep{gelman-2006,Wakefield2007,fong-2010,pcprior}. Part of the hassle in choosing a prior stems from the difficulty of interpreting variance parameters, especially for intrinsic processes, where the standard deviation is to be interpreted as a conditional one \citep{fong-2010, riebler-2016}. On top of that, in models with various terms, the tendency is to set priors independently for each precision parameter, while some authors are beginning to recognize that it might be more practical to think about total variability and how each term in the model contributes to that rather than to concentrate on single variance components separately \citep{Wakefield2007,riebler-2016,Fuglstad:2020,Ventrucci:2020}. 
In the context of disease mapping, \cite{Wakefield2007} proposes using an inverse Gamma prior on the total variability, along with a Beta prior that distributes the variance between a spatially correlated random field and a spatially unstructured effect (the so called BYM model \cite{besag-york-mollie}). Using a similar parametrization, \cite{riebler-2016} present a prior that shrinks towards no spatial effect following the penalized complexity (PC) prior approach of \cite{pcprior}. Outside the disease mapping literature, \cite{Ventrucci:2020} develop a PC prior in one-factor mixed models for the relative contribution of group-specific variability. In a more general context,  \cite{Fuglstad:2020} introduce a  framework for hierarchically distributing the variance in additive models, where, at each level of the total variance decomposition, ignorance or preference about the variance contribution of a term is expressed via a Dirichlet or a PC prior, respectively.
We add to the literature by considering also the temporal dimension in disease mapping models. In particular, all the terms in the model (main effects and interaction) are assumed to follow intrinsic models. This differentiates our work from the literature mentioned above.

In this work, we revisit the spatio-temporal models proposed by  \cite{held:2000}, where the space-time interaction term can be one of four different types, depending on the degree of dependence assumed between time and space. These four types are characterized by different prior assumptions, expressed in terms of a Kronecker product. We propose an intuitive reparametrization that leads to partitioning the generalized variance between the main effects and interaction. The main and interaction effects are not independent, and hence using a joint prior on those terms is preferable.
 We do so by including a mixing parameter that 1) easies interpretation and 2) naturally leads to a prior that is intuitive to elicit. 
 One of the advantages of the Bayesian framework is that whenever information on the disease process is available, it can be encoded into the prior \citep{Robertson:2010}. Often, the epidemiologist might have an intuition on how important the interaction term is in explaining the spatio-temporal variation of a particular disease. However, translating this information in terms of a precision parameter is not trivial at all. We follow the penalized complexity prior (PC) framework of \cite{pcprior} to derive a prior for the mixing parameter that avoids overfitting by construction and allows the user to code any prior information easily. This way we alleviate both problems, by considering an interaction model that not only enhances interpretability but also permits a more intuitive construction of the prior. We call this reparametrized version the \emph{variance partitioning (VP) model}. The proposed methodology is applicable to any of the four space-time interactions described in \cite{held:2000}.

The rest of the paper is organized as follows. Section 2 covers spatio-temporal disease mapping models, with a particular emphasis on the space-time interaction framework by \cite{held:2000}, followed by a brief discussion of priors for variance parameters with special attention to the PC prior approach. In Section 3 the VP model is described in detail and the PC prior for the mixing parameter is presented, while the technical details are relegated to the supplementary material. Section 4 illustrates the proposed model on two case studies, a well known example in the disease mapping literature and an Italian Covid-19 dataset. The paper closes with a discussion in Section 5.

\section{Spatio-temporal disease mapping}\label{sec:spmodels}
Consider data on $n_1$ time points and $n_2$ non-overlapping areas, $y_{ij}$ is the observed number of cases at time $i=1,\ldots,n_1$ and area $j=1,\ldots,n_2$. The most commonly used models for $y_{ij}$ are the binomial and the Poisson; in either case, the model in the linear predictor scale can be written as $\eta_{ij} = \alpha + f_1(i) +f_2(j) + f_{12}(i,j)$, where $f_1(i)$ and $f_2(j)$ represent the main temporal and spatial effects respectively and the function $f_{12}(i,j)$ captures the space-time interaction. The model can be parametrized with random effects as 
\begin{equation}
\eta_{ij} = \alpha + \beta_{1_i} +\beta_{2_j} + \delta_{ij},\label{eq:model_classic}
\end{equation}
where $\bm \beta_1 = (\beta_{1,1}, \ldots,\beta_{1,n_1})^T$ and $\bm \beta_2=(\beta_{2,1}, \ldots,\beta_{2,n_2})^T$ are vectors of random effects describing the temporal and spatial main effect, respectively, and $\bm \delta=\{\delta_{ij}\}, i=1,\ldots,n_1, j=1,\ldots,n_2$
is the vectorized spatio-temporal interaction term. The random effects $\bm \beta_1, \bm \beta_2$  and $\bm \delta$ are typically assumed as smooth processes modelled using intrinsic Gaussian Markov Random Fields (IGMRF, \cite{rue-2005}), a special type of improper GMRF, defined below. Appropriate constraints \citep{Goicoa:2018} need to be imposed to ensure identifiability of the terms in (\ref{eq:model_classic}). The constraints on the interaction term are summarized on Table \ref{tab:types}, while on the temporal and spatial main effects it is enough to impose a sum to zero constraint. As usual, any available covariates can be included in model (\ref{eq:model_classic}) as fixed effects.

\begin{definition}[Improper GMRF] 
Let $\bm Q$ be an $n \times n$ symmetric positive semi-definite (SPSD) matrix with rank $n-p>0$. Then $\bm x=(x_1,\ldots,x_n)^T$ is an improper GMRF of rank $n-p$ with parameters $(\bm\mu,\bm Q)$ if its density is
\[
	\pi(\bm x)=(2\pi)^{\frac{-(n-p)}{2}}(|\bm Q|^*)^{1/2}\exp\left(-\frac{1}{2}(\bm x -\bm \mu)^T \bm Q (\bm x - \bm \mu)\right),
\]
\end{definition}
\noindent
where $|\bm Q|^*$ is the generalized determinant of the precision matrix $\bm Q$.
Improper GMRFs are used as smoothing priors in structured additive regression (STAR) models, a flexible class including generalized linear mixed models, temporally dynamic models, spatial varying coefficient models, etc; for an account of STAR models see \cite{fahrmeir-2013-book} and references therein. 

Following \cite{rue-2005} we define an IGMRF of order 1 as an improper GMRF where $\bm Q \bm 1 = \bm 0$, i.e., the precision matrix is singular with null space spanned by a column vector of ones, $\bm 1_n$ of length $n$.  Popular examples of an IGMRF of order 1 are the first order random walk (RW1), which is a possible option to model the temporal main effect $\bm \beta_1$, and the intrinsic conditional autoregressive (ICAR) model by \cite{besag}, which is often assumed in disease mapping to model the spatial effect $\bm \beta_2$ when smoothing across neighbouring regions is required. 

An IGMRF of order 2 is an improper GMRF whose precision matrix is singular and its null space is spanned by a constant vector $\bm 1_n$ and a linear vector $(1,\ldots,n)^T$. A popular example is the second order random walk (RW2; \cite{lindgren-2008}), popularly used  for modelling smooth covariate effects in STAR models, and often implemented in spatio-temporal disease mapping for modelling the main temporal effect $\bm \beta_1$ when smoothness in the disease risk over time is anticipated.

All the IGMRFs described above have in common that their precision matrix can be written as $\bm Q = \tau \bm R$, where $\tau$ is a precision parameter and $\bm R$ is a known structure matrix that encodes the dependence structure. In particular, for the RW1 
\[
	R_{k,l}= \left\{
	\begin{array}{ll}
	1 & k=l \in \{1,n\}\\
	2 & k=l \in\{2,\ldots,n-1\}\\ 
	-1 & k\sim l\\
	0 & \text{otherwise,}
	\end{array}
	\right. 
\]
where notation $k \sim l$ indicates contiguous time points. For the ICAR, the structure matrix is given by 
\[R_{k,l}= \left\{
	\begin{array}{ll}
	m_k & k=l\\
	-1 & k\sim l\\
	0 & \text{otherwise,}
	\end{array}
	\right.
\]
where $m_k$ is the number of neighbours for region $k$ and notation $k \sim l$ indicates contiguous areas that share a common border.  The structure matrix of a RW2 can be written as $\bm R = \bm D^T \bm D$ where $\bm D$ is a second order difference matrix of dimension $(n-2) \times n$.

It is common in the disease mapping literature to consider one or both main effects $f_1$ and $f_2$ as a sum of structured and unstructured effects, so that model (\ref{eq:model_classic}) becomes
\begin{equation}
\eta_{ij}  =  \alpha + \beta_{1_i}+\epsilon_{1_i} + \beta_{2_j}+\epsilon_{2_j} + \delta_{ij},\label{eq:model_uns+struc}
\end{equation}
where $\bm \epsilon_1\sim N\left(\bm  0, \tau_{\epsilon_1}\bm I_{n_1}\right)$, $\bm \epsilon_2\sim N\left(\bm  0, \tau_{\epsilon_2}\bm I_{n_2}\right)$. Typically, a RW1 or RW2 model is assumed for the temporal effect $\bm \beta_1 \sim N\left(\bm  0,  \tau_1^{-1}\bm R_{1}^{-}\right)$ and an ICAR is assumed for the spatial effect $\bm \beta_2\sim N\left(\bm  0,  \tau_2^{-1}\bm R_{2}^{-}\right)$, where notation $\bm M^-$ indicates the generalized inverse of matrix $\bm M$. The combination of the structured and unstructured spatial terms $\beta_{2_j} + \epsilon_{2_j}$ is commonly known as the BYM model \citep{besag-york-mollie}.

\subsection{Modelling interactions via Kronecker product IGMRFs}\label{sec:Kronecker}

We describe now the interaction term $\bm \delta$ in Eq. (\ref{eq:model_uns+struc}). Smoothness is induced by assuming 
\begin{equation*}
\bm \delta   \sim  N(\bm  0, \tau_{12}^{-1}\bm R_{I}^-),
\end{equation*}
which is a Kronecker product IGMRF with precision $\bm Q= \tau_{12} \bm R_{I}$, i.e. an improper GMRF with precision given by the Kronecker product of two IGMRFs. These models are used for smoothing spatial and spatio-temporal data, and they are the Bayesian equivalent of tensor product spline models \citep{wahba:1978}. \cite{held:2000} envisions four different types of interactions, reported in Table \ref{tab:types}. Interaction type I can be seen as unstructured variation due to unobserved covariates, while interaction types II and III allow for the temporal trend to change from location to location and the spatial trend to change over time, respectively, but in an independent manner. Interaction type IV is the most complex one, assuming that the temporal trend changes with location in a spatially dependent way, or equivalently, that the way in which the spatial trend changes over time is time-dependent. 

\begin{table}[h!]
\small\sf
\centering
\caption{The four types of interactions in spatio-temporal smoothing according to \cite{held:2000}. The IGMRF on the interaction parameter vector $\bm \delta$ has structure $\bm R_I$ given by a Kronecker product; $r_1=1$ or $2$ depending on the order of the RW assumed for the time effect. }
\begin{tabular}{clll}
\hline
type &  $\bm R_{I}$  & {rank}$(\bm R_{I})$ & linear constraints on $\bm \delta$ \\
\hline
I & $\mathbf{I}_{n_2} \otimes \mathbf{I}_{n_1} $& $n_1 n_2$ & not needed\\
II & $\mathbf{I}_{n_2} \otimes \bm R_{1}$& $n_2(n_1-r_1)$ &  $\left[ \bm I_{n_2} \otimes \bm 1_{n_1} \right]^T\bm \delta=\bm 0_{n_2}$\\
III& $\bm R_2 \otimes \bm I_{n_1} $& $(n_2-1)n_1$ &   $\left[  \bm 1_{n_2}\otimes \bm I_{n_1} \right]^T\bm \delta=\bm 0_{n_1}$\\
IV& $\bm R_2 \otimes \bm R_1$& $(n_2-1)(n_1-r_1)$ &   $\left[ \bm I_{n_2} \otimes \bm 1_{n_1} \right]^T\bm \delta=\bm 0_{n_2}; \quad \left[  \bm 1_{n_2}\otimes \bm I_{n_1} \right]^T\bm \delta=\bm 0_{n_1}$ \\
\hline
\end{tabular}
\label{tab:types}
\end{table}

Model (\ref{eq:model_classic}) includes different precision parameters $\tau_1$ and $\tau_2$ for smoothing over time and space and an additional one, $\tau_{12}$, controlling the variance of the interaction term, which yields a model able to capture the smooth spatio-temporal structure underlying the data with high flexibility. 
However, these models have limitations in terms of interpretation of the results, as precision parameters are not informative about the total variance explained by the associated components and the priors are not easy to elicit (see Section \ref{sec-prior precision}). We propose an alternative parametrization to address these issues in Section \ref{sec:vp}.

\subsection{Priors for the precision parameters}\label{sec-prior precision}

There are two main challenges in prior choice for the precision parameters in model (\ref{eq:model_classic}). The first problem regards the so called scaling issue that affects IGMRFs in general; \cite{sorbye-2013} proposed addressing this issue by scaling the precision structure $\bm R$ so that the geometric mean of the diagonal elements in $\bm R^-$ is $1$. In this way, the prior for $\tau$ will roughly encode the same degree of complexity across different types of structures and hence will have the same interpretation. Of particular interest is the spatial case where, after scaling the precision of the ICAR, the  prior for the precision parameter becomes transferable across different applications using different graph structures. 

The second challenge regards the structure of the Kronecker product IGMRF, which can be thought of as an extra layer of flexibility on top of the main effects model. The common practice is to set independent priors on each precision parameter, but this totally disregards the model structure. Popular choices are Gamma for $\tau$, or half-t and uniform on the standard deviation $1/\sqrt{\tau}$ \citep{gelman-2006}. The Gamma prior has repeatedly been pointed out as a poor choice often made by convenience; among the reasons why it should be avoided is that it forces overfitting or underfitting depending on the choice of its parameters \citep{FruhwirthSchnatter-2010, FruhwirthSchnatter-2011, fong-2010, pcprior, ventrucci-rue-2016}. 

In Section \ref{sec:vp} we propose a novel modelling framework where the interaction is seen as a flexible extension of the main effects model, and the prior is set so that the interaction term shrinks to the main effects following the PC prior framework.
Recently, PC priors have been proposed as a way to prevent overfitting, based on four simple principles, that we briefly summarize and illustrate below for the precision parameter $\tau$ of a Gaussian random effect. For further details the reader is referred to \cite{pcprior}.

Let $\pi_1$ denote the density of a model component $\bm w$ with precision parameter $\tau$. This model component can be seen as a flexible extension of a based model with density $\pi_0$ and $\tau=\infty$ (i.e. absence of random effects). The four principles are:
\begin{enumerate}
	\item Parsimony: The prior for $\tau$ should give proper shrinkage to $\tau=\infty$ and decay with increasing complexity of $\pi_1$, so that the simplest model is favoured unless there is evidence for a more flexible one. 

	\item The increased complexity of $\pi_1$ with respect to $\pi_0$ is measured using the Kullback-Leibler divergence \citep[KLD, ][]{kld-1951},
\[
	\text{KLD}(\pi_1||\pi_0)=\int \pi_1(w)\log\left(\frac{\pi_1(w)}{\pi_0(w)}\right)dw.
\]
For ease of interpretation, the KLD is transformed to a unidirectional distance measure
\begin{equation*}\label{eq:KLD}
	d(\tau)=d(\pi_1||\pi_0)=\sqrt{2\text{KLD}(\pi_1||\pi_0)}
\end{equation*}
that can be interpreted as the distance from the flexible model $\pi_1$ to the base model $\pi_0$. 

	\item The PC prior is defined as an exponential distribution on the distance, 
\begin{equation*}
\pi(d(\tau)) = \lambda \exp(-\lambda d(\tau)),
\label{eq:pc}
\end{equation*}	
	with rate $\lambda>0$. 
	The PC prior for $\tau$ follows by a change of variable transformation, leading in this case to a type-2 Gumbel distribution with parameters $(1/2,\lambda)$:
\begin{equation}
\pi(\tau) = \frac{\lambda}{2}\tau^{-3/2} \exp(-\lambda \tau^{-1/2}), \quad \tau>0,\lambda>0.
\label{eq:gumbel2}
\end{equation}		
	\item The parameter $\lambda$ in (\ref{eq:gumbel2}) can be selected by the user based on his prior knowledge of $\tau$ (or an interpretable transformation of it such as the standard deviation). This can be expressed in an intuitive way with a probability statement, e.g. setting $U$ and $a$ such that $\mathbb{P}(1/\sqrt{\tau}>U)=a$, so that $\lambda = -\log(a)/U$. Knowledge on the marginal standard deviation can aid in choosing a sensible value for $U$; \cite{pcprior} provide a practical rule of thumb: once the precision $\tau$ is integrated out, the marginal standard deviation of the random effect for $a=0.01$ is about $0.31U$.
\end{enumerate}

\section{Partitioning the variance between main and interaction}
\label{sec:vp}

We present below the VP model assuming model (\ref{eq:model_classic}), but everything applies straightforwardly to model (\ref{eq:model_uns+struc}) as well; details about the VP version of model (\ref{eq:model_uns+struc}) can be found in Section \ref{sec:examples}.

From model (\ref{eq:model_classic}) it is hard to quantify the relative contribution of the main and interaction components to the total variance, because the involved precision parameters are not interpretable in terms of the variance explained by the associated components. Our proposal is to reparametrize model (\ref{eq:model_classic}) as a weighted sum of two IGMRFs representing the main and interaction components by means of a mixing parameter $\gamma \in [0,1]$. We include a further mixing parameter $\phi \in [0,1]$ to distribute the variance between the temporal and spatial main effects. Assume model (\ref{eq:model_classic}), the reparametrized version of the linear predictor is

\begin{equation}\label{eq:modelrep_grid}
\begin{array}{l}
\eta_{ij}  =  \alpha + \sqrt{\tau^{-1}}\left[\sqrt{1-\gamma} \left(\sqrt{1-\phi}\beta_{1_i} + \sqrt{\phi}\beta_{2_j}\right) + \sqrt{\gamma}\delta_{ij}\right], \\
\bm \beta_1    \sim N\left(\bm  0,  \tilde{\bm R}_1^-\right), \quad  \quad  \bm \beta_2    \sim  N\left(\bm  0, \tilde{\bm R}_2^-\right),  \quad   \quad \bm \delta     \sim  N\left(\bm  0,  \tilde{\bm R}_I^{-}\right),
\end{array}
\end{equation}
where $\tau>0$ is an overall precision parameter, $0<\gamma<1,0<\phi<1$.
We consider a RW1 or a RW2 prior on the temporal main effect $\bm\beta_1$ and an ICAR prior on the spatial main effect $\bm\beta_2$ as specified in Section 2. Note that, differently from model (\ref{eq:model_classic}), the precision structures $\tilde{\bm R}_1, \tilde{\bm R}_2$ have been scaled according to \cite{sorbye-2013}.  The interaction term $\bm \delta$ is modelled as a Kronecker product IGMRF; following  \cite{held:2000} we consider interaction types I, II, III, and IV as described in Table \ref{tab:types}. 

Model (\ref{eq:modelrep_grid}) includes the same vectors of random effects as model (\ref{eq:model_classic}), but in contrast to model (\ref{eq:model_classic}), we now have very intuitive hyperparameters: $\tau$ is the total precision, i.e. $\tau^{-1}$ is the total generalized variance,  and $\gamma$ and $\phi$ are two interpretable mixing parameters. The value of $\gamma$ can be interpreted as the proportion of total variance explained by the interaction $\bm \delta$. The variance explained by the main effects is therefore given by $\tau^{-1}(1-\gamma)$: $1-\phi$ quantifies the proportion of such variance which can be attributed to the temporal random effects $\bm \beta_1$, with $\phi$ being the proportion attributed to the spatial random effects $\bm \beta_2$. 

We need to assign priors to the overall precision parameter $\tau$ and the mixing parameters $\gamma$ and $\phi$. In the next section we focus on the prior for $\gamma$, and leave prior choice for the remaining parameters to Section \ref{sec:examples}.

\subsection{A Penalized Complexity prior for $\gamma$}
\label{sec:pc}

Our choice of a PC prior for $\gamma$ follows naturally from the model reparametrization in Eq. (\ref{eq:modelrep_grid}) and provides a way of eliciting the prior in a very intuitive way. Furthermore, it avoids overfitting by construction hence guaranteeing a parismonious model. Our PC prior for $\gamma$ (see Result~\ref{res1} below) 
is based on the assumption that the interaction model in  (\ref{eq:modelrep_grid}) shrinks to the main effects model $(\bm \beta_1+\bm \beta_2)$. 

\begin{result}
\label{res1}
Let us assume a model of the form (\ref{eq:modelrep_grid}), for all types of interaction in Table~\ref{tab:types}:
\begin{enumerate}
	\item The distance from the base model is
		$$	d(\gamma) \simeq  \sqrt{\gamma}, \quad  0<\gamma<1 $$
	\item The PC prior for $\gamma$ with base model $\gamma=0$ is
	\begin{eqnarray}
	\pi(\gamma)=\frac{\theta\exp(-\theta\sqrt{\gamma})}{2\sqrt{\gamma}(1-\exp(-\theta))} \quad \quad 0 < \gamma <  1, \theta>0.
\label{eq:pcprior_gamma}
\end{eqnarray}
\end{enumerate}
The proof can be found in Supplemental material \ref{appendix:proof1}-\ref{appendix:proof3}. 
\end{result}

\begin{figure}
\centerline{
\includegraphics[width=.5\textwidth]{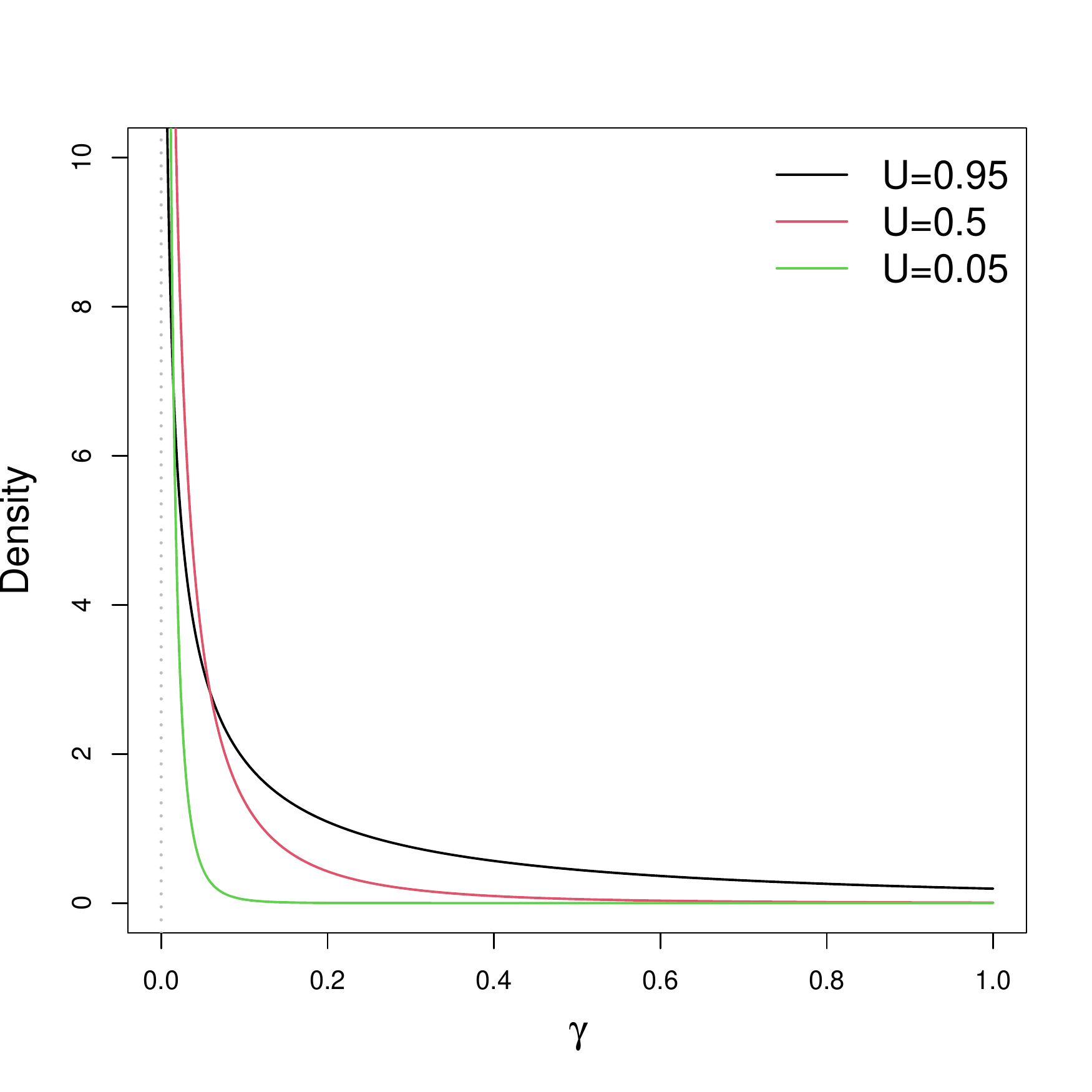}
\includegraphics[width=.5\textwidth]{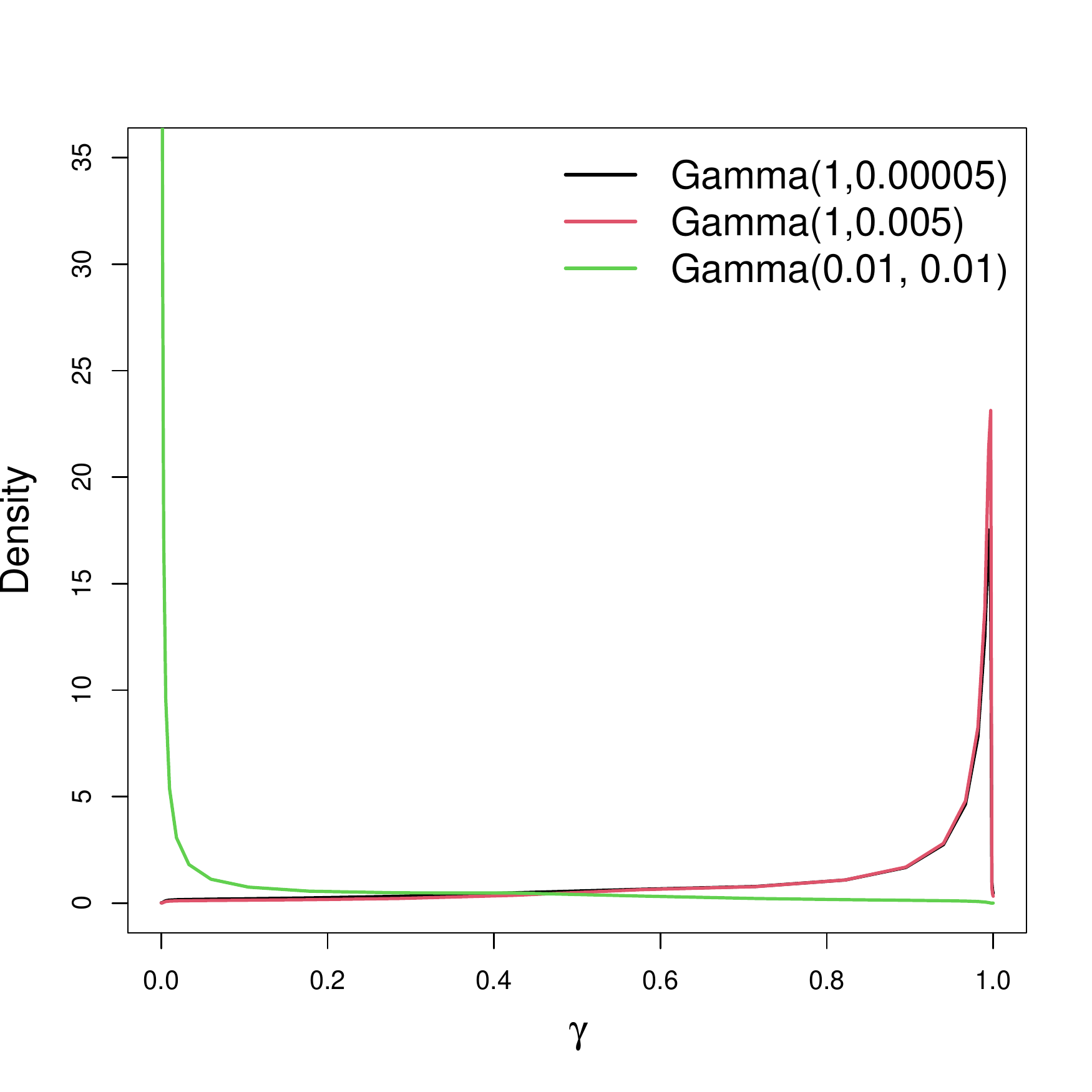}
}
\caption{Left panel: PC prior $\pi(\gamma)$ using $a=0.99$ and three different values for $U$. Right panel: implied prior on $\gamma$ when a Gamma prior is used on all three precision parameters $\tau_1,\tau_2,\tau_{12}$.}
\label{fig:pcprior_gamma}
\end{figure}

The scaling of the PC prior for $\gamma$, i.e. the choice of $\theta$ in Eq.~(\ref{eq:pcprior_gamma}), is done by defining the probability of a tail event on $\gamma$. The parameter $\theta$ controls the strength of penalisation for deviating from the base model; the higher the $\theta$ the greater the penalty.
We suggest setting $U$ and $a$ such that $\mathbb{P}(\gamma < U)=a$; this way $\theta$ is obtained by numerically solving:
\[
	\frac{1-\exp(-\theta\sqrt{U})}{1-\exp(-\theta)}=a, \quad \quad a >\sqrt{U}.
\]
Note that it is not possible to assign equal weight to the main and interaction terms in the model, i.e. $U=a=0.5$ because of the constraint $a >\sqrt{U}$.  However, we can always encode a fair amount of uncertainty into the prior by choosing $a$ close to 1 and large values of $U$. In the left panel of Figure \ref{fig:pcprior_gamma}, $\theta$ is obtained using $a=0.99$ and three different values for $U$. A large $U$ allows for more flexibility as the corresponding density curve decreases steadly towards zero as $\gamma$ increases, while for a small value of $U$ the density curve drops towards zero quite sharply, strongly penalizing any deviation from the base model. For comparison, the right panel in Figure\ref{fig:pcprior_gamma} shows the prior on $\gamma$ that corresponds to using a Gamma prior on all three precision parameters in model (\ref{eq:model_classic}) for three different parameter choices. The figure illustrates how the resulting prior on $\gamma$ depends strongly on the chosen values for the Gamma parameters, going from one extreme to the other in terms of prior weight on the base model.

Results from a simulation study reported in Supplemental material \ref{appendix_sim} indicate that the posterior mean estimates of $\gamma$ are reasonably close to the true value under different scenarios. We have observed stable results for several choices of $U$, unless one defines on purpose an unflexible prior, where most of the probability mass is placed near the base model (e.g. when adopting $a=0.99$ and a small $U=0.05$). Results are comparable to those obtained using a Uniform prior on $\gamma$ unless there is no interaction (i.e. $\gamma=0$), in which case the uniform leads to greater bias when the population at risk is small.


\section{Examples}\label{sec:examples}


As introduced in Section \ref{sec:spmodels}, model (\ref{eq:model_uns+struc}) is more common in practice and indeed it is the model adopted in this section for both case-studies.
In the case of structured and unstructured main effects, another set of parameters $\psi_1$ and $\psi_2$ can be included to further distribute the variance, so that model (\ref{eq:modelrep_grid}) becomes:
\begin{multline}
\eta_{ij}  =  \alpha + \sqrt{\tau^{-1}}\Bigl(\sqrt{1-\gamma} \left(\sqrt{1-
\phi}\left(\sqrt{1-\psi_1}\beta_{1_i}+\sqrt{\psi_1}\epsilon_{1_i}\right) + 
  \right.\\
  \left.
  \sqrt{\phi}\left(\sqrt{1-\psi_2}\beta_{2_j} +\sqrt{\psi_2}\epsilon_{2_j} \right) \right) 
  + \sqrt{\gamma}\delta_{ij} \Bigr), 
\label{eq:modelrep_grid_unst+struct}
\end{multline}
where $\tau>0$, $0<\gamma<1,0<\phi<1, 0<\psi_1<1, 0<\psi_2<1$, $\bm \epsilon_1\sim N\left(\bm  0, \bm I_{n_1}\right)$, $\bm \epsilon_2\sim N\left(\bm  0, \bm I_{n_2}\right)$ and $\bm \beta_1$, $\bm \beta_2$ and $\bm \delta$ as in model (\ref{eq:modelrep_grid}). Result~\ref{res1} about the PC for $\gamma$ still holds; see Supplemental material  \ref{appendix:proof3}.

Note that the parameters in model (\ref{eq:modelrep_grid_unst+struct}) are identifiable as the model is just a reparametrized version of the classic space-time interaction model (\ref{eq:model_uns+struc}), where each random effect has its corresponding precision parameter. The number of parameters is exactly the same; in fact, it can be shown that there is a one-to-one mapping between the parameters of both versions of the model.
As in model (\ref{eq:model_classic}), appropriate constraints need to be imposed to ensure identifiability of the terms in (\ref{eq:modelrep_grid_unst+struct}). The constraints on the interaction term are summarized on Table \ref{tab:types}, while on the temporal and spatial structured main effects it is enough to impose a sum to zero constraint.

In the next two examples, we use the PC prior in Eq. (\ref{eq:gumbel2}) for $\tau$ and the PC prior in Eq. (\ref{eq:pcprior_gamma}) for $\gamma$.
Regarding $\phi$, $\psi_1$ and $\psi_2$, we simply choose a uniform on (0,1) as a prior for each of them, but other choices are possible. In fact, a PC prior could also be used for $\phi$ following the work by \cite{Fuglstad:2020}, 
who also consider the use of a Dirichlet prior where the base model attributes equal weights to each component, thus expressing ignorance about how the variance is distributed. Similarly, one could use a PC prior on each $\psi_1$ and $\psi_2$ as in \cite{riebler-2016}, considering as base model the absence of structured effects. 

All the VP models presented in the next two examples were run using \RINLA \citep{rue-inla}, see code in Supplemental material \ref{appendix:Rcode}. 

\subsection{Ohio lung cancer}

We illustrate our model using the Ohio lung cancer data  \citep{held:2000, held:1998, waller:1997} which is available at \url{http://www.biostat.umn.edu/~brad/data2.html}. These data report yearly counts of lung cancer deaths for white males from 1968 to 1988, in the 88 counties of Ohio. Figure~\ref{fig:ohio_data} left panel displays the time series of mortality rate for all counties. Our aim is not to find the best model for this data, but to show what our approach can add in terms of interpretability of the results compared to a classical analysis as performed in \cite{held:2000}. 

\subsubsection{Model}
Let $y_{ij}$ be the number of deaths at time $i=1,\ldots,21$  in county $j=1,\ldots,88$ and pop$_j$ be the population at risk in county $j$, we consider the model proposed in \cite{held:2000} assuming structured and unstructured effects for both space and time main effects, plus a space-time interaction term. The classical parameterization in  \cite{held:2000} follows,
\begin{eqnarray}
y_{ij} &\sim & \text{Bin}(\text{pop}_j, \exp(\eta_{ij})/\exp(1+\eta_{ij})),\nonumber\\
\eta_{ij} & = &\alpha + \underbrace{\beta_{1_i}+\epsilon_{1_i} + \beta_{2_j}+\epsilon_{2_j}}_{\texttt{main}} + \underbrace{\delta_{ij}}_{\texttt{int}},
\label{eq:model_ohio_cl}
\end{eqnarray}
where the main effects are modelled as:
\begin{eqnarray*}
\bm \epsilon_1\sim N\left(\bm  0, \tau_{\epsilon_1}^{-1}\bm I_{n_1}\right); \quad
\bm \epsilon_2\sim N\left(\bm  0, \tau_{\epsilon_2}^{-1}\bm I_{n_2}\right); \\
\bm \beta_1 \sim N\left(\bm  0,  \tau_{1}^{-1} {\bm R}_{1}^{-}\right); \quad
\bm \beta_2\sim N\left(\bm  0,  \tau_{2}^{-1} {\bm R}_{2}^{-}\right). 
\end{eqnarray*}
where ${\bm  R}_{1}$ and ${\bm  R}_{2}$ are the unscaled structure matrices of a RW1 (for time) and an ICAR (for space). The space-time interaction is modelled by a Kronecker product IGMRF built on the precision matrices of the structured components $\bm \beta_1$ and $\bm \beta_2$. 
All the interaction types in Table~\ref{tab:types} are considered in the following analysis.  This model would require priors for the precision hyperparameters $\tau_{\epsilon_1}, \tau_{\epsilon_2}, \tau_{1}, \tau_{2}, \tau_{12}$. 

Instead of working with the above model, we assume the VP model in Eq.~(\ref{eq:modelrep_grid_unst+struct}),  with scaled structure matrices, with the priors for $\tau, \gamma, \phi, \psi_1, \psi_2$ stated in Section \ref{sec:examples}. For $\tau$'s PC prior we set $a=0.01$ and $U=1/0.31$ following the rule of thumb described in Section \ref{sec-prior precision}. The PC prior for $\gamma$  is scaled by imposing $U=0.5, a=0.99$; results (not shown here) were stable for varying $U=\{0.05, 0.5, 0.95\}$.

\subsubsection{Results}

Table~\ref{tab:ohio_model_selection2} reports various model selection criteria for the VP model, for interaction types I, II, III and IV, namely DIC \citep{dic}, WAIC \citep{waic}, leave-one-out log score (LOOLS), computed as $-\sum_{i=1}^n\log\pi(y_i|y_{-i})$, and the log-marginal likelihood (logMLIK), $\pi(y|\mathcal{M})$, which quantifies the likelihood of the data $y$ under a given model $\mathcal{M}$. 
PC priors enhance the marginal likelihood as a simple and effective tool for \emph{fair} model comparison, when the compared models have similar structure and only differ on a particular component \citep{sorbye-fractional-2018, Ventrucci:2020}.  Assume $\mathcal{M}_1$ and $\mathcal{M}_2$ are the interaction type I and II models, respectively: these models are the same except for a different type of interaction. The Bayes factor \citep{kass1995bayes} is defined as 
\begin{equation}
 K=\frac{\pi(y|\mathcal{M}_1)}{\pi(y|\mathcal{M}_2)}=\frac{\pi(\mathcal{M}_1|y)}{\pi(\mathcal{M}_2|y)}\frac{\pi(\mathcal{M}_2)}{\pi(\mathcal{M}_1)}.
 \label{eq:bf}
\end{equation}
The scale parameter $\theta$ of the PC prior for $\gamma$, which controls the decay rate from the base model (the model with no interaction), has to be chosen for both $\mathcal{M}_1$ and $\mathcal{M}_2$. We can handle this choice conveniently by setting the same $\theta$ for $\mathcal{M}_1$ and $\mathcal{M}_2$, which implies that $\pi(\mathcal{M}_2)/\pi(\mathcal{M}_1)$ in Eq.(\ref{eq:bf}) cancels out and the Bayes factor turns out to be the ratio of the posterior odds. For all the four interaction models in Table~\ref{tab:ohio_model_selection2} we follow this strategy and set the same decay rate for the PC prior on $\gamma$. The advantage is that the \emph{a priori} contribution of the interaction to the total (generalized) variance is the same, no matter what interaction type is assumed; this is a desirable feature when having to choose among different types of interaction models the one that best fits the data. Therefore, we suggest comparing the logMLIK values for model choice purposes here; furthermore, DIC is known to favour complexity in models with many random effects \citep{riebler:2010}.

From Table~\ref{tab:ohio_model_selection2} we see that DIC, WAIC and LOOLS point to type II (followed by type IV) as the best model; a similar conclusion based on DIC was found in  \cite{held:2000}. Interestingly, logMLIK is largest for type I which indicates that the model with main effects plus an individual-level random effects capturing unstructured variation may be a better description of the Ohio data.  

\begin{table}[ht]
\small\sf
\centering
\caption{Model comparison criteria (computed using \RINLA) for the VP model, under the four interaction types.} 
\begin{tabular}{lllll}
  \hline
{\bf interaction type}    & logMLIK & DIC (deviance; $p_D$) & WAIC  & LOOLS  \\ 
\hline
 I  &  -5623.52 & 10945.23 (10722.68; 222.55) & 10957.75 & 5489.75\\ 
 II  &  -6759.84 & 10916.00 (10739.09; 176.91) & 10931.3   & 5469.43\\
 III  &  -6098.89 & 10957.86 (10792.68; 165.18) & 10980.99   & 5496.28\\
 IV  &  -7200.13 & 10919.23 (10755.06; 164.17) & 10934.82   & 5470.89\\
   \hline
\end{tabular}
\label{tab:ohio_model_selection2}
\end{table}

\begin{figure}
\centerline{\includegraphics[width=0.5\textwidth]{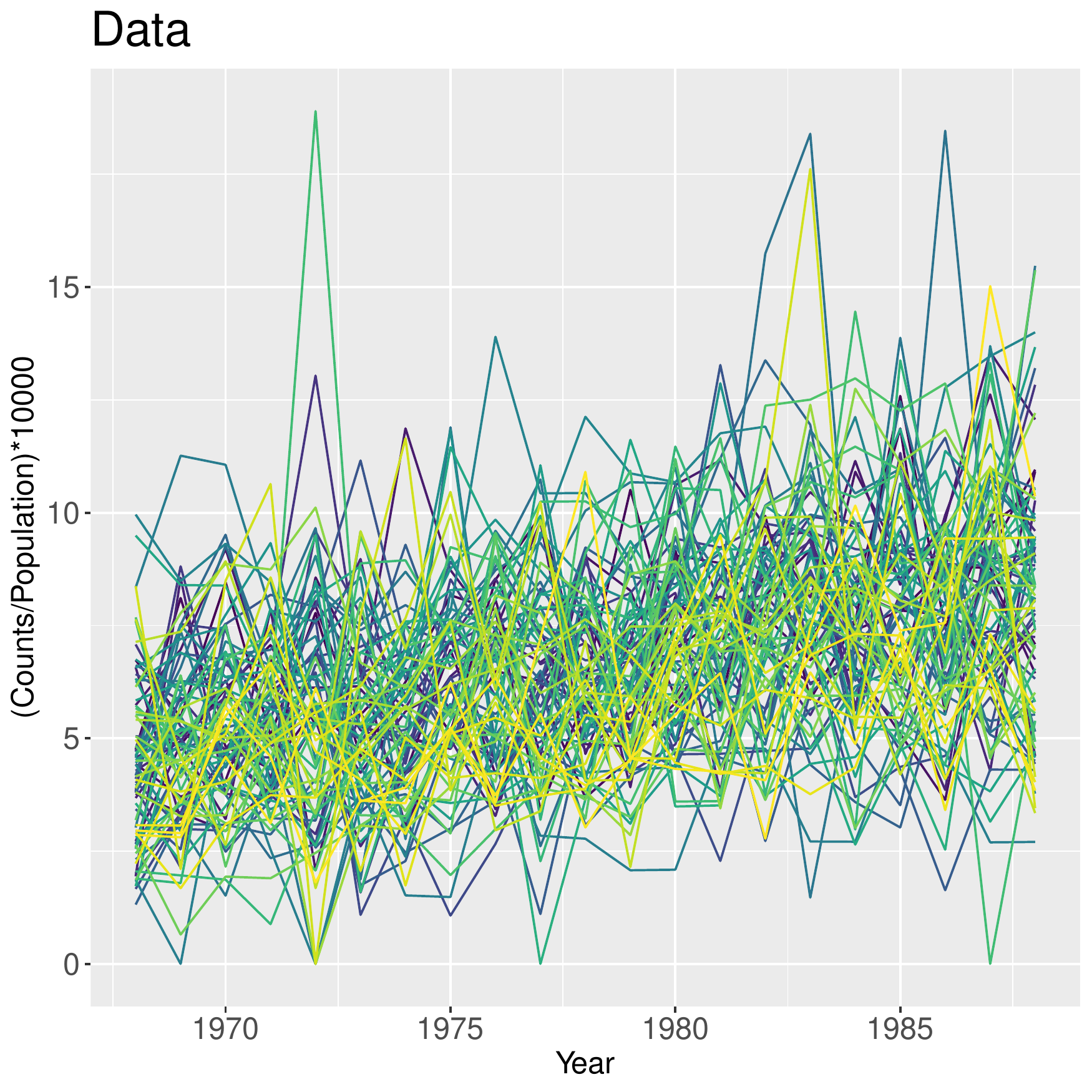}
\includegraphics[width=0.5\textwidth]{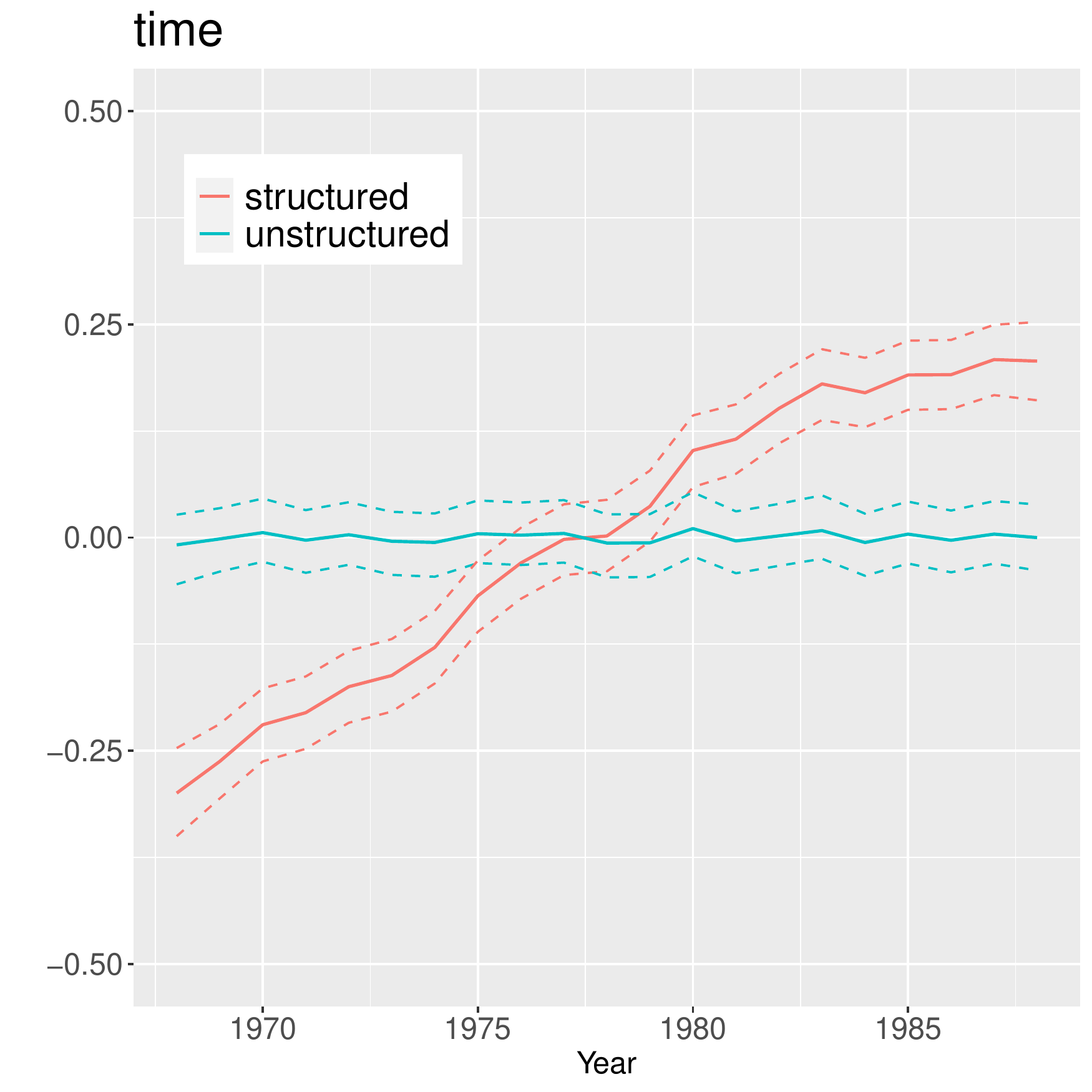}}
\centerline{
\includegraphics[width=0.5\textwidth]{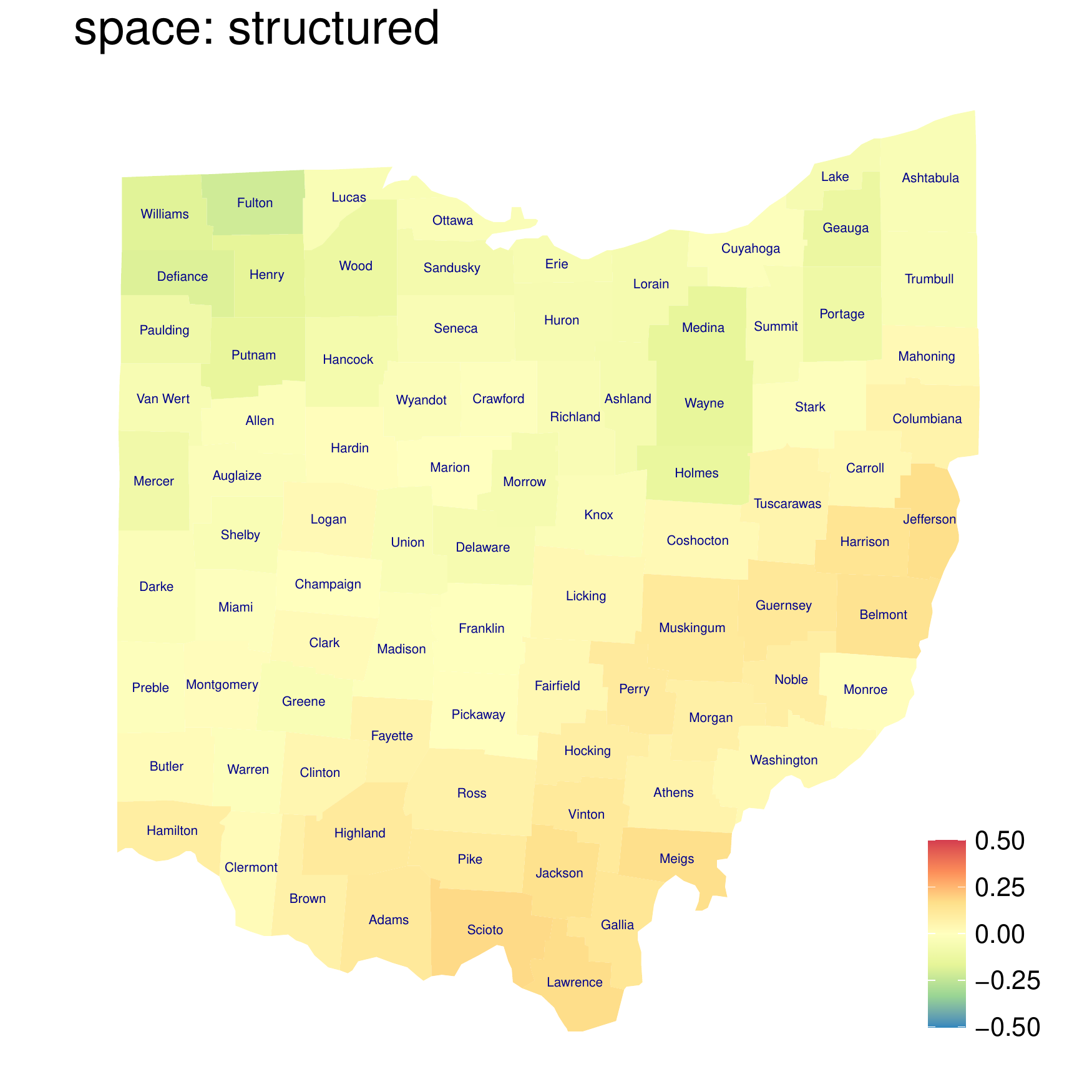}
\includegraphics[width=0.5\textwidth]{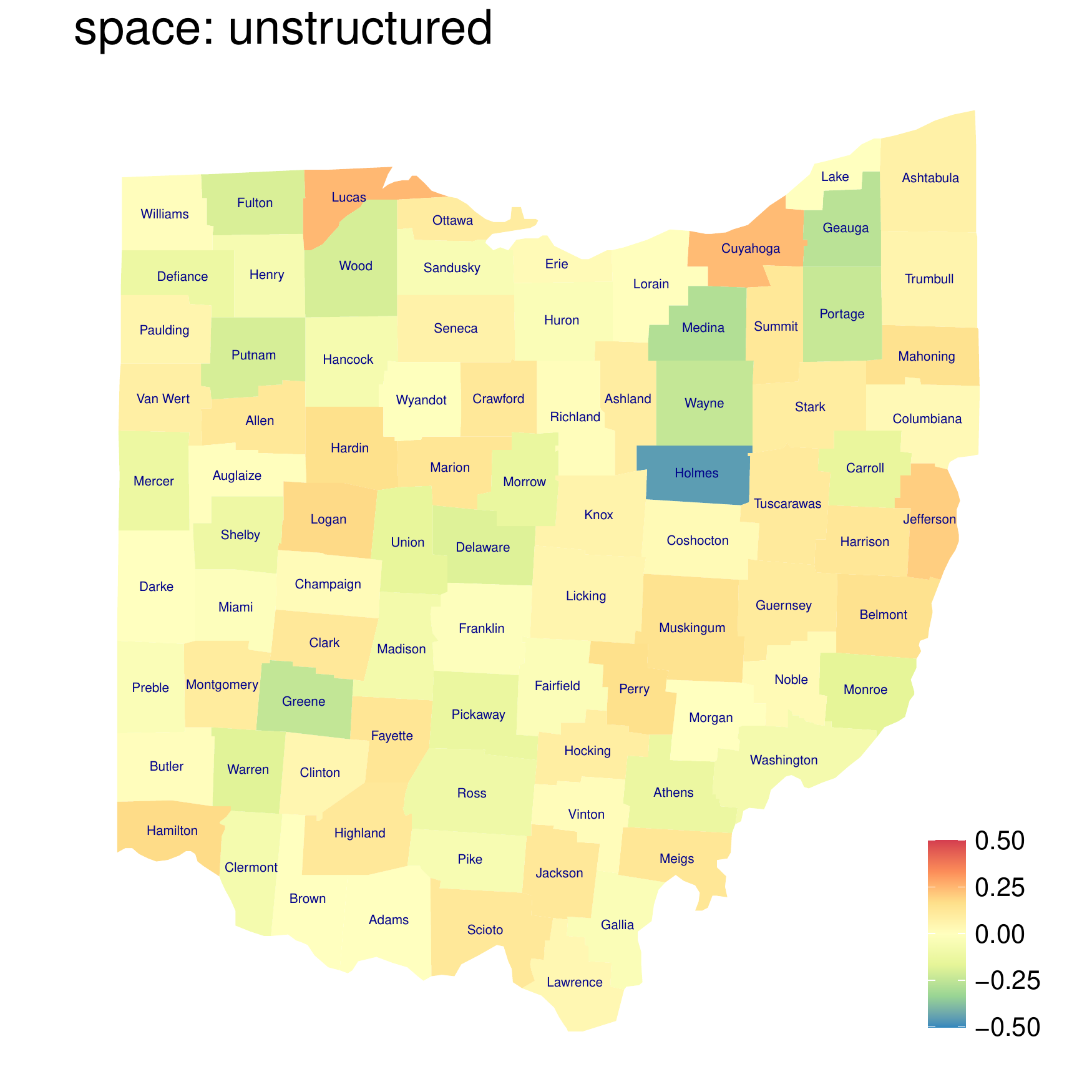}
}
\caption{Top left panel: time series of lung cancer (white males) disease rates per 10000 population at risk, for the 88 counties in the Ohio dataset. Top right panel:  temporally structured and temporally unstructured components for type I interaction model, in the scale of the linear predictor. Bottom left and right panels show, respectively, the spatially structured and unstructured components for type I interaction model, in the scale of the linear predictor.}
\label{fig:ohio_data}
\end{figure}

In order to show now the gain of using our approach compared to a classical analysis we start by discussing some plots obtained for type I interaction about the main effects. The top right panel in Figure~\ref{fig:ohio_data} displays the estimated main temporal effect, in the scale of the linear predictor, decomposed into its structured and unstructured (iid) components. The unstructured effects looks very flat compared to the structured ones which is probably responsible for most of the temporal variation in the relative risk. The relative risk increases roughly linearly in time, with a less steep increase towards the end of the time window.  The bottom panels in Figure~\ref{fig:ohio_data} display the estimated structured (left) and iid (right) spatial effects in the scale of the linear predictor. Here the unstructured effect shows larger variability than the structured one which shows a very smooth spatial gradient from north-west to south-east. A visual inspection of this sort, also possible when the classical model is used, gives useful insights into the spatial and temporal patterns in the data. However, it does not allow proper quantification of the variance attributable to the various sources (main, interaction, spatial and temporal effects, etc), while this quantification is readily available from our VP model.

Table~\ref{tab:ohio:vptable} reports the mean (with $2.5$ and $97.5$ quantiles between brackets) of the posterior distribution of the mixing parameters $\gamma, \phi, \psi_1, \psi_2$, from which we can understand quantitatively the contribution of the main sources of variation. In particular, rows 1 and 2 report the total variance partitioned into interaction versus main effects; rows 3 and 4 quantify how the variance attributable to the main effects is partitioned into space and time; rows 5 to 6 and 7 to 8 give the variance partitioning for the structured versus iid for space and time, respectively. The main findings about the variation of the spatio-temporal mortality risk pattern are as follows. First, the estimated contribution of the interaction is about $4.8\%$, which means that the main effects are responsible for most of the variability in mortality risk with the interaction playing a minor role in describing this data. This is reasonable for non-infectious diseases such a cancer \citep{abellan:2008}. Second, space is responsible for about $87.5\%$ of the variation in risk explained by the main effects, which is hard to grasp from only looking at the plot of the main temporal and spatial effects in Figure~\ref{fig:ohio_data}. This result highlights the fact that lung cancer in Ohio had, in the period of time considered, larger variability over space than time which could be informative for policy makers and epidemiologists and may contribute to generate hypothesis on the role played by possible environmental risk factors in the region.
Third, within the main spatial and temporal effects, the structured component is predominant for time, while the iid component is predominant for space. However, these estimated contributions, and in particular the latter, are affected by greater uncertainty than the previous estimates, indicating that the data are less informative about the posterior for $\psi_1$ and $\psi_2$ than they are for $\gamma$ and $\phi$. These findings are stable across the different types of interactions (see Supplemental material \ref{app:more_about_ohio}).

\begin{table}[ht]
\small\sf
\centering
\caption{Variance partitioning table for Ohio lung cancer, type I interaction. The column named \emph{contribution} reports the posterior mean of the hyper-parameters displayed in the column named \emph{estimator}, with 0.025 and 0.975 posterior quantiles between brackets. All values are in a $(0,1)$ interval and indicate the proportional contribution of the model component  \emph{level 2} to the variance explained by the model component \emph{level 1}. }
\begin{tabular}{|ll|lc|}
\hline
\multicolumn{2}{|c|}{\bf Model component} &  \multicolumn{2}{|c|}{\bf Variance Partitioning}   \\
{\bf level 1}  &{\bf level 2} & {\bf estimator} & {\bf contribution} \\
  \hline
\texttt{main+int} &\texttt{main }  & $1-\hat\gamma$ &  0.952 (0.913, 0.979)  \\
&\texttt{int}      & $\hat\gamma$ & 0.048 (0.021, 0.087)   \\
  \hline
\texttt{main}&\texttt{space} & $\hat\phi$ & 0.875 (0.765, 0.946)  \\
&\texttt{time}    & $1-\hat\phi$ & 0.125 (0.054, 0.235)   \\
  \hline
\texttt{time}&\texttt{iid} & $\hat\psi_1$ & 0.069 (0.010, 0.229)  \\
&\texttt{str} & $1-\hat\psi_1$ & 0.931 (0.771, 0.990)    \\
  \hline
\texttt{space}&\texttt{iid} & $\hat\psi_2$ & 0.658 (0.273, 0.925)  \\
&\texttt{str} & $1-\hat\psi_2$ & 0.342 (0.075, 0.727)   \\
 \hline 
\end{tabular}
\label{tab:ohio:vptable}
\end{table}

\subsection{Covid-19 in Italy}
\label{sec:examples_covid}
We use the VP model to study Covid-19 incidence variations across space and time in Italy. Data cover all of the 107 Italian provinces and span a period of time that goes from the onset of the pandemic on 24$^{th}$ February 2020  to late July 2021 for a total of 70 weeks; the full dataset is made available by the Italian National Institute of Health through the website \url{https://github.com/pcm-dpc/COVID-19}. Data are originally available on a daily basis, but we aggregate them by week to smooth out artefactual patterns mainly due to delays in reporting new cases. The final dataset consists of weekly counts of new Covid-19 cases $y_{ij}$, for week $i=1, ..70$ and province $j=1,\ldots,107$, and the population at risk for each province pop$_j$.

\begin{figure}
\includegraphics[width=0.31\textwidth]{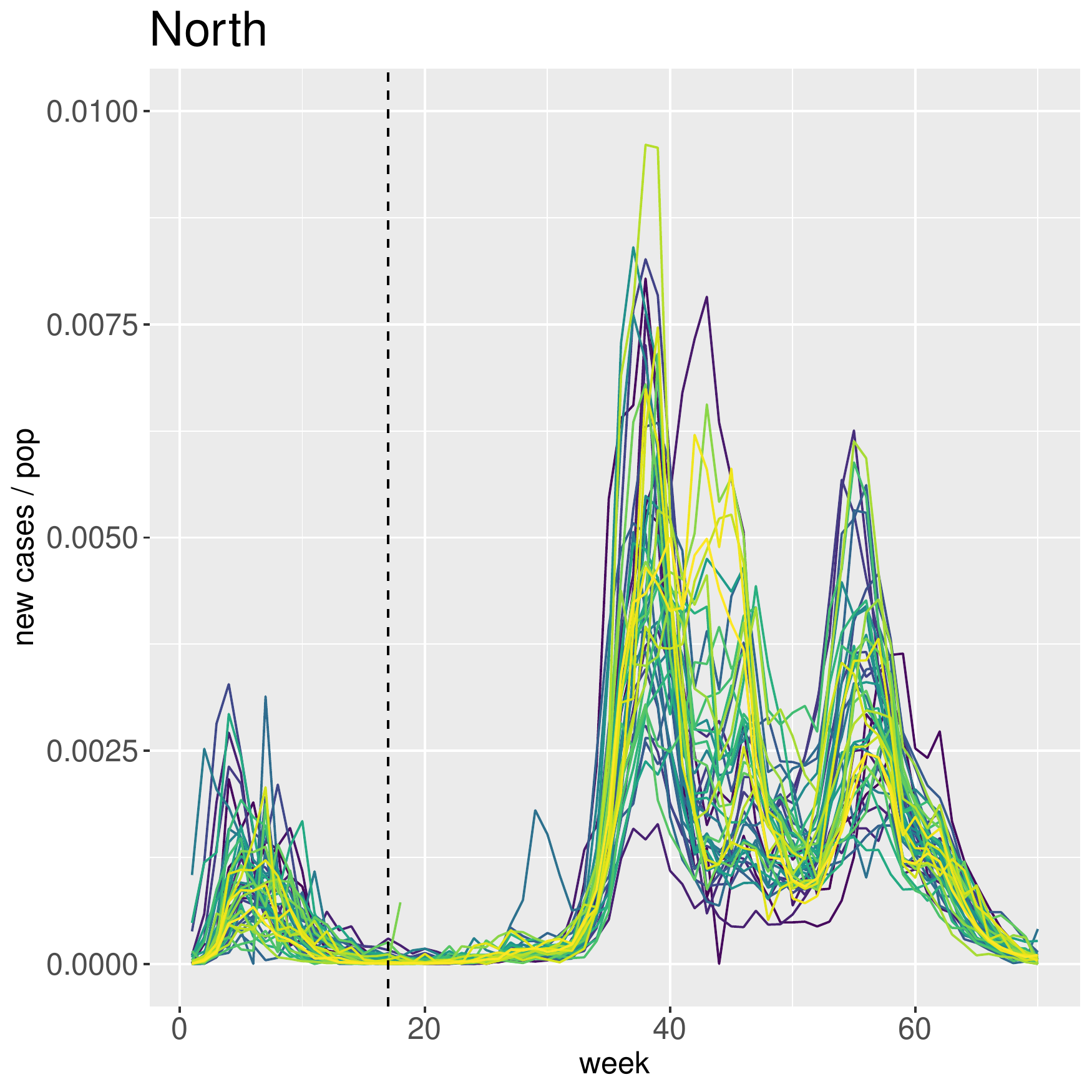}
\includegraphics[width=0.31\textwidth]{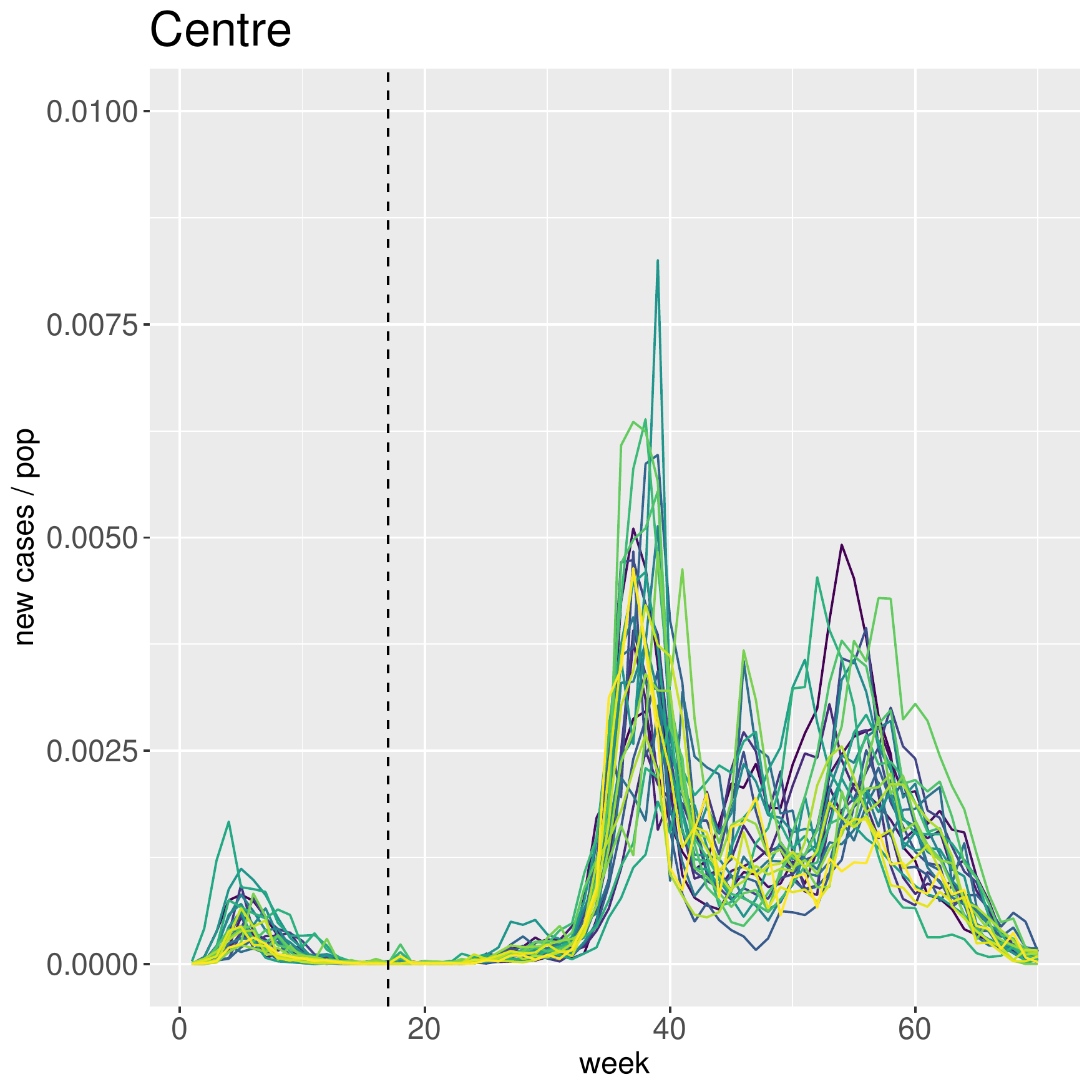}
\includegraphics[width=0.31\textwidth]{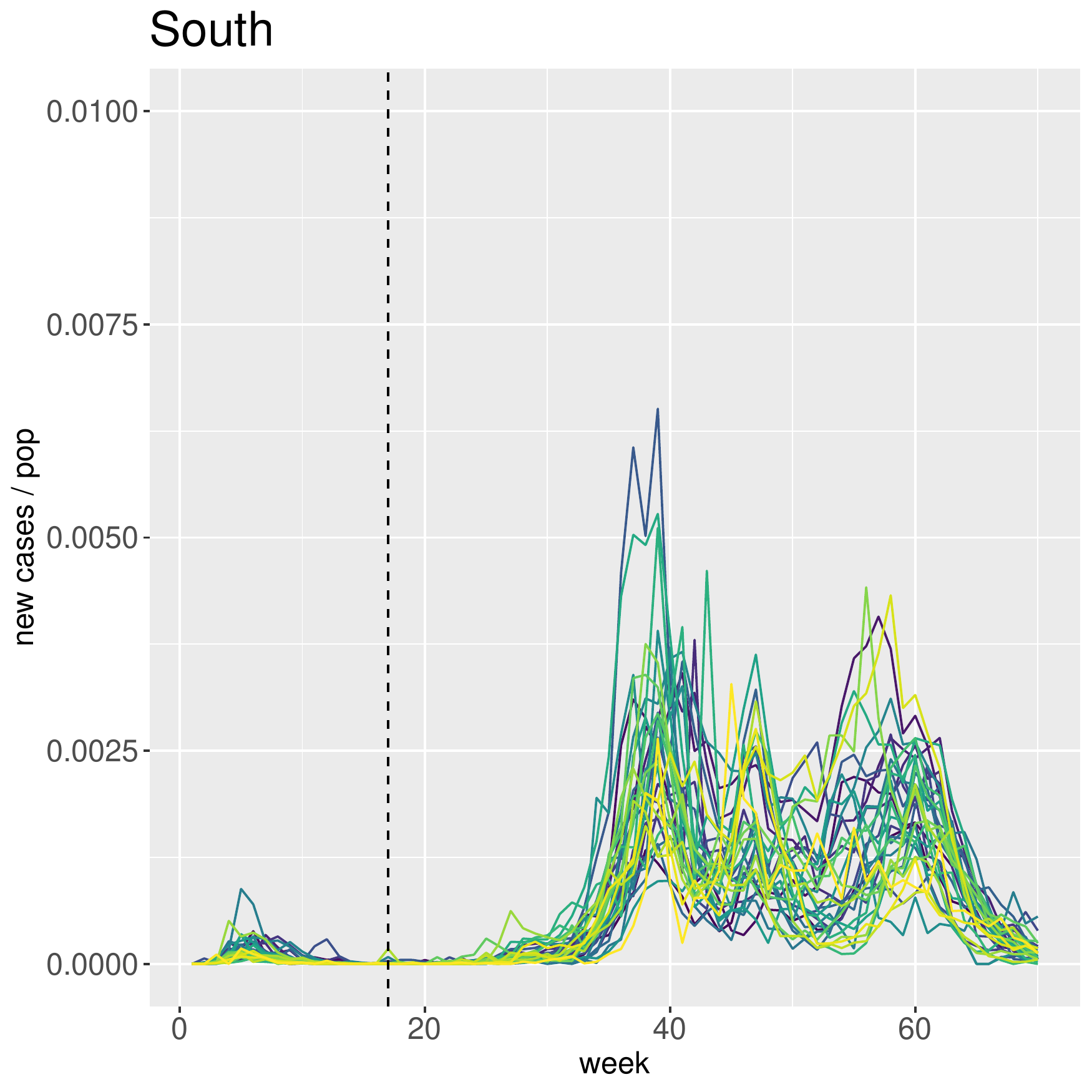}
\caption{Weekly Covid-19 incidence rates in the North (left panel), Centre (central panel) and South (left panel) of Italy. The vertical dashed line marks the separation between the first (W1) and second (W2) wave.}
\label{fig:covid_data}
\end{figure}

Our goal is to analyze the sources of variation in Covid-19 incidence rates in a scale between 0 and 1, which is easy to interpret and visualize and provides a clear idea of the contribution of each source. We follow the ideas in \cite{Picado:2007} in considering the interaction term as a measure of local heterogeneity, which can be seen as an indirect measure of how effective the control measures are. Hence a primary interest is to quantify the contribution of the interaction to the total variability, i.e. the posterior estimate for $\gamma$. Our second interest is to investigate changes in the estimated local heterogeneity across geographical macro-regions and time windows. We run two analysis: in the first one we fit the VP to the full dataset, in the second one we run the same VP model to separate subsets of the data which are constructed using combinations of geographical area, with levels north (N), centre (C) and south (S), and pandemic wave, with levels W1 and W2. The first wave (W1) covers the first 18 weeks and roughly indicates the national lock down period, while the second wave (W2) covers the rest of the time frame and indicates the period where restriction measures were set at a regional level. Data are displayed in Figure \ref{fig:covid_data}.

\subsubsection{Model}
We consider the binomial model in Eq. (\ref{eq:model_ohio_cl}), where structured and unstructured random effects are specified for both space and time as main effects. We model the temporally structured effects as a RW1 (as we do not anticipate smoothness) and the spatially structured effects as an ICAR, and assume a type IV space-time interaction to capture potential complex space-time patterns which are not explained by the main space and time components. In this particular example, the spatial main effect may reflect differences on the public health policy strategies adopted in each area (for example different testing rates across provinces). Again, we avoid the classic parametrization and take advantage of the VP approach described in (\ref{eq:modelrep_grid_unst+struct}). Doing so, we can elicit the prior easily and describe the various sources of variability in the data in an intuitive way in terms of the mixing parameters $\gamma, \phi, \psi_1, \psi_2$.

Available information on the nature of the disease can be used to aid in parameter choice for the PC priors on $\gamma$ and $\tau$.
Since we know that we are dealing with a contagious disease which evolves over time possibly in a different manner across provinces we anticipate a relevant contribution of the interaction term. Thus we choose $\theta$ in Eq. (\ref{eq:pcprior_gamma}) by setting $U=0.95, a=0.99$, which implies a large probability that $\gamma < 0.95$. In choosing the scale parameter $\lambda$ of the PC prior for $\tau$ in Eq. (\ref{eq:gumbel2}) we consider the scale of the logit transformed incidence rates and use the rule of thumb described in \cite{pcprior}, imposing a marginal standard deviation equal to 2 for the incidence rates in the linear predictor (logit) scale. 

\subsubsection{Results}

\begin{figure}
\centerline{
\includegraphics[width=0.33\textwidth]{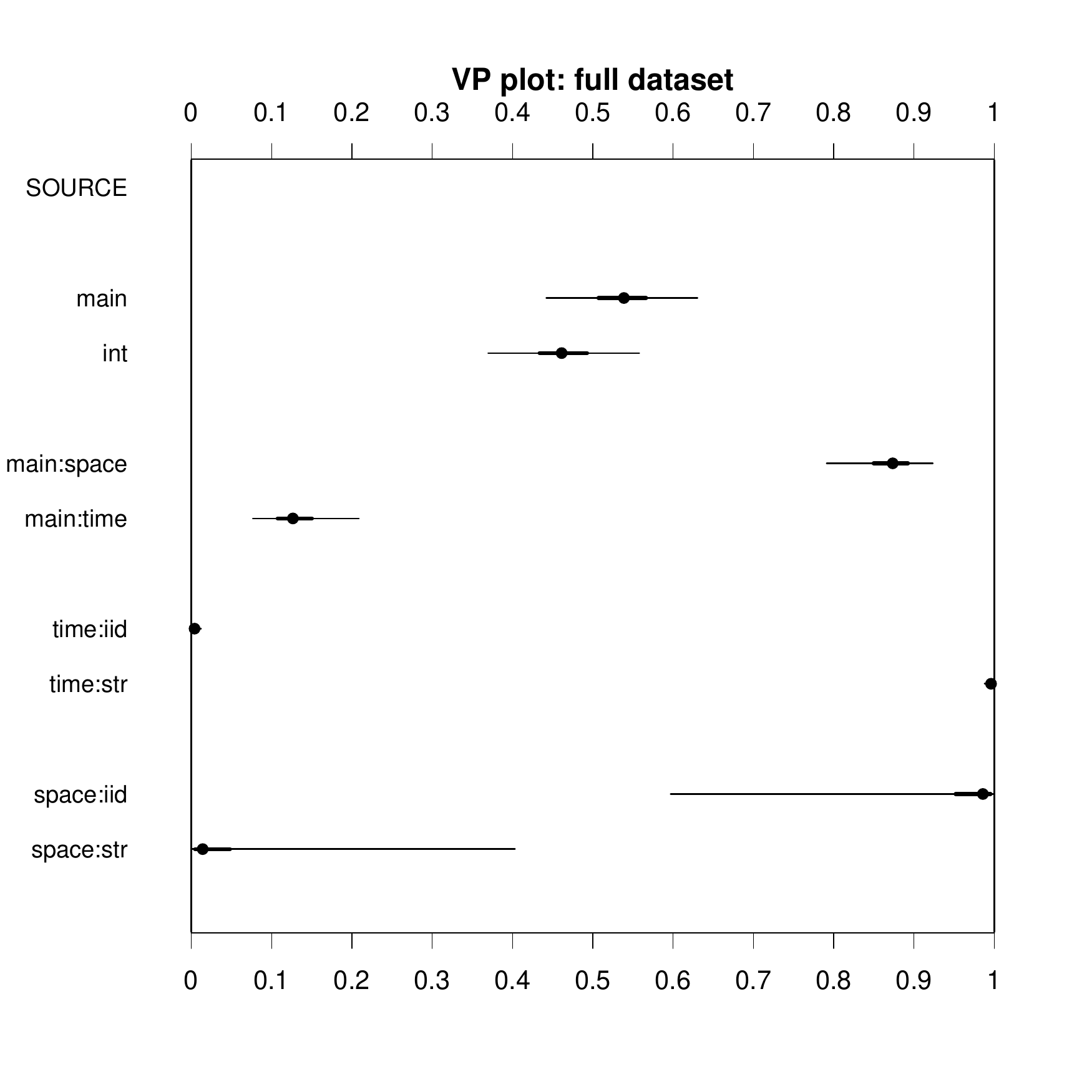}
\includegraphics[width=0.33\textwidth]{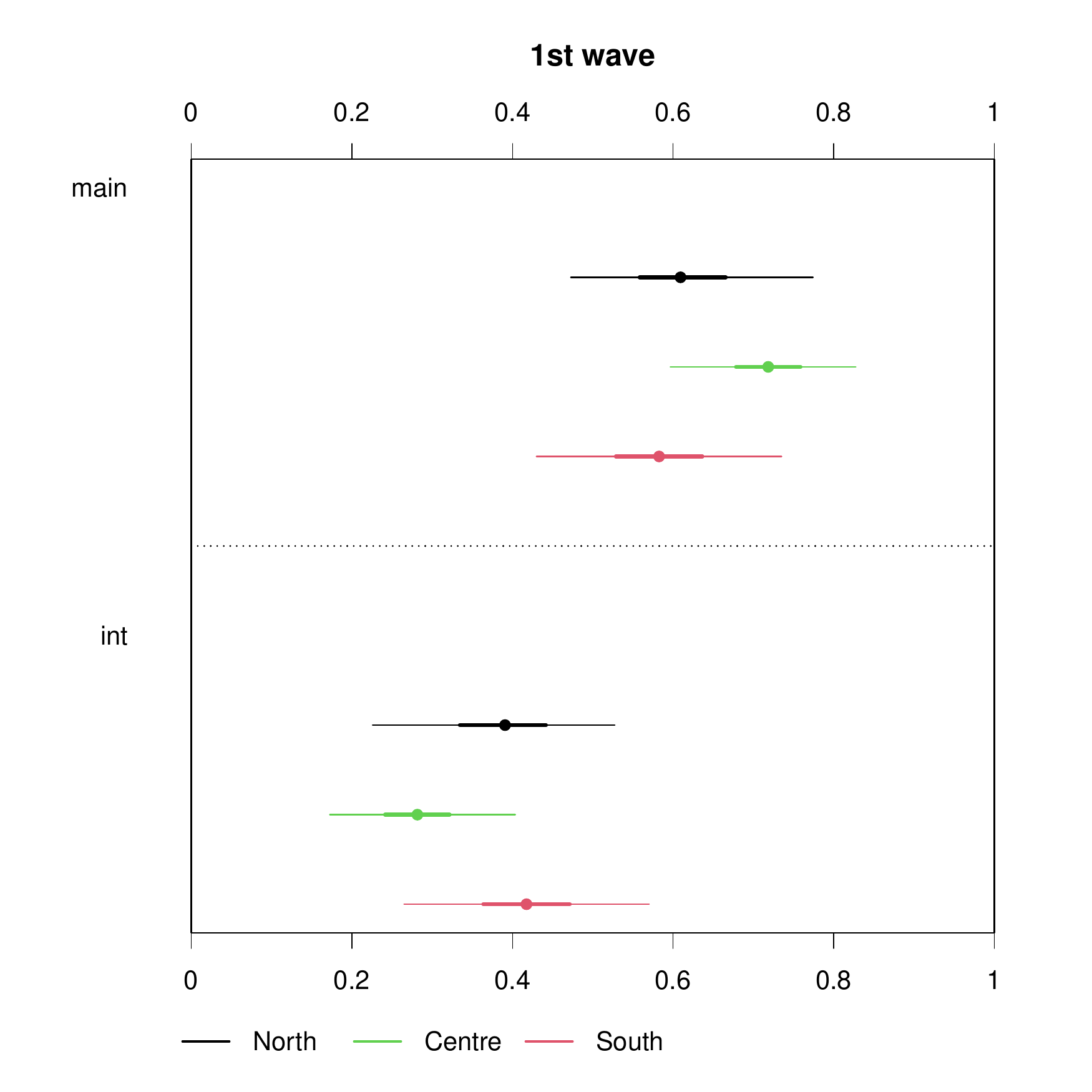}
\includegraphics[width=0.33\textwidth]{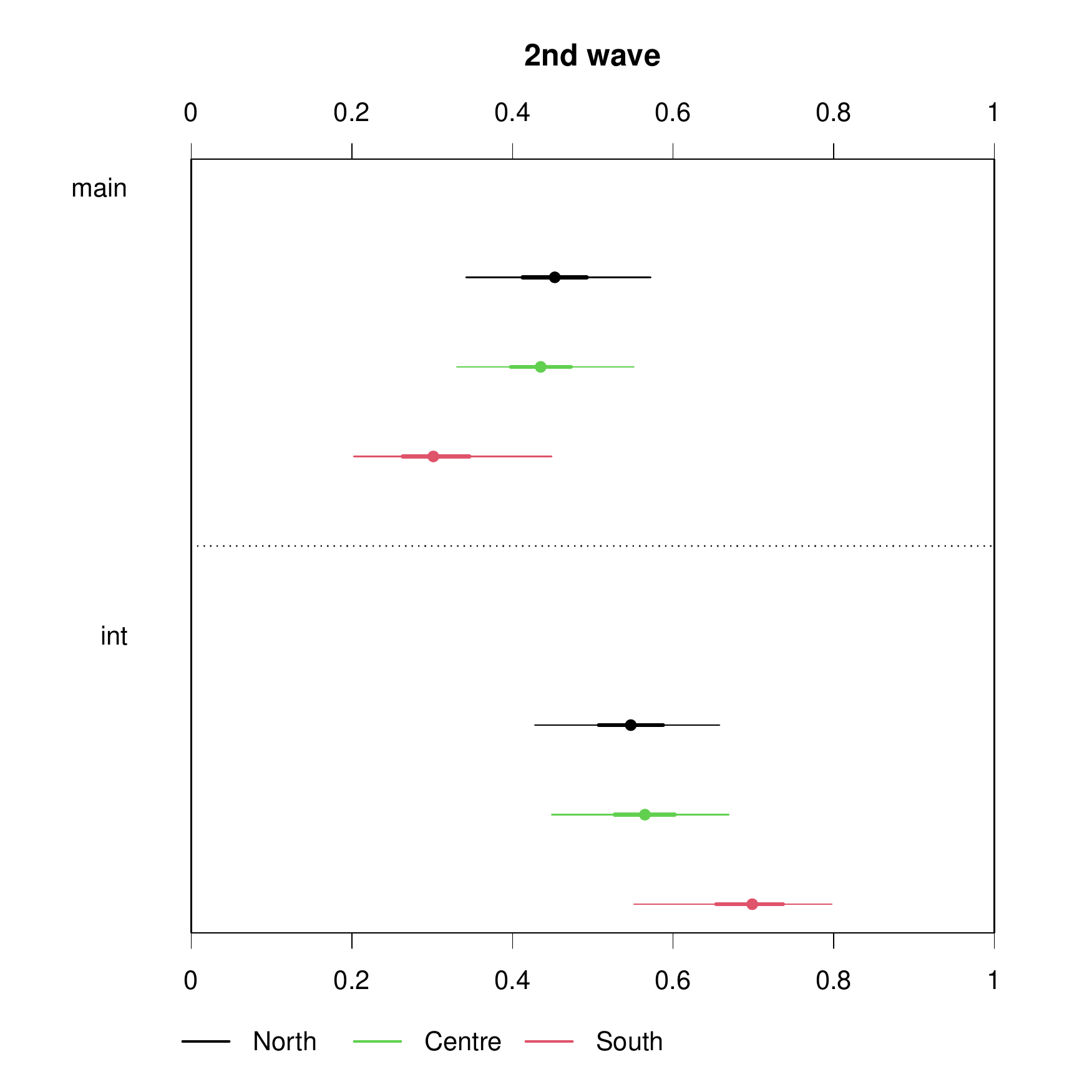}}
\caption{Variance partitioning plot for Covid-19 full dataset (left panel), first wave (middle panel) and second wave (right panel). The middle and right panels allow comparison across northern (black), central (green) and southern (red) areas in Italy.}
\label{fig:covid_vpplot_compare}
\end{figure}

The left panel in Figure~\ref{fig:covid_vpplot_compare} reports the variance partitioning plot for the full Covid-19 dataset; this plot is just a graphical version of the variance partitioning table that was presented in Table~\ref{tab:ohio:vptable} for the Ohio lung cancer data. This plot resembles the graphs in \cite{Gelman:anova} that summarize anova results in terms of estimated standard deviation for each bunch of random effects in the model. Our variance partitioning plot follows the same idea but represents the contribution of each source in a scale $(0,1)$. The main effects acount for the greatest proportion of the total variation. Within the main effects, the variability in incidence rates is mostly driven by the spatial component, in particular by the unstructured part of it (although the corresponding posterior estimates are highly uncertain), while for the temporal part is the structured component that explains most of the variability.

The middle and right panels in Figure~\ref{fig:covid_vpplot_compare} report the variance partitioning plot for the models fitted to different subsets of the full dataset to investigate whether the spatio-temporal pattern in Covid-19 cases is consistent or not across geographical areas (N, C, S) and pandemic waves (W1, W2). It is interesting to see that the impact of the interaction term is greater in the second wave than in the first one for all three areas, suggesting greater local heterogeneity during the second wave. This could reflect the fact that restricition measures went from being national in the first wave to being regional in the second one, so we expect greater heterogeneity over space during the latter. Within the first wave, the main effects are responsible for a greater proportion of variation in all three areas, but that attributable to the interaction is slightly greater in the South, followed by the North and then the Centre.

\section{Discussion}
In this paper, we revisit spatio-temporal disease mapping, with particular attention to the interaction models discussed in \cite{held:2000},  and propose a new model that allows variance partitioning among the main effects and the space-time interaction. When defining priors on the hyperparameters that control complexity of each intrinsic GMRF component, it is important to bear in mind that the main effects belong to the null space of the interaction term. This means that the interaction can naturally be regarded as an extension of the model including the main effects alone. This idea leads to a model reparametrization where a mixing parameter $\gamma$ balances out the contribution of the main and interaction effects to the total variance. The proposed approach implicitly defines a joint prior on the precision parameters of the various terms in the classic parametrization of the model. 

The advantages of this reparametrization are twofold; on the one hand, prior choice can be made in an intuitive manner using a PC prior, avoiding the issue of eliciting priors on hard-to-interpret precision parameters. In space-time disease mapping, the nature of the disease can provide useful information to elicit the prior; for example, for non-infectious diseases such as the one considered in the first case study most of the variation is expected to be explained by the main effects \citep{abellan:2008}. This knowledge can be easily passed onto the PC prior for the mixing parameter $\gamma$, while coding this information into a precision parameter in the classic parametrization would be far from easy.  On the other hand, the posterior for $\gamma$ becomes a useful tool to investigate variations in disease risk on a very practical scale and can provide useful insights into epidemiological interpretations. We have illustrated the use of the VP model in two examples; the variance partitioning tables and plots summarize the contribution of the different sources of variation in terms of proportion of explained (generalized) variance.

In a broader perspective, our work falls within the framework of variance distributing models as introduced by \cite{Fuglstad:2020}, and adds to the literature in considering intrinsic GMRF models. The variance partitioning approach proposed here may be adopted in all those applications where intrinsic GMRFs are meant as tools to perform smoothing in more than one dimension; for instance in the analysis of grid-data such as those arising from agricultural field trials or spatio-temporal data from environmental studies and ecological surveys.

\newpage
\appendix

\section{Proofs} \label{appendix:proofs}
For ease of presentation, we first prove \textbf{Result 1} for model (4), Section 3 of our paper, type IV interaction in Appendix \ref{appendix:proof1} and then show that it is also valid for types I, II and III in Appendix \ref{appendix:proof2}. Details regarding the proof for the model including unstuctured and structured main effects, model (6) Section 4 of our paper, can be found in Appendix \ref{appendix:proof3}. Throughout the proof, we assume a RW2 model on the temporal random effect and an ICAR on the spatial one. The modification of the proof when a RW1 model is used on the temporal effect is straightforward.

\subsection{Proof of Result 1 for type IV interaction} \label{appendix:proof1}

Model (1) in our paper can be written in general form  (in the linear predictor scale) as 
\begin{equation}
\bm\eta = \alpha \bm 1_n + \sqrt{\tau^{-1}} \left(\sqrt{1-\gamma} \bm \omega_0 + \sqrt{\gamma} \bm \omega_1 \right),
\label{eq:rwr_reparameterization_gral}
\end{equation}
where $\tau>0$ is the precision parameter, $0 < \gamma < 1$ is the mixing parameter, $\bm \omega_0$, $\bm \omega_1$ are $n$-dimensional IGMRFs with precision matrices $\bm Q_0$ and $\bm Q_1$ respectively, with
\[
	\bm Q_0^{-}=(1-\phi)(\bm 1_{n_2} \otimes \bm I_{n_1})\tilde{\bm R}_1^{-}(\bm 1_{n_2} \otimes \bm I_{n_1})^T +\phi ( \bm I_{n_2} \otimes \bm 1_{n_1})\tilde{\bm R}_2^- ( \bm I_{n_2} \otimes \bm 1_{n_1})^T
\]
and
\[
	\bm Q_1=\tilde{\bm R_2}\otimes \tilde{\bm R_1}
\]
%
where $\tilde{\bm R_1}$ and $\tilde{\bm R_2}$ are the scaled structure matrices of a RW2 and an ICAR, respectively. Note that rank$(\tilde{\bm R_1})=n_1-1$ and rank$(\tilde{\bm R_2})=n_2-2$, so it follows that rank$(\bm Q_1)=n_1n_2-n_2-2n_1+2$ and $\text{rank}(\bm Q_0)=n_1+n_2-3$. For ease of presentation, we simplify the notation and denote $n=n_1n_2$, $r=2n_1+n_2-2$, so that rank$(\bm Q_1)=n-r$. It is immediate to see that rank of $\bm Q_0$ is smaller than the rank deficiency of $\bm Q_1$, i.e.:
\[
	n_1+n_2 -3 \leq 2n_1+n_2-2 \Leftrightarrow \quad n_1\leq 2n_1+1,
\]
so that rank$(\bm Q_0)=r-l$, where $l\geq 0$ is the difference between rank$(\bm Q_0$) and $r$. For ease of presentation, we can assume $l=0$ (note that if $l\neq 0$ then the adjustment of the proof is straightforward).\\

Consider $\tau=1$ without loss of generality. To derive the PC prior for $\gamma$ we will study the limiting behaviour of $\text{KLD}(\pi_1|| \pi_0)$ for $\gamma=\gamma_0 \rightarrow 0$ under the base model. The distributions $\pi_1$ and $\pi_0$ are defined as follows:
\begin{eqnarray*}
\pi_1 \sim N_1(0,\bm \Sigma_1) & \text{with} & \bm\Sigma_1 = (1-\gamma)\bm Q_0^{-} + \gamma \bm Q_1^{-}\\
\pi_0 \sim N_0(0,\bm\Sigma_0) & \text{with} & \bm\Sigma_0 = (1-\gamma_0)\bm Q_0^{-} + \gamma_0 \bm Q_1^{-}
\end{eqnarray*}

The KLD is given by: 
\begin{eqnarray}
	\text{KLD}(\pi_1|| \pi_0) & = & \frac{1}{2}\left(\text{trace}(\bm{\Sigma_0^{-}\Sigma_1})-(n-r)-\log\frac{|\bm \Sigma_1|}{|\bm \Sigma_0|}\right).\label{eq:kld_gresult}
\end{eqnarray}

Expression (\ref{eq:kld_gresult}) can be computed easily if we consider the eigendecomposition of the matrices $\bm Q_0=\bm V_{\bm Q_0} \bm \Lambda_{\bm Q_0} \bm V_{\bm Q_0}^T$ and $\bm Q_1= \bm V_{\bm Q_1} \bm \Lambda_{\bm Q_1} \bm V_{\bm Q_1}^T$, with
\begin{equation}
\bm\Lambda_{\bm Q_0}=\text{diag}(\tilde{\lambda}_1,\tilde{\lambda}_2,\ldots,\tilde{\lambda}_{r},\underbrace{0,\ldots,0}_{n-r}) \quad \quad ; \quad \quad \bm\Lambda_{\bm Q_1}=\text{diag}(\underbrace{0,\ldots,0}_{r},\lambda'_{r+1}, \ldots,\lambda'_n),
\label{eq:eigenval_AB}
\end{equation}
\begin{equation}
\bm V_{\bm Q_0}=[\bm {e_1}, \bm e_2, \ldots, \bm e_{r},\bm e_{r+1},\ldots,\bm e_n] \quad \quad ; \quad \quad \bm V_{\bm Q_1}=[\hat{\bm e}_1,\ldots,\hat{\bm e}_r, \hat{\bm e}_{r+1},\ldots, \hat{\bm e}_{n}].
\label{eq:eigenvec_AB}
\end{equation}
where $\bm\Lambda_{\bm Q_0}$, $\bm\Lambda_{\bm Q_1}$ represent the diagonal matrix of eigenvalues and $\bm V_{\bm Q_0}$ and $\bm V_{\bm Q_1}$ the matrices whose columns are the associated eigenvectors. A common eigenvector basis $\bm V$ can be formed as
\[\bm V=[\bm e_1,\bm e_2,\ldots,\bm e_r,\hat{\bm e}_{r+1} ,\ldots,\hat{\bm e}_n ],\] 
so that $\bm Q_0=\bm V\bm\Lambda_{\bm Q_0}\bm V^T$ and $\bm Q_1=\bm V\bm\Lambda_{\bm Q_1}\bm V^T$.
If $l\neq 0$ then there would be a set of eigenvectors that are associated to zero eigenvalues in both matrices $\bm Q_0$ and $\bm Q_1$ contemporarily, so the common basis can still be formed.\\

Matrices $\bm \Sigma_0^{-}$ and $\bm \Sigma_1$ can be re-expressed as
  
\[
	\bm{\Sigma_0^{-}} = \left\{  \bm V \left[(1-\gamma_0)\bm\Lambda_{\bm Q_0}^{-1} + \gamma_0 \bm\Lambda_{\bm Q_1}^{-1}\right]\bm V^T\right\}^{-1}= \bm V \left[(1-\gamma_0)\bm\Lambda_{\bm Q_0}^{-1} + \gamma_0 \bm\Lambda_{\bm Q_1}^{-1}\right]^{-1}\bm V^T
\]
 and 
\[
	\bm{\Sigma_1}=\bm V \left((1-\gamma)\bm\Lambda_{\bm Q_0}^{-1} + \gamma\bm\Lambda_{\bm Q_0}^{-1}\right)\bm V^T,
\]
where  $\bm\Lambda_{\bm Q_0}^{-1}$ and $\bm\Lambda_{\bm Q_1}^{-1}$  are diagonal matrices with elements $\lambda_i$ and $\hat{\lambda}_i$. Note that $\bm Q_0$ and $\bm Q_1$ are singular;  following \cite{pcprior} appendix A2, $\lambda_i = 1/\tilde{\lambda}_i$ if $\tilde{\lambda}_i>0$ and $\lambda_i=0$ when $\tilde{\lambda}_i=0$. Analogously, $\hat{\lambda}_i = 1/\lambda'_i$ if $\lambda'_i>0$ and $\hat{\lambda}_i=0$ when $\lambda'_i=0$. \\

First, we compute $\text{trace}(\bm{\Sigma_0^{-1}\Sigma_1})$, for which we need the diagonal $\text{diag}{(\bm{\Sigma_0^{-1}\Sigma_1})}$. Let us define \[\bm D(\gamma)=\text{diag}\left((1-\gamma)\lambda_i+\gamma\hat{\lambda}_i\right)_{i=1,\ldots,n}\]  we can re-express the diagonal as
\[ \text{diag}(\bm{\Sigma_0^{-1}\Sigma_1})=\bm V \bm D(\gamma_0)^{-1}\bm D(\gamma)\bm V^T.\]

The trace simplifies to
\begin{eqnarray}
\text{tr}(\bm V \bm D(\gamma_0)^{-1}\bm D(\gamma)\bm V^T) & = & \text{tr}(\bm V^T \bm V \bm D(\gamma_0)^{-1}\bm D(\gamma))\nonumber\\
& = & \text{tr}(\bm D(\gamma_0)^{-1}\bm D(\gamma))\nonumber\\
& = & \sum_{i=1}^n\frac{(1-\gamma)\lambda_i+\gamma\hat{\lambda_i}}{(1-\gamma_0)\lambda_i+\gamma_0\hat{\lambda_i}}\nonumber\\
& = & \sum_{i=1}^n\alpha(\gamma,\gamma_0)_i\nonumber
\end{eqnarray}

(note that if $l\neq 0$, then we would sum over all indices $i\neq r-l+j$ for $j=1,\ldots,l$).\\

Second, we compute $\log\frac{|\bm \Sigma_1|}{|\bm \Sigma_0|} $ in (\ref{eq:kld_gresult}):
\begin{eqnarray}
 \log|\bm \Sigma_1|-\log|\bm \Sigma_0| & = & \sum_{i=1}^n\left[\log\left((1-\gamma)\lambda_i+\gamma\hat{\lambda_i}\right)-\log\left((1-\gamma_0)\lambda_i+\gamma_0\hat{\lambda_i}\right)\right]\nonumber\\
&=& 
\sum_{i=1}^n\log\left(\frac{(1-\gamma)\lambda_i+\gamma\hat{\lambda_i}}{(1-\gamma_0)\lambda_i+\gamma_0\hat{\lambda_i}}\right)\nonumber\\
&=&
\sum_{i=1}^n\log\alpha(\gamma,\gamma_0)_i
\end{eqnarray}

It results:
\begin{eqnarray}
	\text{KLD}(\pi_1|| \pi_0) 
		&=& \frac{1}{2}\left(\sum_{i=1}^n\alpha(\gamma,\gamma_0)_i-(n-r)-\sum_{i=1}^n\log\alpha(\gamma,\gamma_0)_i\right). \label{eq:kld_gresult_2}
\end{eqnarray}

Below we compute the term $\alpha(\gamma,\gamma_0)_i$ for $i=1,\ldots,r$ and $i=r+1,\ldots,n$:
\begin{itemize}
	\item $i=1,\ldots,r$ ($\hat{\lambda}_i=0$):
	\[
		\alpha(\gamma,\gamma_0)_i=\frac{\frac{1-\gamma}{1-\gamma_0}\lambda_i+\frac{\gamma}{1-\gamma_0}0}{\lambda_i+\frac{\gamma_0}{1-\gamma_0}0}=\frac{1-\gamma}{1-\gamma_0}
	\]
	\item $i=r+1,\ldots,n$ ($\lambda_i=0$):
	\[
	\alpha(\gamma,\gamma_0)_i=\frac{\frac{1-\gamma}{1-\gamma_0}0+\frac{\gamma}{1-\gamma_0}\hat{\lambda_i}}{0+\frac{\gamma_0}{1-\gamma_0}\hat{\lambda_i}}=\frac{\gamma}{\gamma_0}
\]
\end{itemize}
Note that the eigenvalues of $\bm Q_0$ and $\bm Q_1$ turn out to be irrelevant for computing the KLD, as they cancel out in the $\alpha(\gamma,\gamma_0)_i$ terms above. Finally, the KLD is:
\begin{equation}
	\text{KLD}(\pi_1|| \pi_0)  =\frac{1}{2}\left[r\frac{1-\gamma}{1-\gamma_0}+(n-r)\frac{\gamma}{\gamma_0}-(n-r)-r\log\frac{1-\gamma}{1-\gamma_0} -(n-r)\log\frac{\gamma}{\gamma_0}\right].
\label{eq:kld}
\end{equation}

For $\gamma_0\rightarrow 0$ and $\gamma_0\ll\gamma<1$ the dominant term in expression (\ref{eq:kld}) is $(n-r)\frac{\gamma}{\gamma_0}$. Therefore, the distance from the base model, measured as $d(\gamma)=\sqrt{2KLD}$, is
\begin{eqnarray*}
	d(\gamma) & =&\lim_{\gamma_0\rightarrow 0}\sqrt{r\frac{1-\gamma}{1-\gamma_0}+(n-r)\frac{\gamma}{\gamma_0}-(n-r)-r\log\frac{1-\gamma}{1-\gamma_0} +(n-r)\log\frac{\gamma}{\gamma_0}}\\
& \simeq & \sqrt{(n-r)\frac{\gamma}{\gamma_0}} =  c\sqrt{\gamma},
\end{eqnarray*}
for a constant $c>0$ that does not depend on $\gamma$. Since $0 \leq d(\gamma) \leq c$, assigning a truncated exponential with rate $\lambda$ on $d(\gamma)$ we have
\[
	\pi(d(\gamma))=\frac{\lambda \exp(-\lambda c\sqrt{\gamma})}{1-\exp(-\lambda c)}, \quad \quad 0 \leq d(\gamma) \leq c, \quad \lambda >0.
\]
Applying a change of variable and reparametrizing $\theta=\lambda c$ leads to the PC prior for $\gamma$:
\[
\pi(\gamma)=\frac{\theta\exp(-\theta\sqrt{\gamma})}{2\sqrt{\gamma}(1-\exp(-\theta))} \quad \quad 0 < \gamma <  1, \theta>0
\]
which completes the proof.

\subsection{Proof of Result 1 for interaction types I, II and III} \label{appendix:proof2}

From Appendix \ref{appendix:proof1}, it is clear that the proof works provided that a common eigenbasis can be found for matrices $\bm Q_0$ (which is the same as in Appendix \ref{appendix:proof1}) and $\bm Q_1$ (that changes depending on the type of interaction). We first illustrate that this is case for interaction types I, II and III, to then show that the KLD remains the unchanged.\\

\noindent\textbf{\underline{Interaction type I}}\\
For the type I interaction, $\bm Q_1=\bm I_{n_{2}} \otimes \bm I_{n_{1}}$ so it has a single eigenvalue equal to 1 with multiplicity $n_1n_2$. Given that any vector of $\mathbb{R}^{n_1n_2}$ is  an eigenvector of $\bm Q_1$, it is enough to use the eigenvectors from the eigendecomposition of $\bm Q_0$ as a common eigenbasis. \\

\noindent\textbf{\underline{Interaction type II}}\\
For the type II interaction, 
$\bm Q_1=\bm I_{n_{2}} \otimes \tilde{\bm R}_1$ has $2n_2$ eigenvectors associated to null eigenvalues, and $n_2(n_1-2)$ eigenvectors associated to non-null eigenvalues, that come from the tensor product of non-null eigenvectors from the matrices $\bm I_{n_2}$ and $\bm R_1$.
Let $e^{\bm R_1}_1,\ldots, e^{\bm R_1}_{n_1-2}$ be the eigenvectors associated to non-null eigenvalues of $\bm R_1$; the first $n_1-2$ eigenvectors associated to non-null eigenvalues of the matrix $\bm Q_0$ are:
\begin{equation}\label{eigen_Q0_typeII}
	\bm 1_{n_2}\otimes e^{\bm R_1}_1,\ldots, \bm 1_{n_2}\otimes e^{\bm R_1}_{n_1-2}
\end{equation}
while the first $n_1-2$ eigenvectors associated to non-null eigenvalues of the matrix $\bm Q_1$ are:
\[
	\bm e_{1}\otimes e^{\bm R_1}_1,\ldots, \bm e_{1}\otimes e^{\bm R_1}_{n_1-2}
\]
where $\bm e_{1}$ is the first eigenvector of the identity matrix $\bm I_{n_2}$. 
We can eigen decompose the identity matrix using the eigenbasis for $\bm R_2$, so that $\bm e_{1} = \bm 1_{n_2}$; this guarantees that a common matrix of eigenvectors $\bm V$ can be found. In particular, it would be formed of the $n_1+n_2-3$ non null eigenvectors from $\bm Q_0$ and the $n_1n_2-2n_2$ non null eigenvectors from $\bm Q_1$. Note that these two collection of vectors will have $n_1 -n_2-3$ vectors in common from the eigenvectors in (\ref{eigen_Q0_typeII}) if $n_1 > n_2 + 3$.\\

\noindent\textbf{\underline{Interaction type III}}\\
In the type III interaction, 
$\bm Q_1=\tilde{\bm R}_2 \otimes \bm I_{n_1}$ has $n_1$ eigenvectors associated to null eigenvalues and $n_1n_2 - n_1$ eigenvectors with non-null eigenvalues.
In particular, let $e^{\bm R_2}_1,\ldots, e^{\bm R_2}_{n_2-1}$ be the eigenvectors associated to non-null eigenvalues of $\bm R_2$; the following are $n_2-1$ eigenvectors associated to non-null eigenvalues of the matrix $\bm Q_0$:
\begin{equation}\label{eigen_Q0_typeIII}
	 e^{\bm R_2}_1\otimes\bm 1_{n_1},\ldots,  e^{\bm R_2}_{n_2-1}\otimes\bm 1_{n_1}
\end{equation}
while for matrix $\bm Q_1$ we find the following $n_2-1$ eigenvectors associated to non-null eigenvalues :
\[
	 e^{\bm R_2}_1\otimes\bm e_{1},\ldots,  e^{\bm R_2}_{n_2-1}\otimes\bm e_{1}
\]
where $\bm e_{1}$ is the first eigenvector of the identity matrix $\bm I_{n_1}$. Similarly to the type II interaction, we can use the eigenbasis for $\bm R_1$ to eigen decompose $\bm I_{n_1}$ so that a common eigenbasis can be found. It would be formed of the $n_1+n_2-3$ non null eigenvectors from $\bm Q_0$ and the $n_1n_2-n_1$ non-null eigenvectors from $\bm Q_1$. Note that these two collection of vectors will have $n_2-3$ vectors in common from the eigenvectors in (\ref{eigen_Q0_typeIII}) if $n_2 > 3$.\\

Regarding the KLD, which is calculated based on the eigenvalues of $\bm Q_0$ and $\bm Q_1$, whenever the rank of $\bm Q_0$ is not smaller than the rank defficiency of $\bm Q_1$, there will be a number of pairs of eigenvalues that are not zero contemporarily. This number is equal to $n_1+n_2-3$ in the type I, $n_1 -n_2-3$ in the type II and $n_2-3$ in the type III interaction. Nevertherless, the contribution of the corresponding term $\alpha(\gamma,\gamma_0)_i$ in the KLD is minimal and the dominant term when $\gamma_0 \rightarrow 0$ remains the same as shown in Appendix \ref{appendix:proof1} for the type IV interaction, so the PC prior does not change.

\subsection{Model with structured and unstructured main effects}\label{appendix:proof3}
In the case of structured and unstructured main effects, matrix $\bm Q_0^{-}$:
\begin{eqnarray*}
	\bm Q_0^{-} & = & (1-\phi)(\bm 1_{n_2}\otimes \bm I_{n_1})\left((1-\psi_1)\tilde{\bm R}_1^- +\psi_1 \bm I_{n_1}\right)(\bm 1_{n_2}\otimes \bm I_{n_1})^T + \\
	& & \phi (\bm I_{n_2}\otimes \bm 1_{n_1})\left((1-\psi_2)\tilde{\bm R}_2^- +\psi_2 \bm I_{n_2}\right)(\bm I_{n_2}\otimes \bm 1_{n_1})^T
\end{eqnarray*}
and $\text{rank}(\bm Q_0)\leq n_1 +n_2$.\\

\noindent
\underline{\textbf{Interaction type IV}}\\
Following the proof in Appendix \ref{appendix:proof1}, it is enough to show that $\text{rank}(\bm Q_0)\leq 2n_2 + n_1 -2$. Given that the rank of $\bm Q_0$ is at most $n_1+n_2$, the rank condition is true provided that $0 \leq n_2-2$, i.e. that there are at least 2 spatial locations, which is always true in practice.\\

\noindent \underline{\textbf{Interaction types I,II, III}}\\
For interaction types I, II and III it is still possible to find a common eigenbasis, as adding a constant to the diagonal of a matrix does not change its eigenvectors. The eigenvalues do change though, so now the number of eigenvalues that are not zero contemporarily in $\bm Q_0$ and $\bm Q_1$ (whenever the rank of $\bm Q_0$ is not smaller than the rank defficiency of $\bm Q_1$) are $n_1+n_2-1$ for type I, $n_1-n_2-1$ for type II and $n_2-1$ for type III, and the dominant term in the KLD remains the same as before.

\section{Simulation study}
\label{appendix_sim}

We run a simulation study to investigate the performance of the VP model when using the PC prior for $\gamma$ proposed in Section 3.1 Eq. (5) in our paper. We generate datasets based on the space and time patterns estimated from the Covid-19 data described in Section 4.2 in our paper; to limit the computational burden we select a subset of the the full dataset (north provinces, wave 1) with $n_1=17$ weeks and $n_2=47$ provinces. Assume $i$ and $j$ are indices for weeks and provinces, respectively, we simulate data as 
\begin{eqnarray}
y_{ij} & \sim & \text{Bin}(pop_j, \mu_{ij}),\\
logit(\mu_{ij}) & = & \sqrt{1/\tau}\left\{\sqrt{1-\gamma}\left[\sqrt{1-\phi} \hat \beta_{1_i} + \sqrt{\phi} \hat \beta_{2_j}\right] + \sqrt{\gamma} \hat \delta_{i,j}\right\},
\label{eq:model_sim}
\end{eqnarray}
where $pop_j$ is the population in province $j$, $\mu_{ij}$ the Covid-19 incidence rate at week $i$ in province $j$. The vectors $\bm \hat \beta_1 = (\hat\beta_{1,1}, \ldots,\hat\beta_{1,n_1})^T$, $\bm \hat\beta_2=(\hat\beta_{2,1}, \ldots,\hat\beta_{2,n_2})^T$ and $\bm \hat\delta=\{\hat\delta_{ij}\}, i=1,\ldots,n_1, j=1,\ldots,n_2$ contain the posterior means for, time, space and space-time random effects, respectively. These estimates come from the VP model (Eq. (4) in Section 3  in our paper) fitted to the Covid-19 data, north provinces wave 1, by assuming a type IV interaction (see the top panels in Figure \ref{fig:sim_scenarios} for the time and space main effects). We further assume $\tau=1, \phi=0.5$ and keep them fixed throughout the simulation study, while letting the mixing parameter $\gamma$ vary, in order to create different scenarios according to the contribution of the interaction to the total (generalized) variance. 

Our goals are: 1) to check how well the true $\gamma$ is recovered when estimated using our VP model (Eq. (4), Section 3 in our paper)  - where, as an estimator for $\gamma$ we take the posterior mean; 
2) to assess sensitivity to the choice of $\theta$, the scaling parameter for the PC prior on $\gamma$.  

\subsection{Simulation study scenarios}

\begin{figure}
\includegraphics[width=0.45\textwidth]{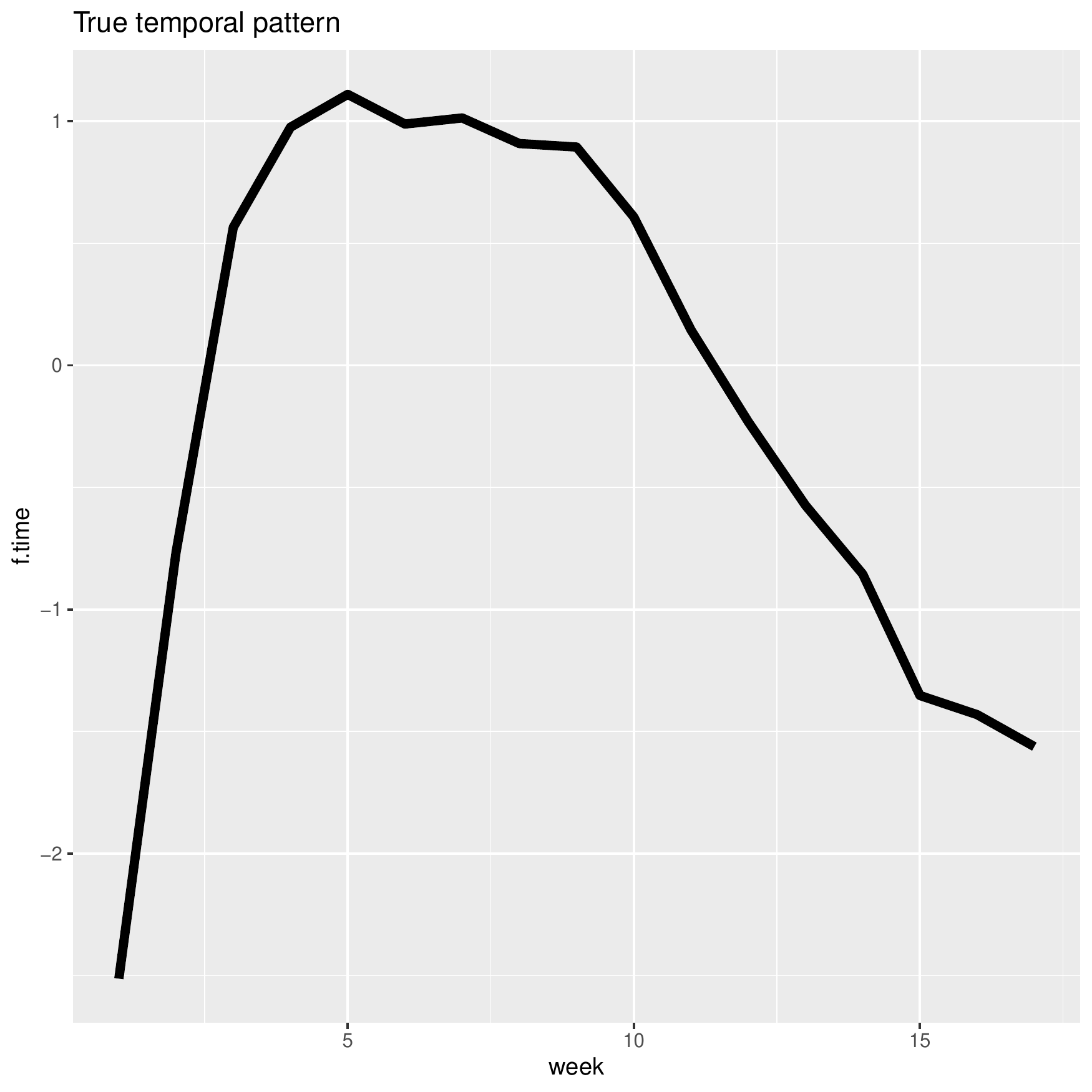}
\includegraphics[width=0.45\textwidth]{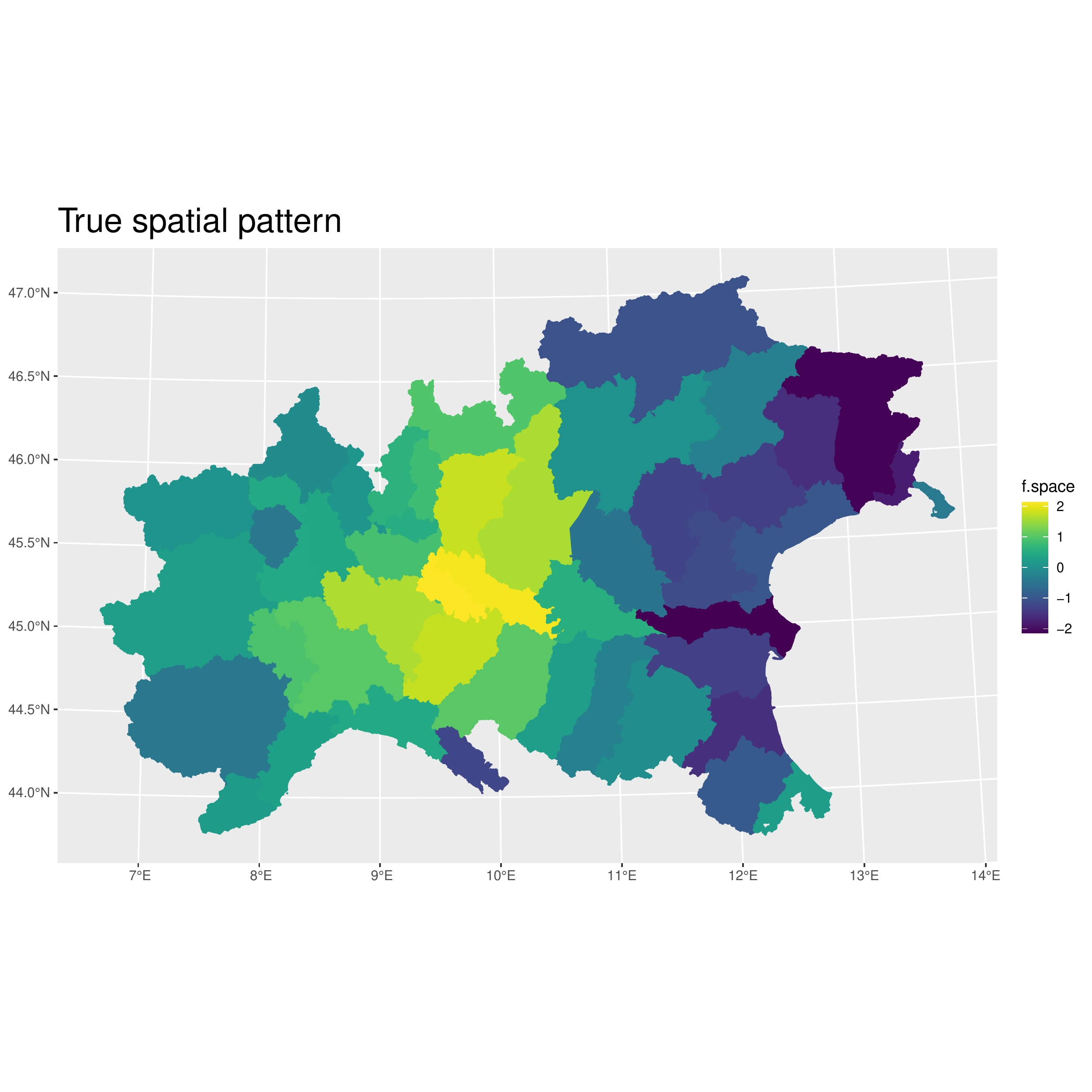}\\
\includegraphics[width=0.45\textwidth]{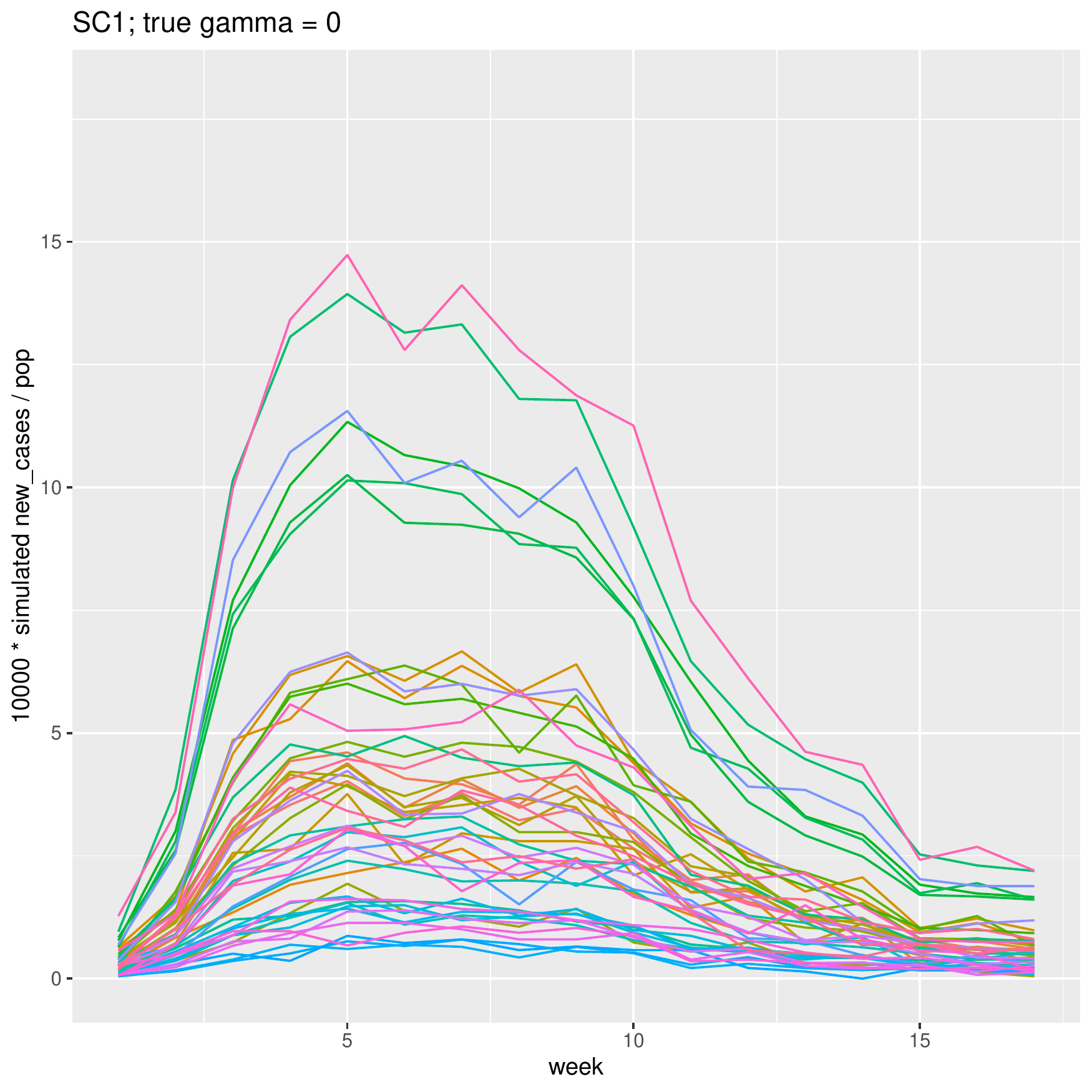}
\includegraphics[width=0.45\textwidth]{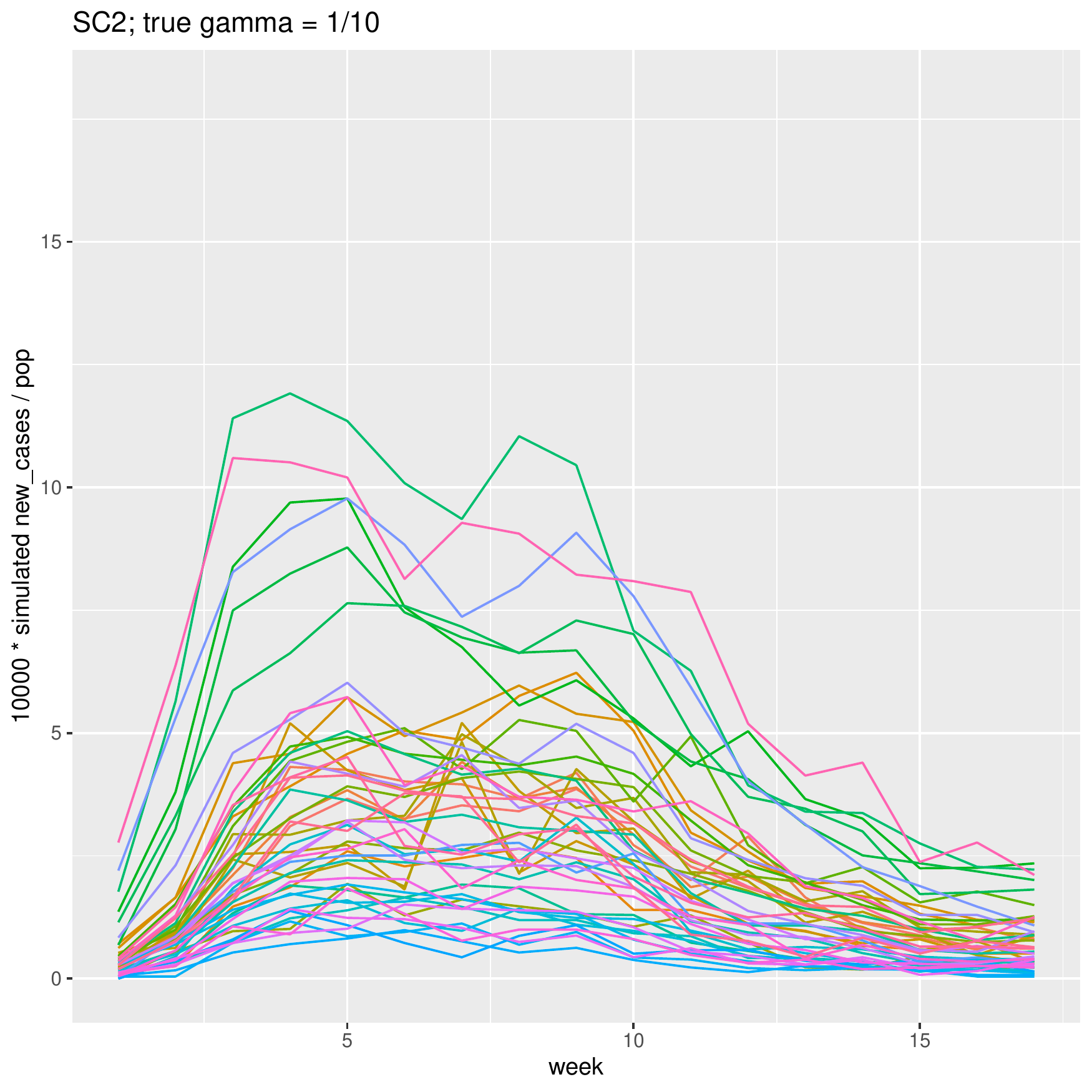}\\
\includegraphics[width=0.45\textwidth]{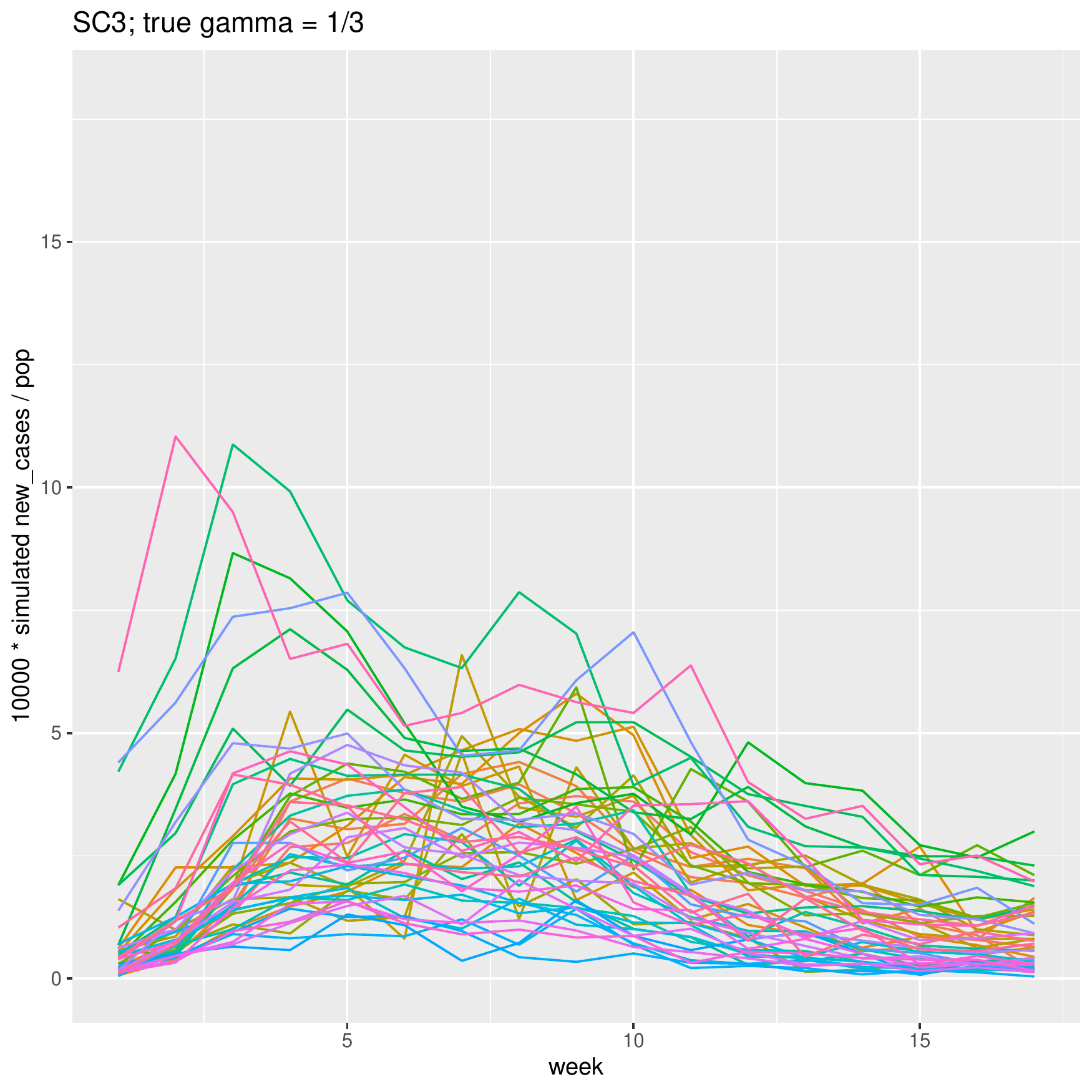}
\includegraphics[width=0.45\textwidth]{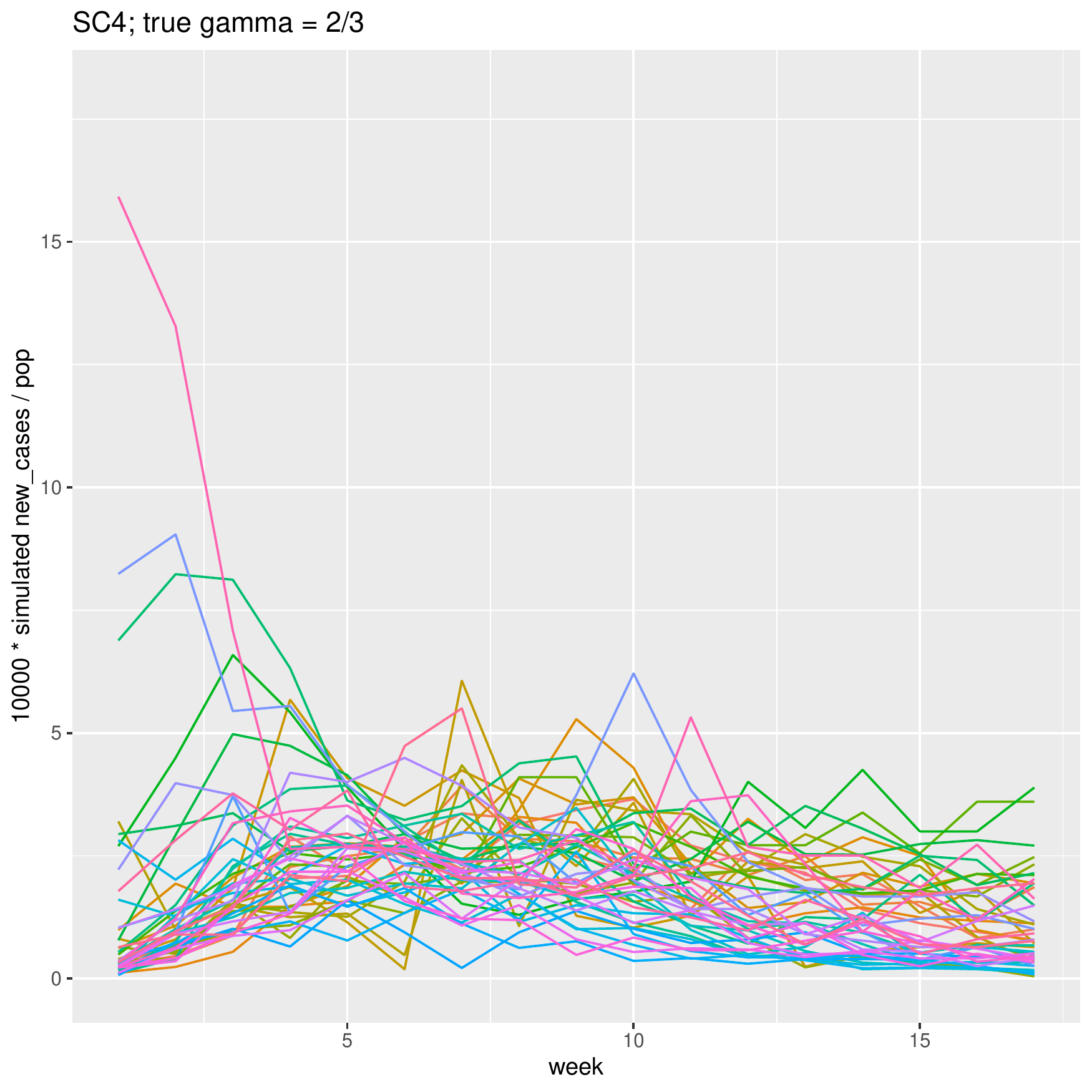}
\caption{Simulation scenarios. Top panels: plots of the main effects for time and space. The central and bottom panels display one simulated dataset under each of the four scenarios (SC1, SC2, SC3, SC4) varying according to the strength of the interaction $\gamma$.}
\label{fig:sim_scenarios}
\end{figure}

The following scenarios are considered regarding the contribution of the interaction to the total variance:
\begin{itemize}
\item SC1: $\gamma=0$ (additive model, no interaction); 
\item SC2: $\gamma=1/10$  (low interaction);
\item SC3: $\gamma=1/3$ (moderate interaction);
\item SC4: $\gamma=2/3$  (strong interaction).
\end{itemize}
Scenario SC1 ($\gamma=0$) assumes an additive model where the time pattern remains the same across provinces. Scenarios SC2 and SC3 represent cases of, respectively, low and moderate interaction. SC4 is intended as a limiting case where the interaction between space and time main effects is very strong; as we can see from Figure \ref{fig:sim_scenarios} (bottom right panel), where one simulated dataset under SC4 is displayed, the temporal pattern can vary substantially across provinces and some of them show a decreasing trend at the beginning of the first wave period, which is clearly unrealistic for Covid-19 disease. 

We consider different scenarios by letting the number of trials of the Binomial model, $pop_j,  j=1, \ldots,n_2$, vary. The following three sample size (i.e. population at risk) levels are considered:
\begin{itemize}
\item Actual sample size: the population in province $j$ is taken as $pop_j$;
\item Smaller sample size: the population in province $j$ is taken as $pop_j/10$;
\item Larger sample size: the population in province $j$ is taken as $pop_j \cdot 10$.
\end{itemize}
The second scenario represents a smaller sample size case, where the data carries less information about $\gamma$ thus we expect less accuracy in the model estimates;  analogously, the third scenario represents a case where data are more informative about $\gamma$, hence we expect the model to provide improved estimates in this case.

We simulated 100 datasets under SC1, SC2, SC3 and SC4, for each of the three different sample size levels described above. The VP model (Eq. (4), Section 3 in our paper) was fitted to each dataset assuming a RW1 as the time main effect, an ICAR as the space main effect and a type IV space-time interaction. All the computations were done using \RINLA. 

The VP model was fitted under 4 different prior choices for $\gamma$:
\begin{itemize}
\item prior 1: PC$(U=0.05, a=0.99)$;
\item prior 2: PC$(U=0.5, a=0.99)$;
\item prior 3: PC$(U=0.95, a=0.99)$;
\item prior 4: Uniform$(0,1)$.
\end{itemize}

The first three priors consider the different scalings of the PC prior displayed in Figure~1 of our paper. This way, robustness of the results for changing $U$ can be tested. The values $U=\{0.05,0.5,0.95\}$ reflect, respectively, an unflexible, moderate and flexible prior on the space-time interaction random effects. We also estimated the model using a uniform prior for $\gamma$. 


Regarding $\tau$ and $\phi$, we assigned a Gumbel type 2 PC prior on $\tau$ and a Uniform$(0,1)$ on $\phi$ to express ignorance about the variance contribution of space (and time). We considered two different scalings of the PC prior on $\tau$ ($U=2/0.31$ and $U=100/0.31$) but they did not make any difference on posterior estimates.

\subsection{Simulation study results}
\begin{figure}[H]
\includegraphics[width=0.32\textwidth]{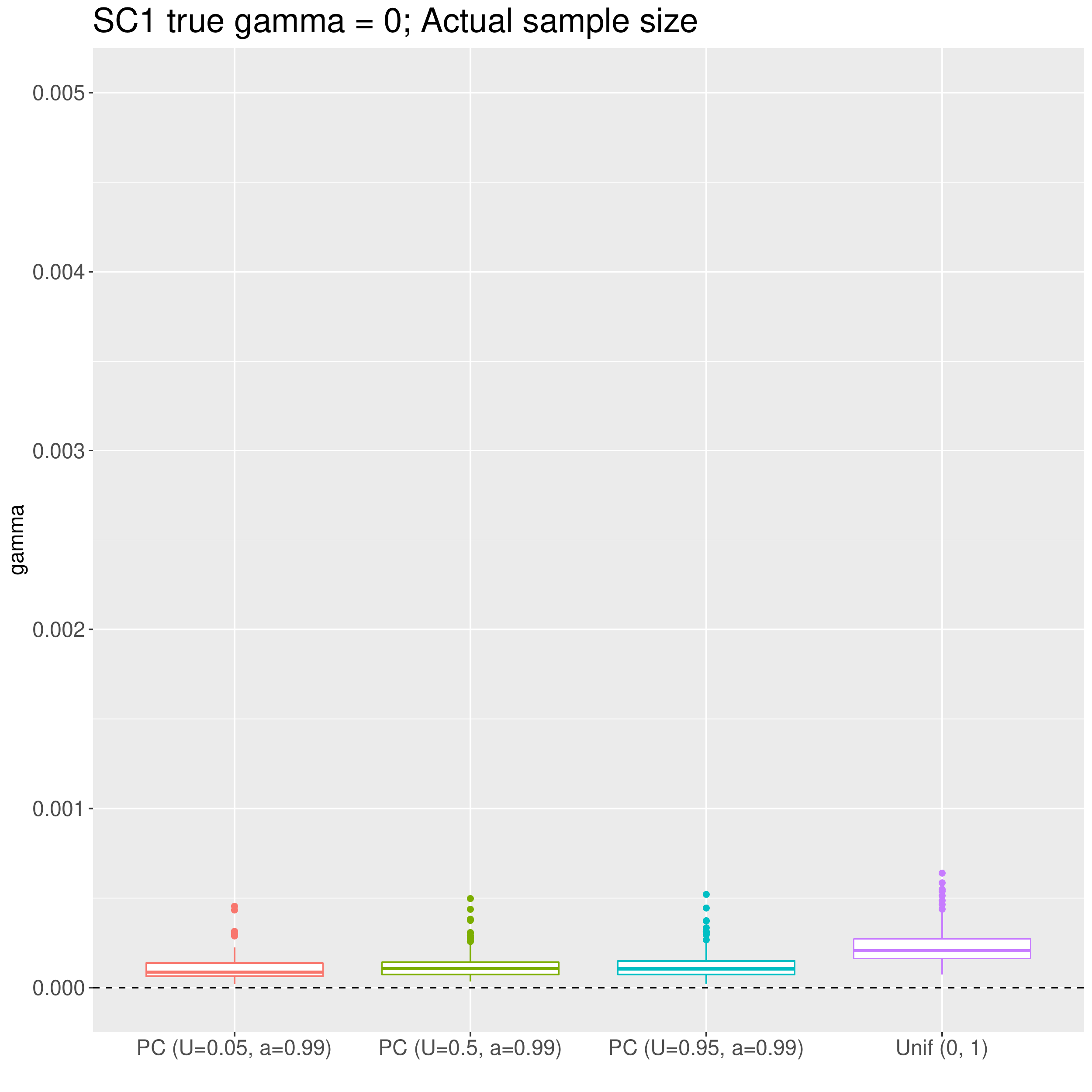}
\includegraphics[width=0.32\textwidth]{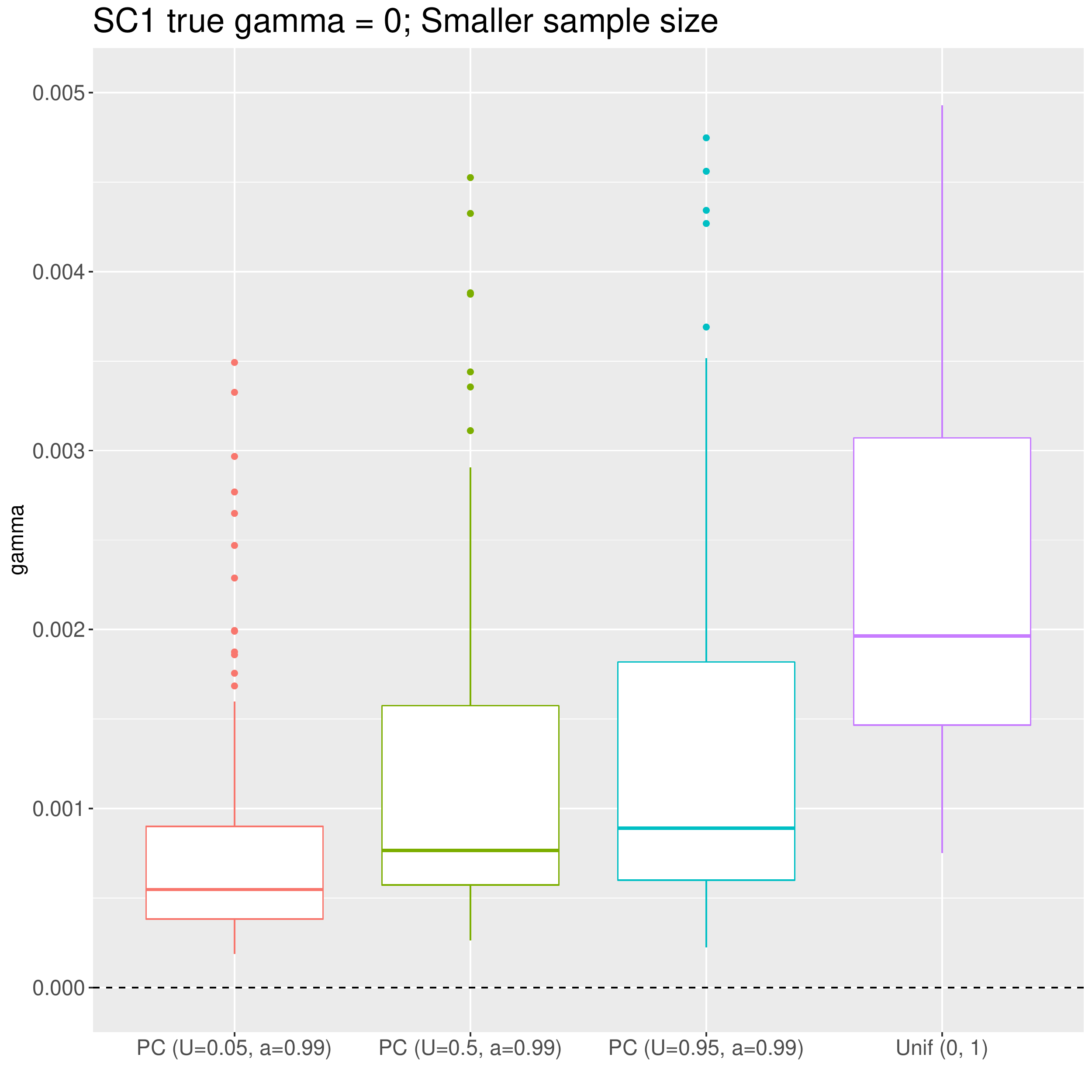}
\includegraphics[width=0.32\textwidth]{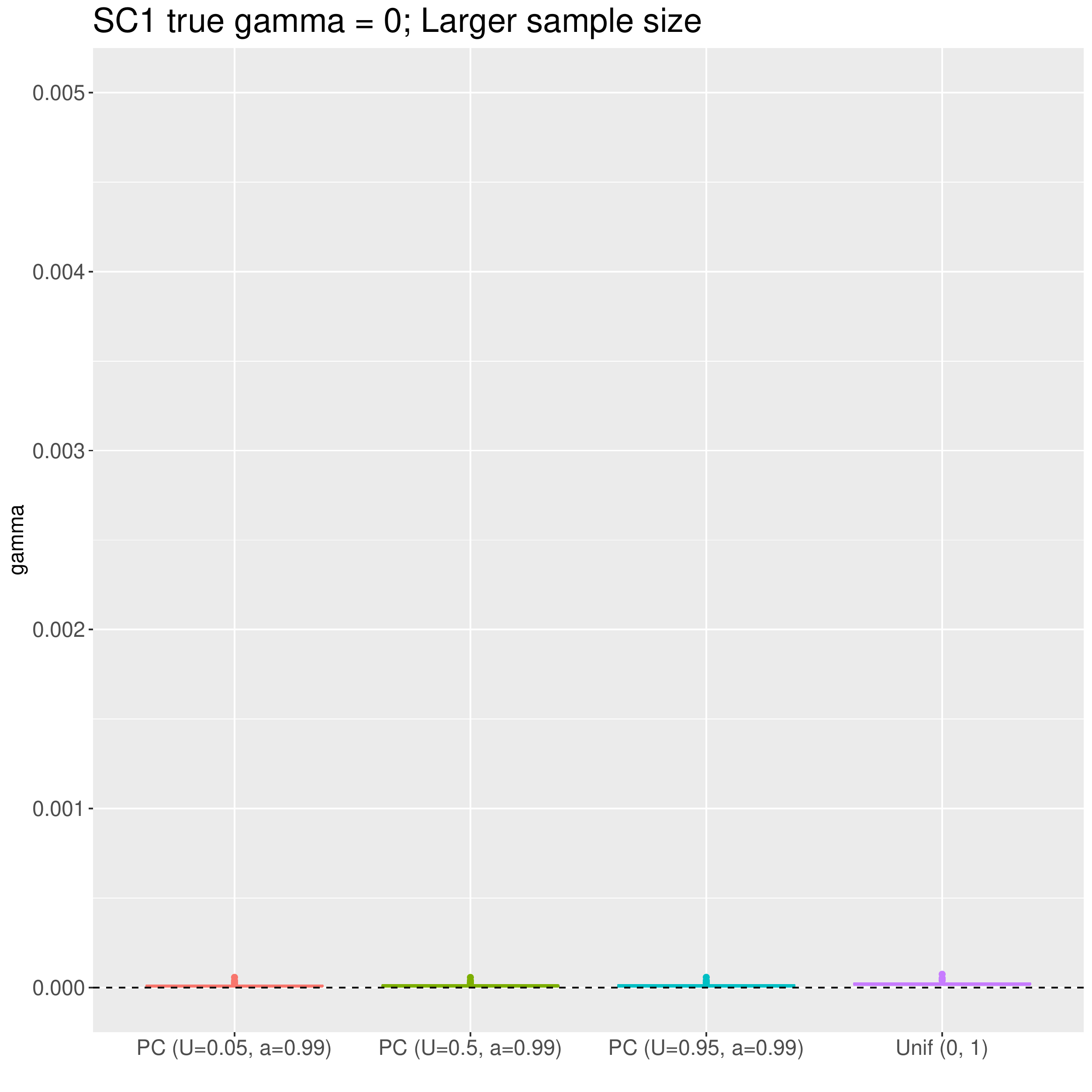}
\caption{Simulation results for the mixing parameter $\gamma$, scenario SC1; true $\gamma=0$.}
\label{fig:sim_gamma_SC1}
\end{figure}

\begin{figure}[H]
\includegraphics[width=0.32\textwidth]{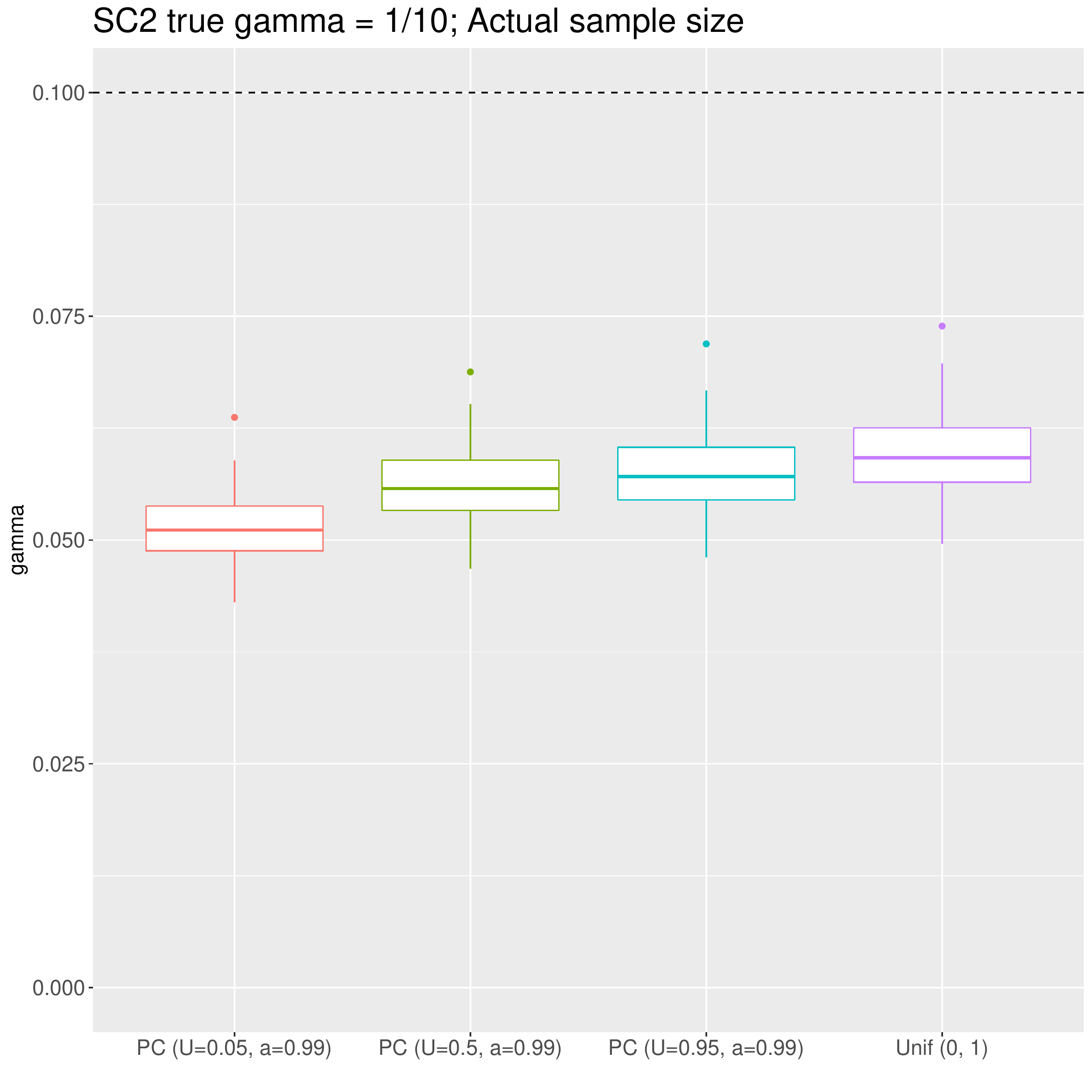}
\includegraphics[width=0.32\textwidth]{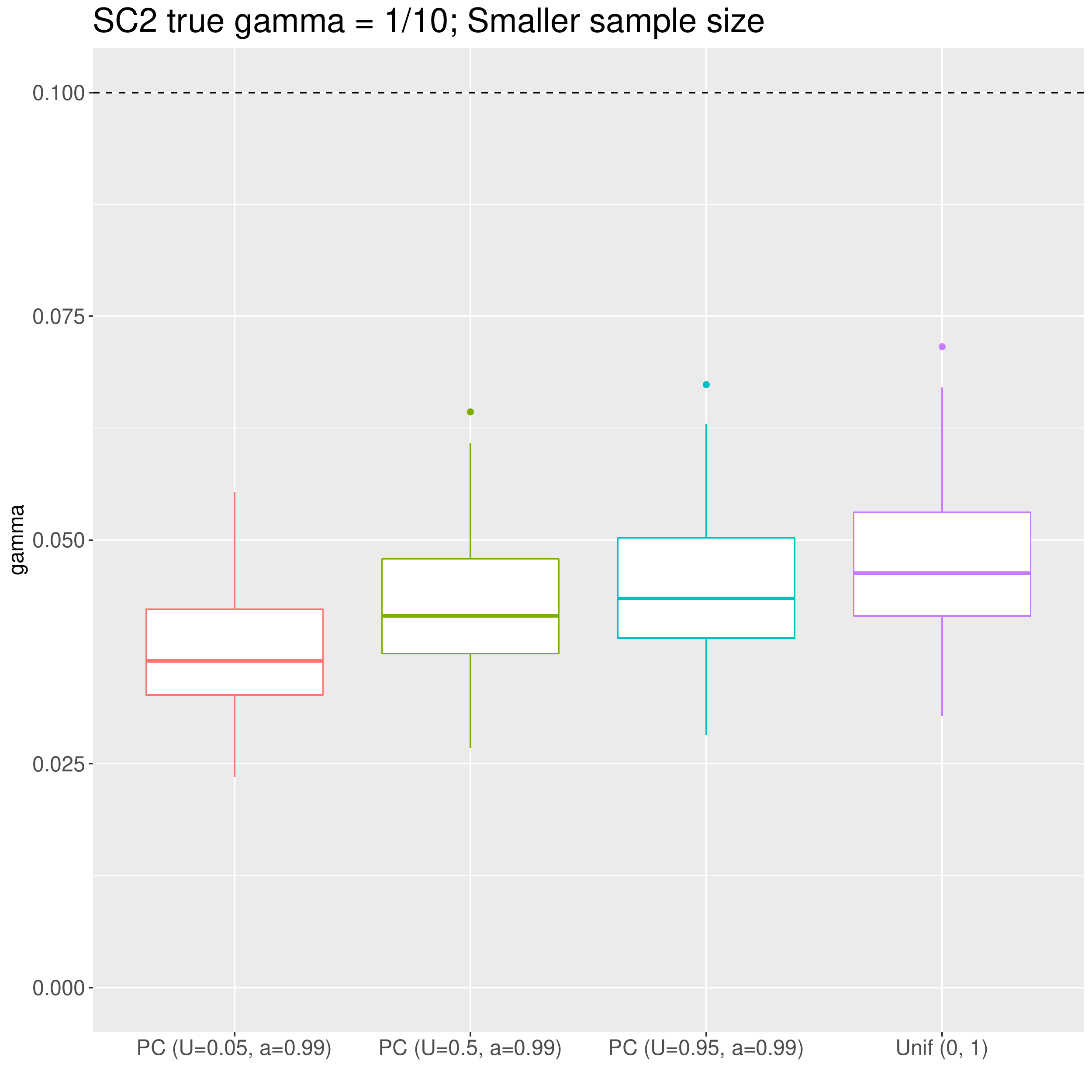}
\includegraphics[width=0.32\textwidth]{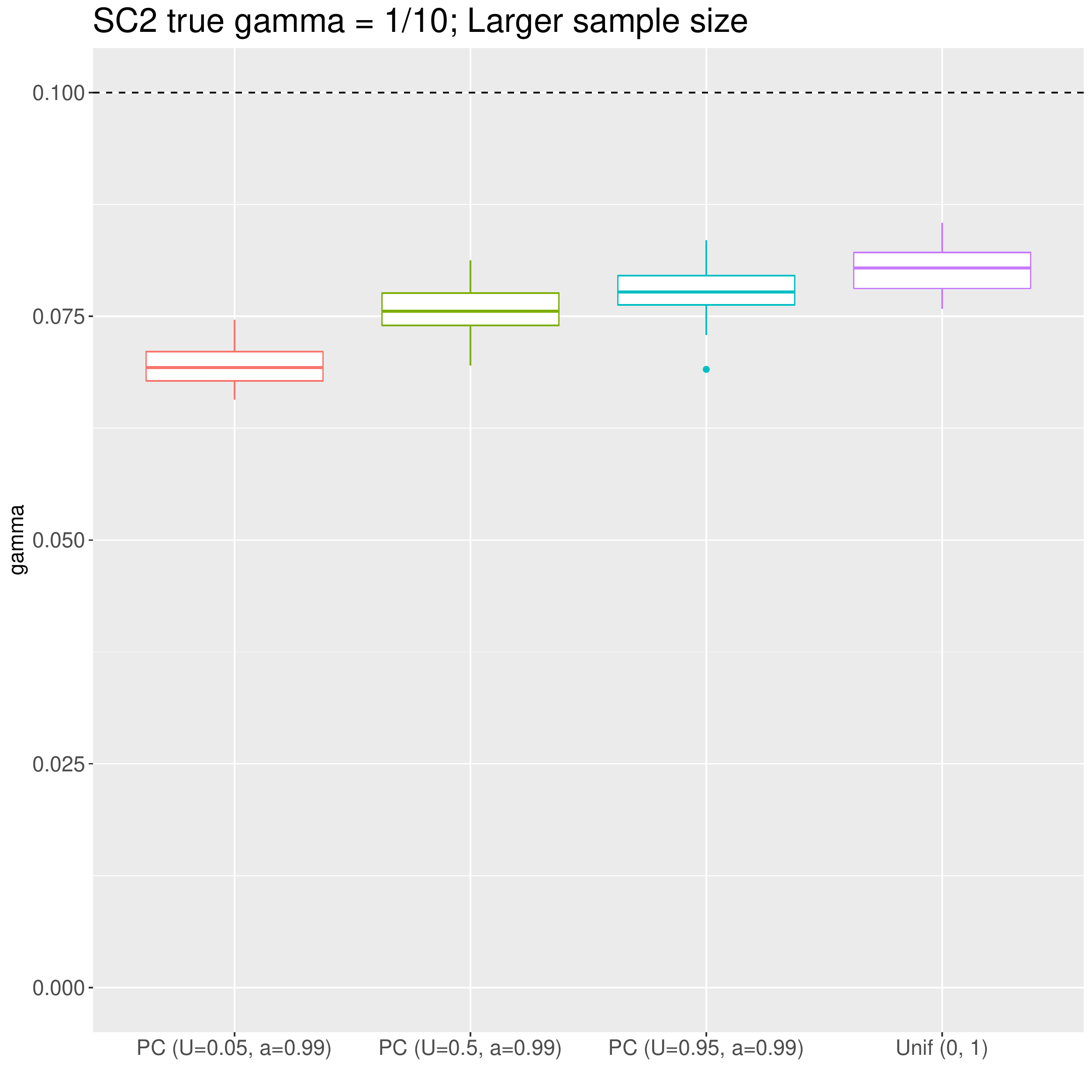}
\caption{Simulation results for the mixing parameter $\gamma$, scenario SC2; true $\gamma=1/10$.}
\label{fig:sim_gamma_SC2}
\end{figure}

\begin{figure}[H]
\includegraphics[width=0.32\textwidth]{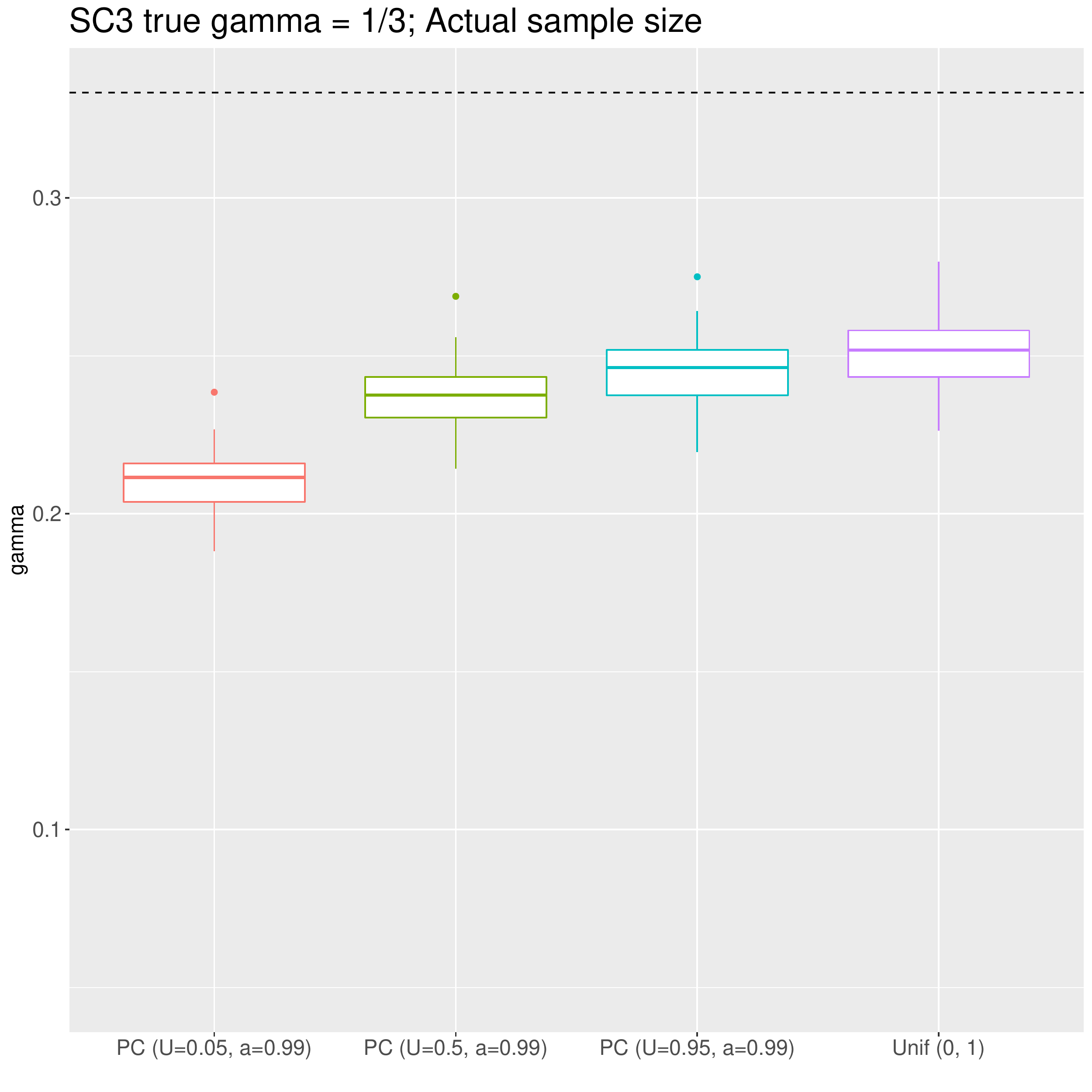}
\includegraphics[width=0.32\textwidth]{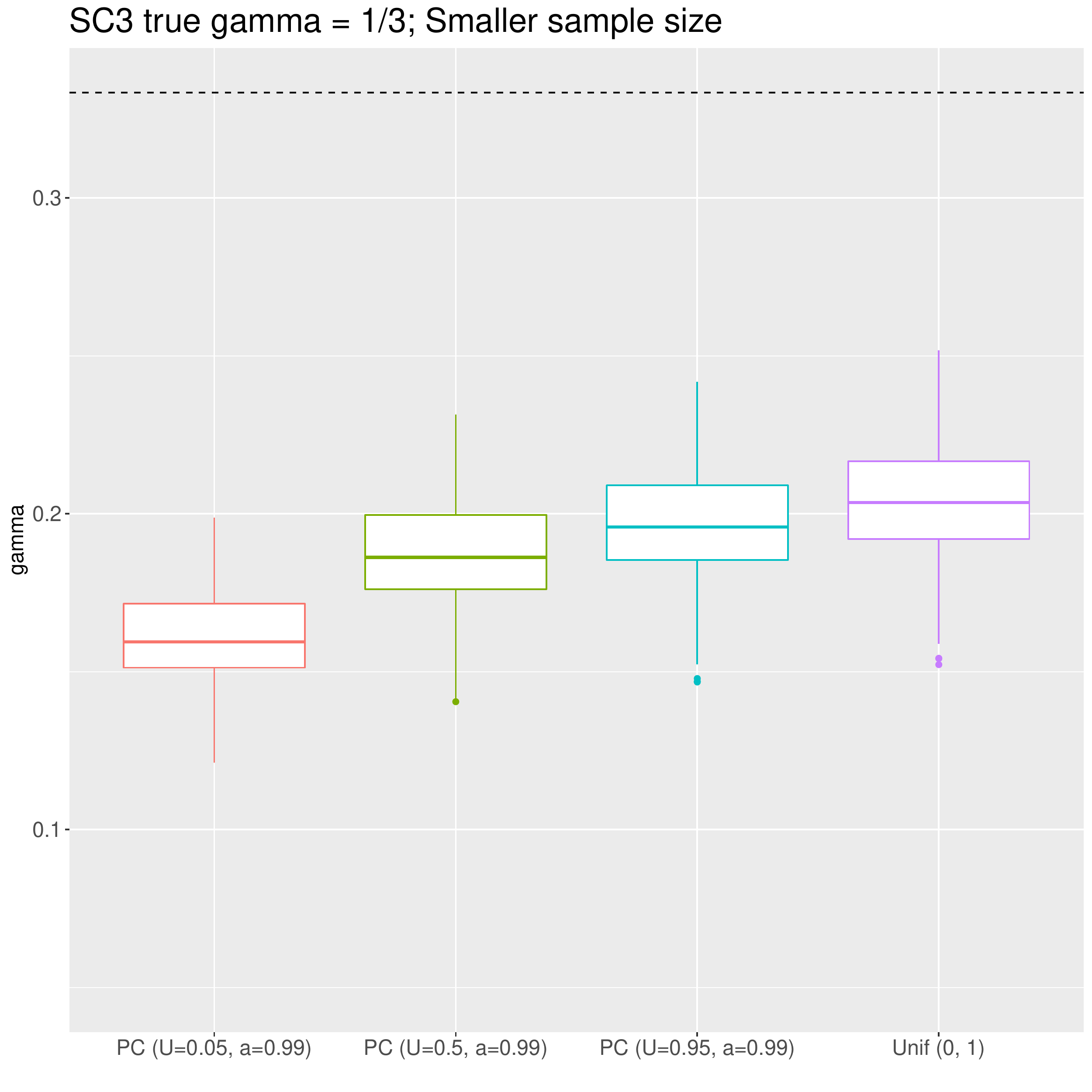}
\includegraphics[width=0.32\textwidth]{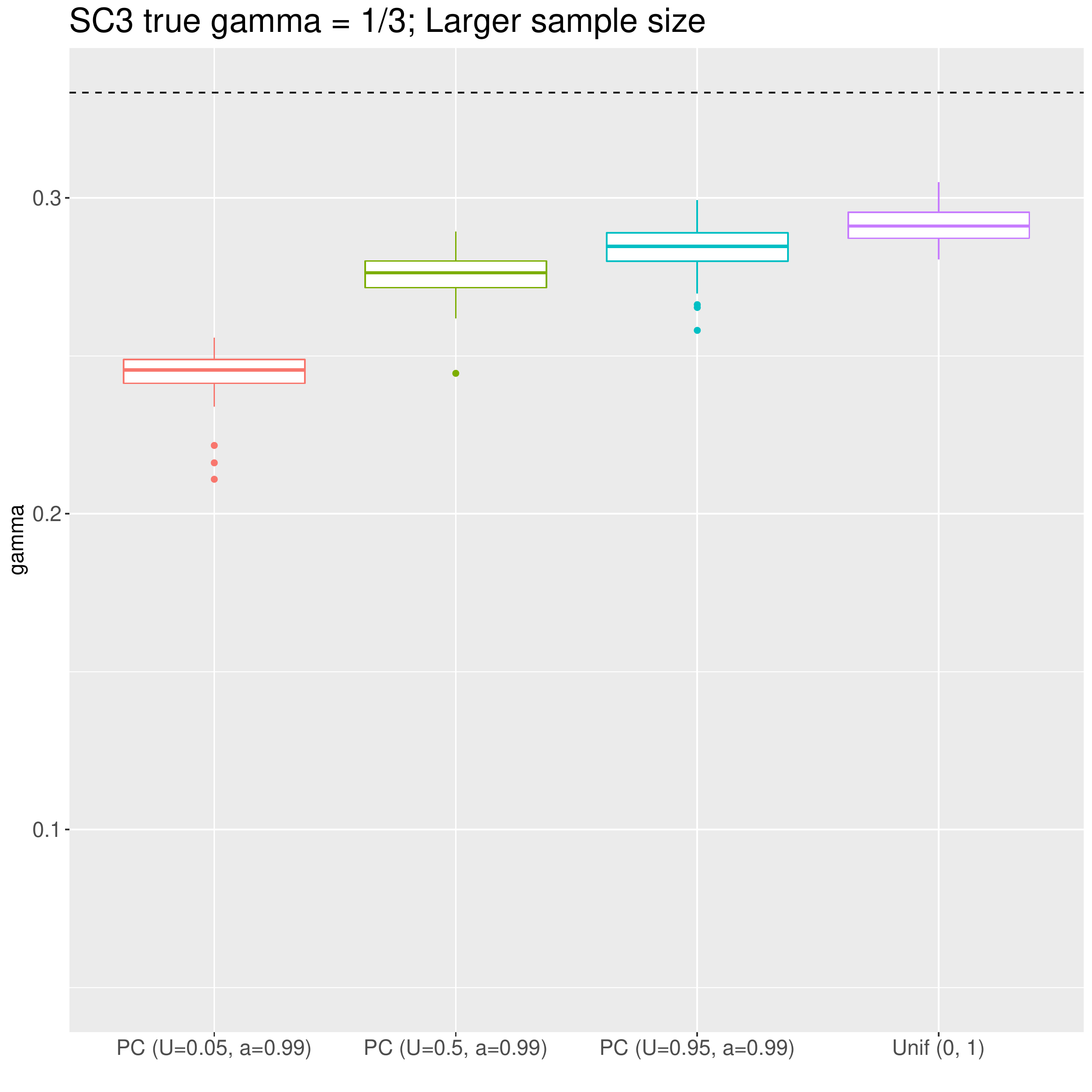}
\caption{Simulation results for the mixing parameter $\gamma$, scenario SC3; true $\gamma=1/3$.}
\label{fig:sim_gamma_SC3}
\end{figure}

\begin{figure}[H]
\includegraphics[width=0.32\textwidth]{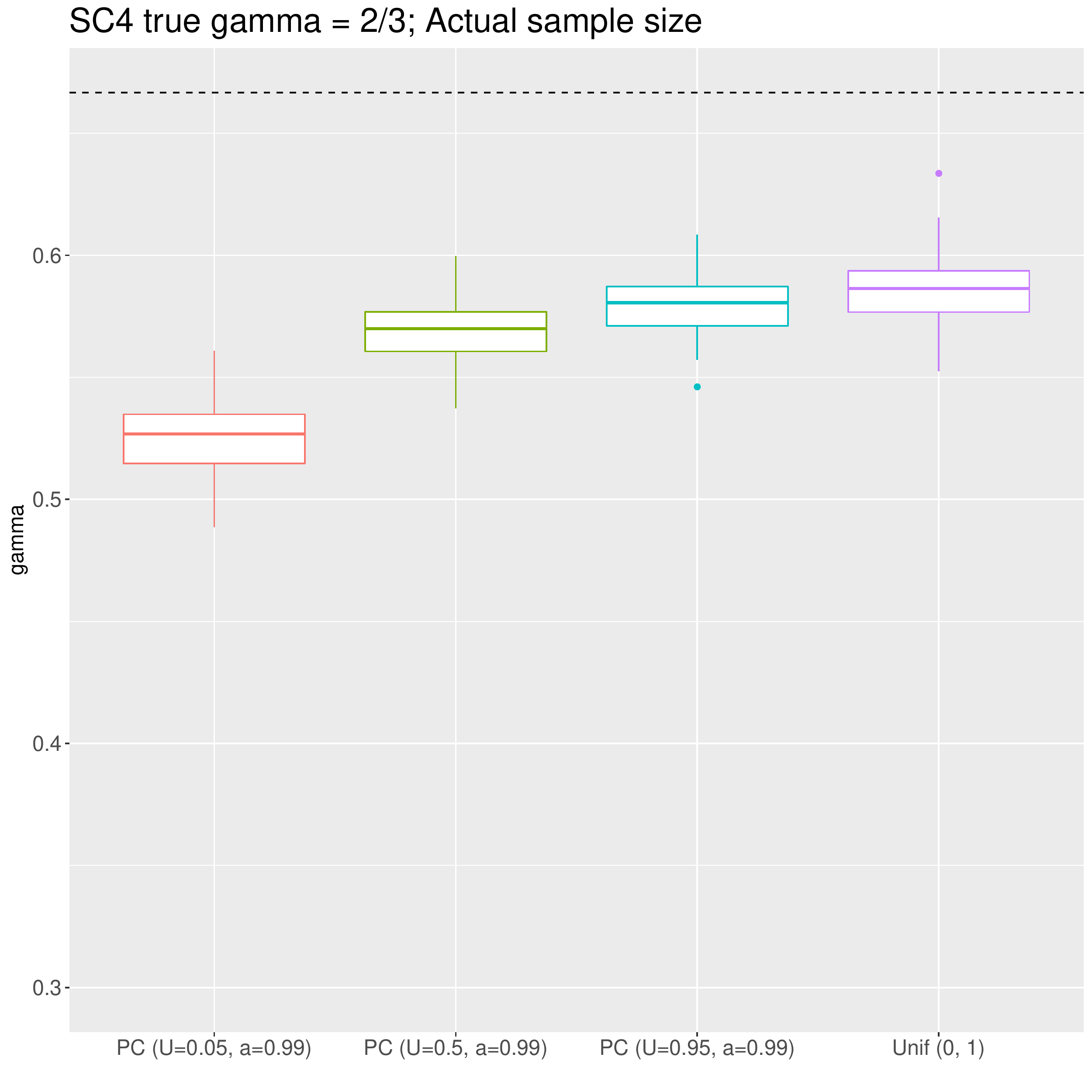}
\includegraphics[width=0.32\textwidth]{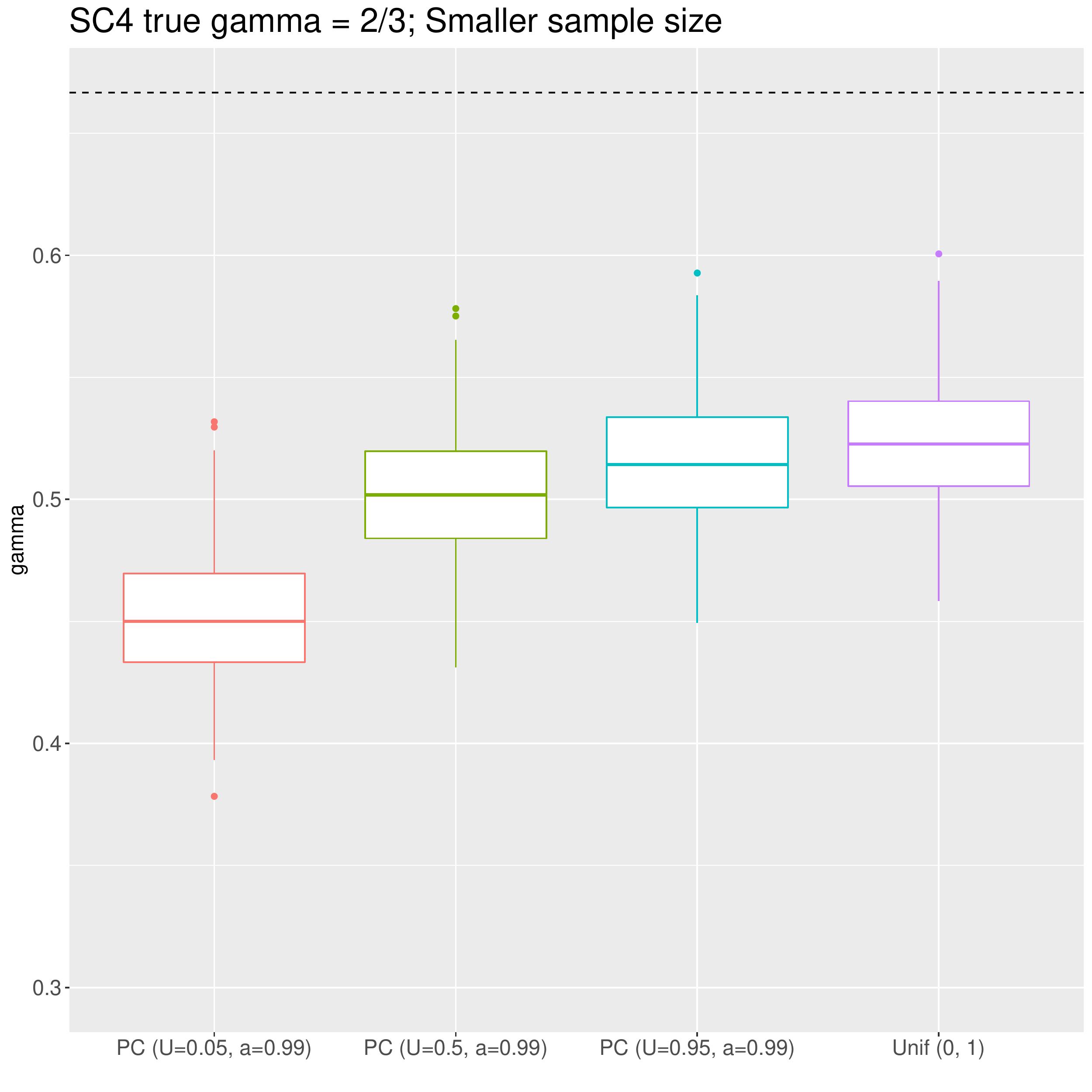}
\includegraphics[width=0.32\textwidth]{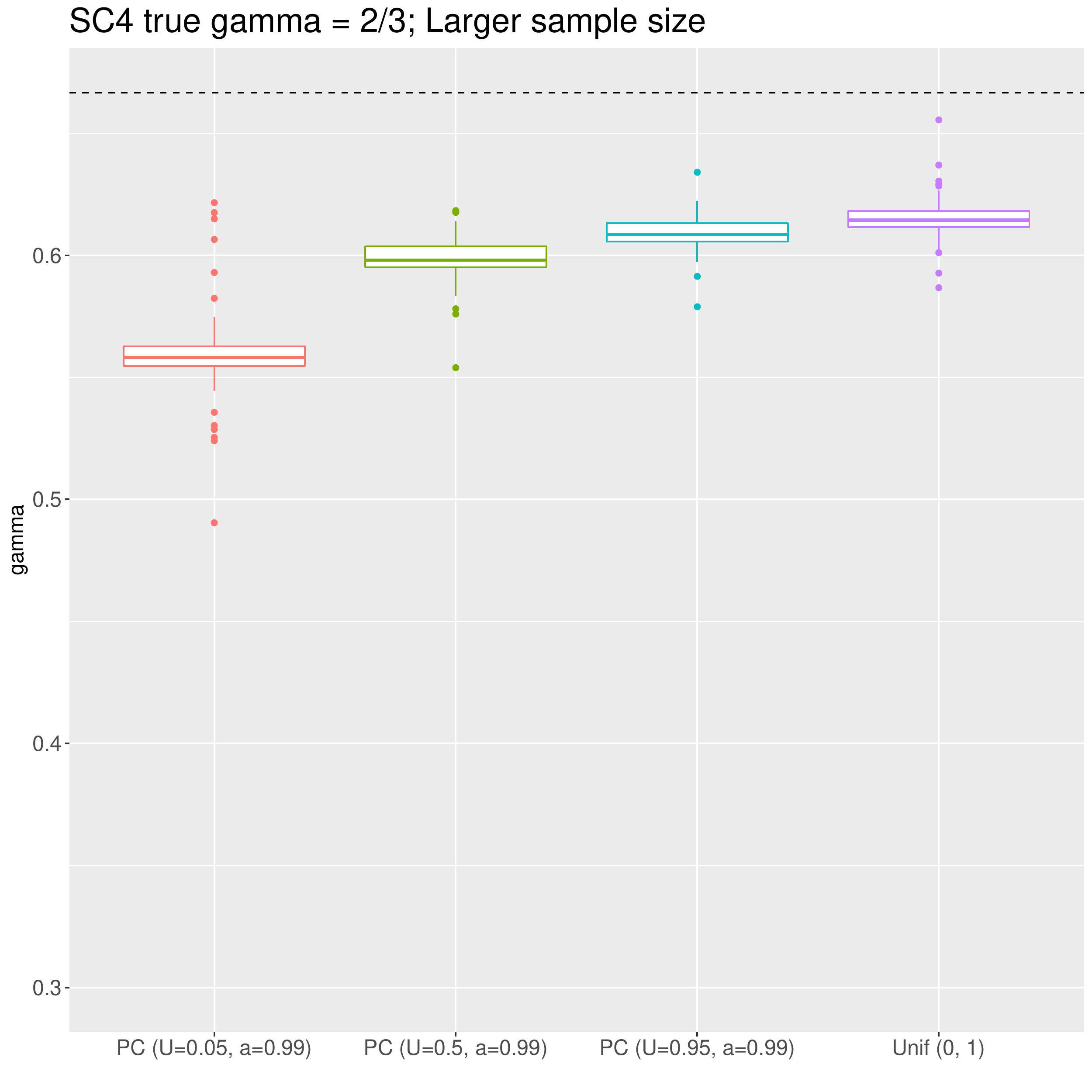}
\caption{Simulation results for the mixing parameter $\gamma$, scenario SC4; true $\gamma=2/3$.}
\label{fig:sim_gamma_SC4}
\end{figure}

Figures \ref{fig:sim_gamma_SC1} to \ref{fig:sim_gamma_SC4} report the boxplots of the posterior mean of $\gamma$ obtained by fitting the VP model to the 100 simulated datasets under the four scenarios SC1, SC2, SC3 and SC4. The horizontal dashed line represents the true $\gamma$ set by simulation in each scenario.  Each figure has three panels that refer to actual (left), smaller (central) and larger (right) sample size cases. The four boxplots in each panel correspond to different priors on $\gamma$: the PC priors with scalings $U=\{0.05,0.5,0.95\}$ and the Uniform prior. 

Regarding SC1 (Figure \ref{fig:sim_gamma_SC1}), where the true $\gamma$ is zero, we see that the Uniform prior implies a larger bias than the PC prior choices do, which is presumably due to the Uniform being prone to overfitting. 
This behaviour is more evident in the small sample size case, as a result of the data being less informative about the proportion of variance explained by the interaction. In scenarios where the true $\gamma>0$ (i.e. SC2, SC3 and SC4) we generally observe a negative bias under all prior choices, however the bias is smaller as the sample size increases.

Regarding the first aim of the study, i.e. checking the ability to recover the true $\gamma$ set by simulation, we can conclude that estimation of $\gamma$ is reasonable in all cases. We would like to emphasize that while the bias achieved under the uniform prior is always slightly smaller than the bias obtained by the PC priors in scenarios SC2, SC3 and SC4, it becomes much larger in SC1 because of the tendency to overfitting of the uniform. This highlights the fundamental advantage of PC priors which avoid overfitting by default as they shrink to the base model $\gamma=0$ by construction. 

Regarding our second aim, i.e. studying sensitivity of the results to the choice of $\theta$, we notice that as long as the unflexible choice of $U=0.05$ is avoided, the mixing $\gamma$ is estimated fairly well using the moderate and flexible choices, $U=0.5$ or $0.95$. In particular, $U=0.5$ or $U=0.95$ return comparable estimates of the mixing parameter $\gamma$ under all scenarios. From these results, we suggest that in absence of strong prior information on $\gamma$ the choice of a PC prior with $U=0.95, a=0.99$ is a reasonable weakly informative prior on $\gamma$ that allows flexibility and at the same time avoids model overfitting. 

As regards estimation of $\phi$ and $\tau$, results (not reported here) show that the true values $\phi=0.5$ and $\tau=1$ are accurately estimated in all scenarios by all priors.


\newpage

\section{Additional material on Ohio and Covid-19 examples} \label{appendix:additional_ohio}
\label{app:more_about_ohio}

\begin{figure}[h]
\includegraphics[width=0.45\textwidth]{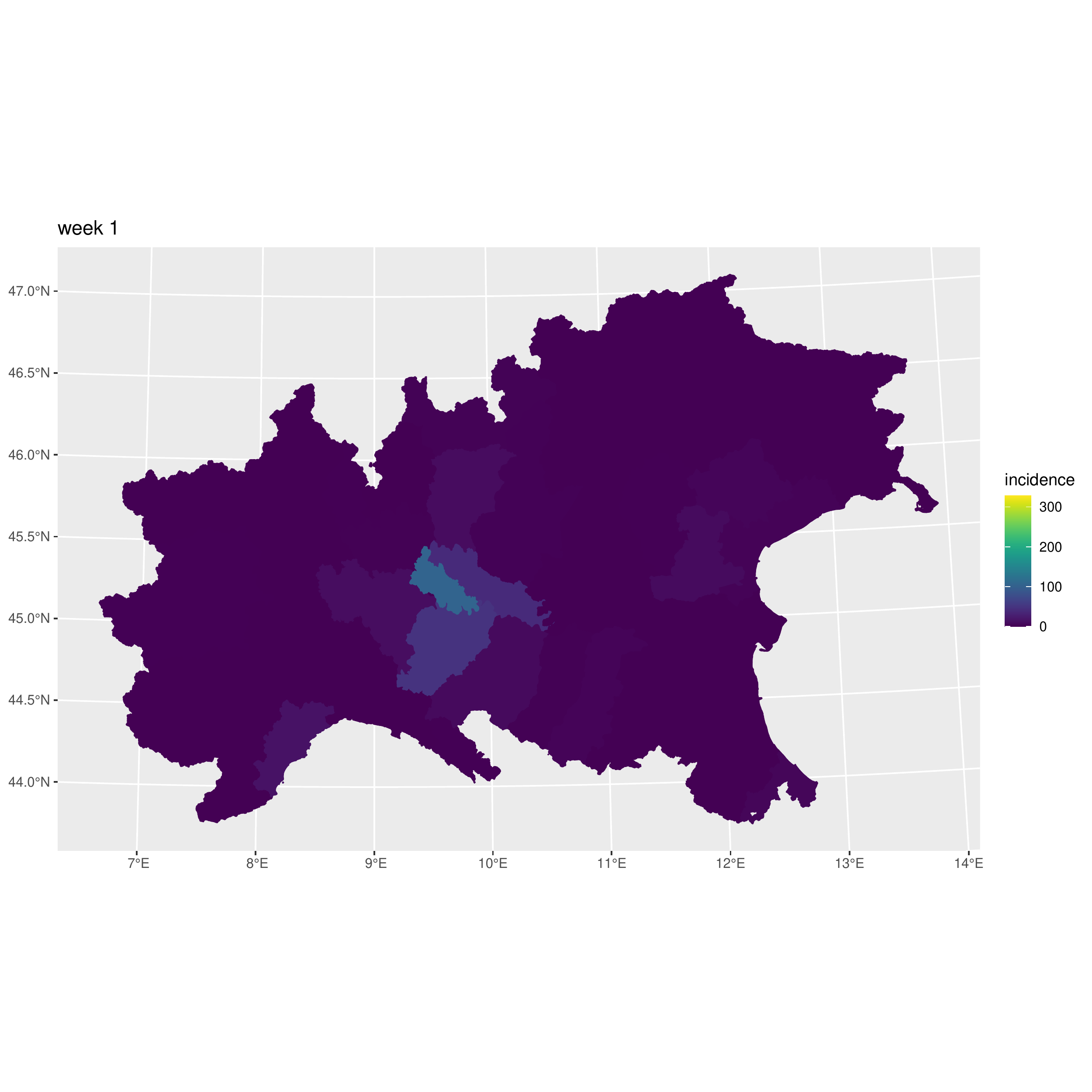}
\includegraphics[width=0.45\textwidth]{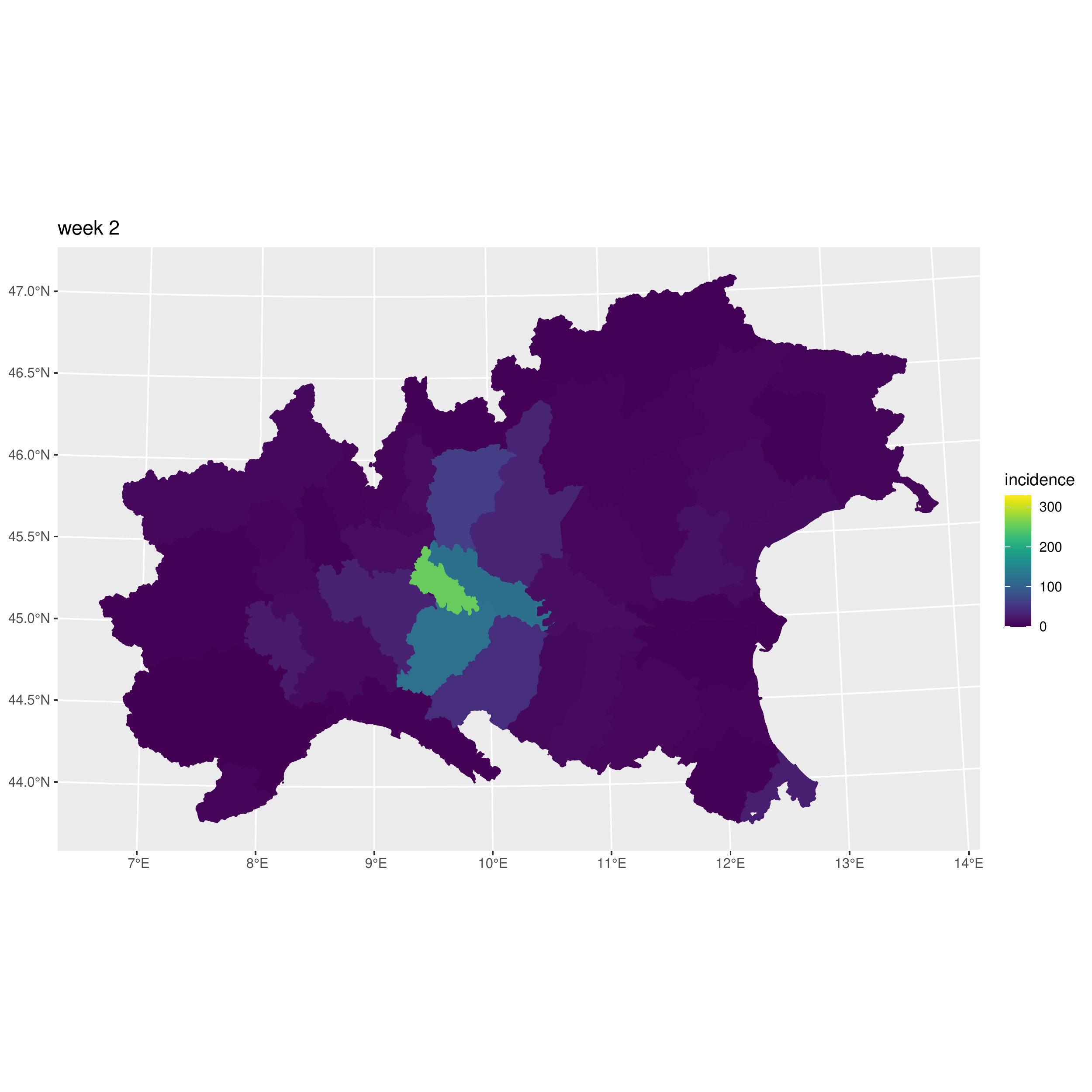}\\
\includegraphics[width=0.45\textwidth]{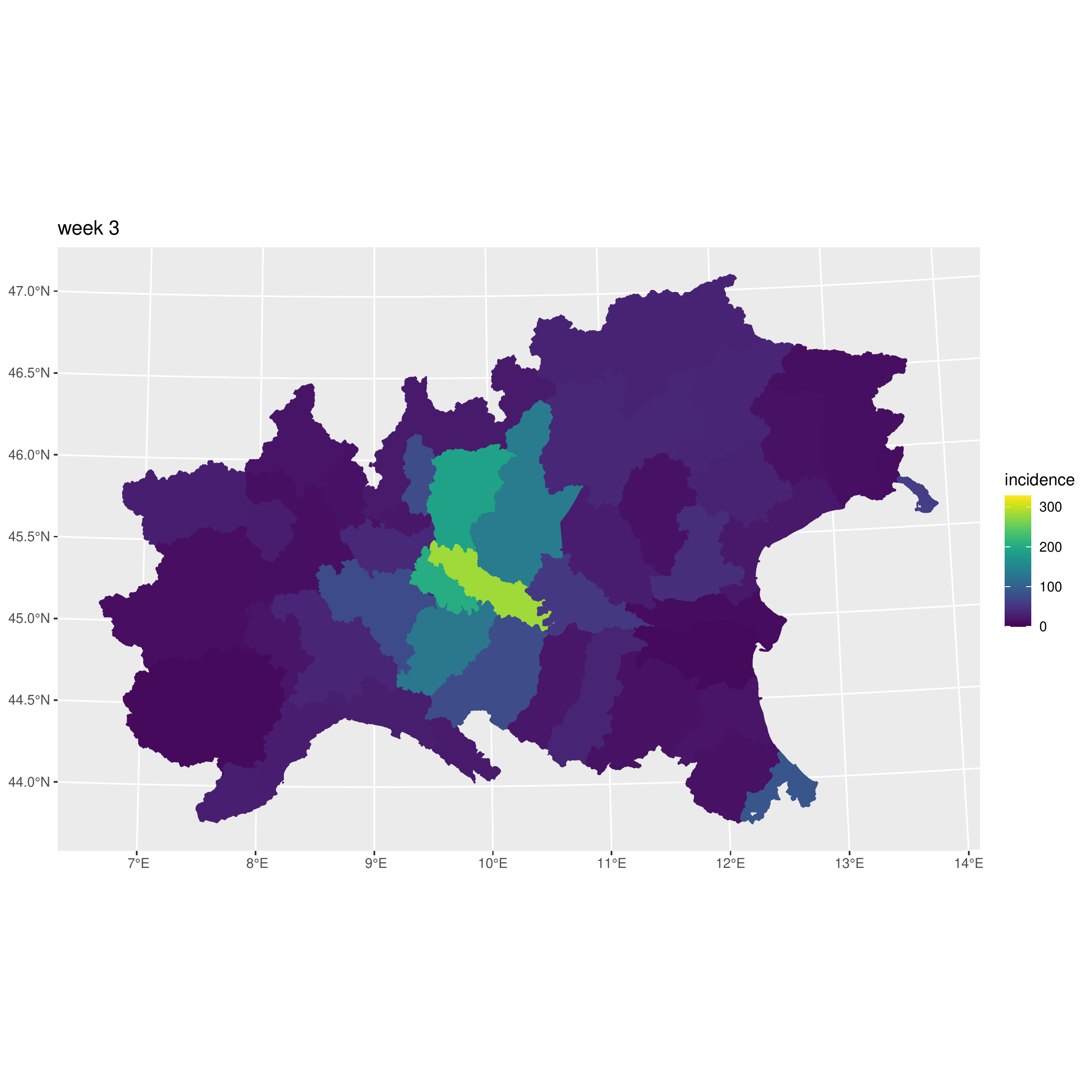}
\includegraphics[width=0.45\textwidth]{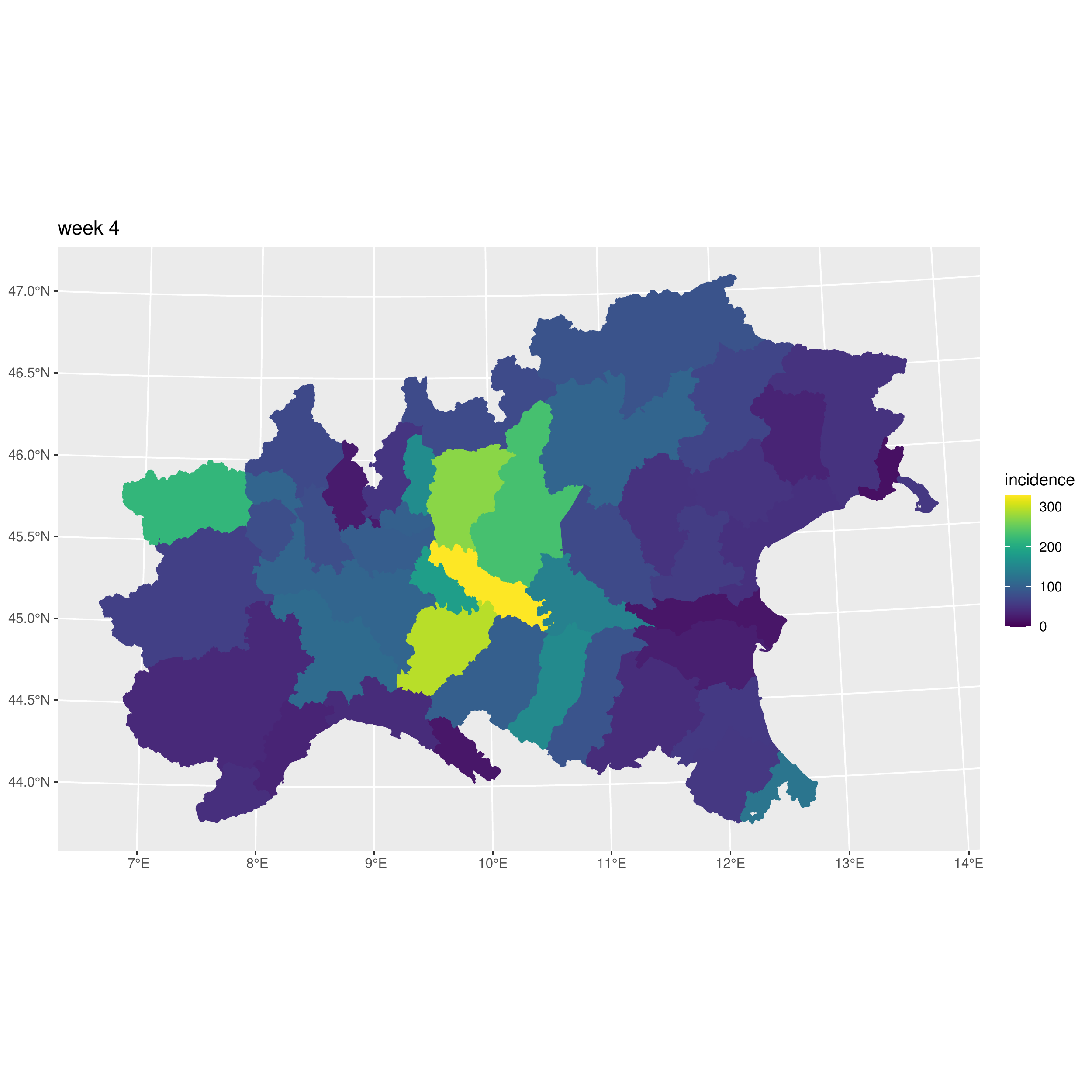}\\
\includegraphics[width=0.45\textwidth]{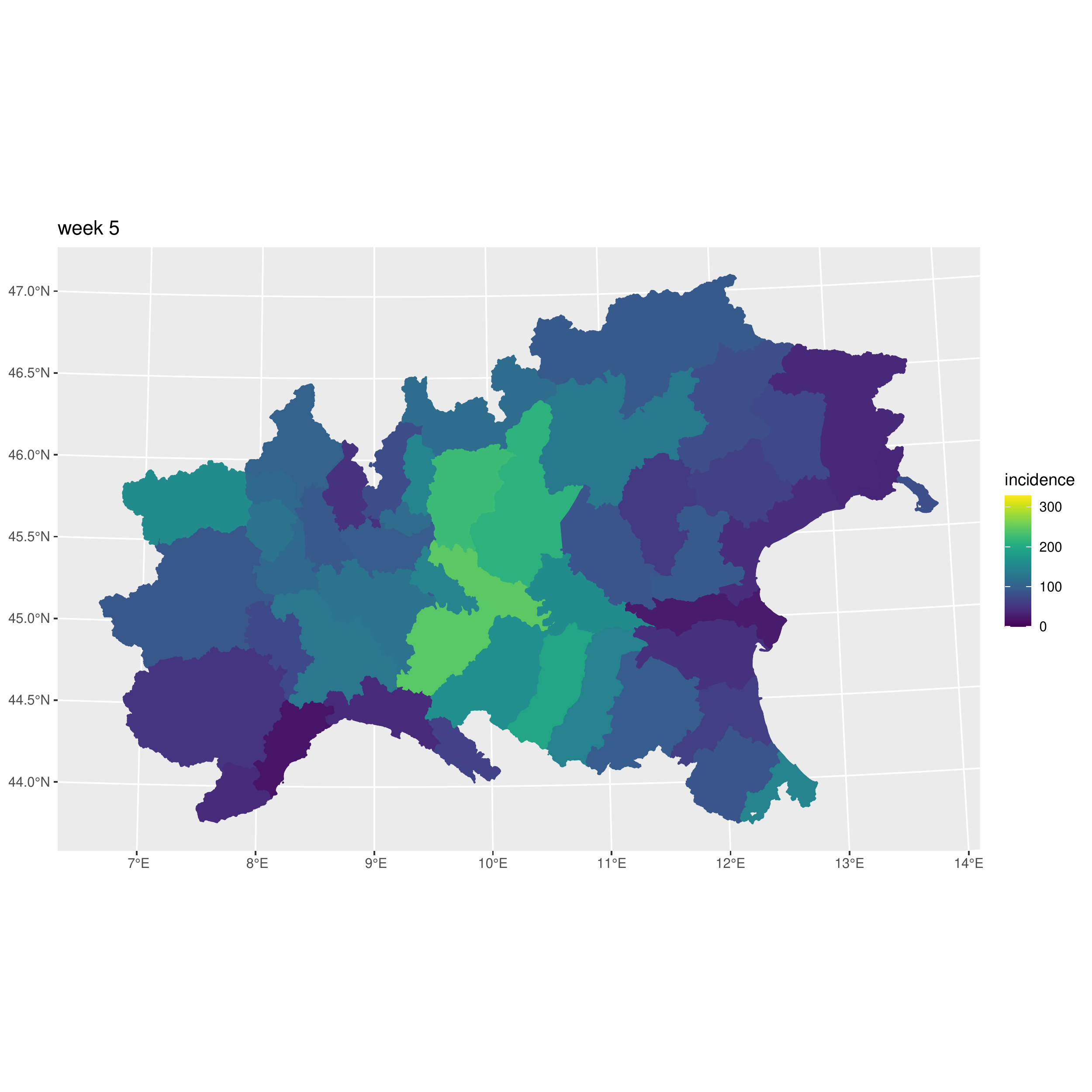}
\includegraphics[width=0.45\textwidth]{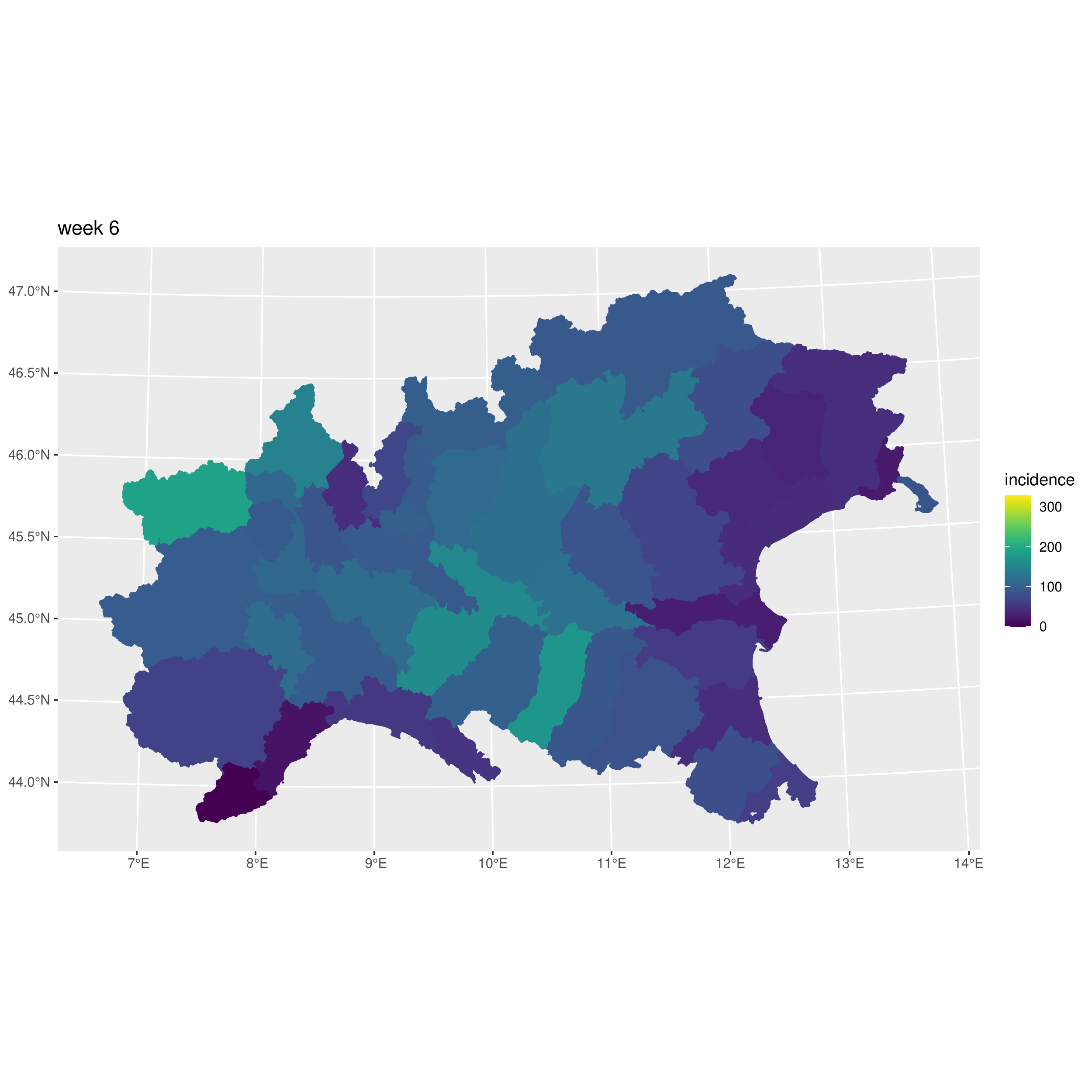}
\caption{Observed weekly cases of Covid-19  per 100000 residents in North Italy during first wave, from week 1 to week 6.}\label{fig:covid_vpplot_space_time}
\end{figure}

\begin{figure}[h]
\includegraphics[width=0.45\textwidth]{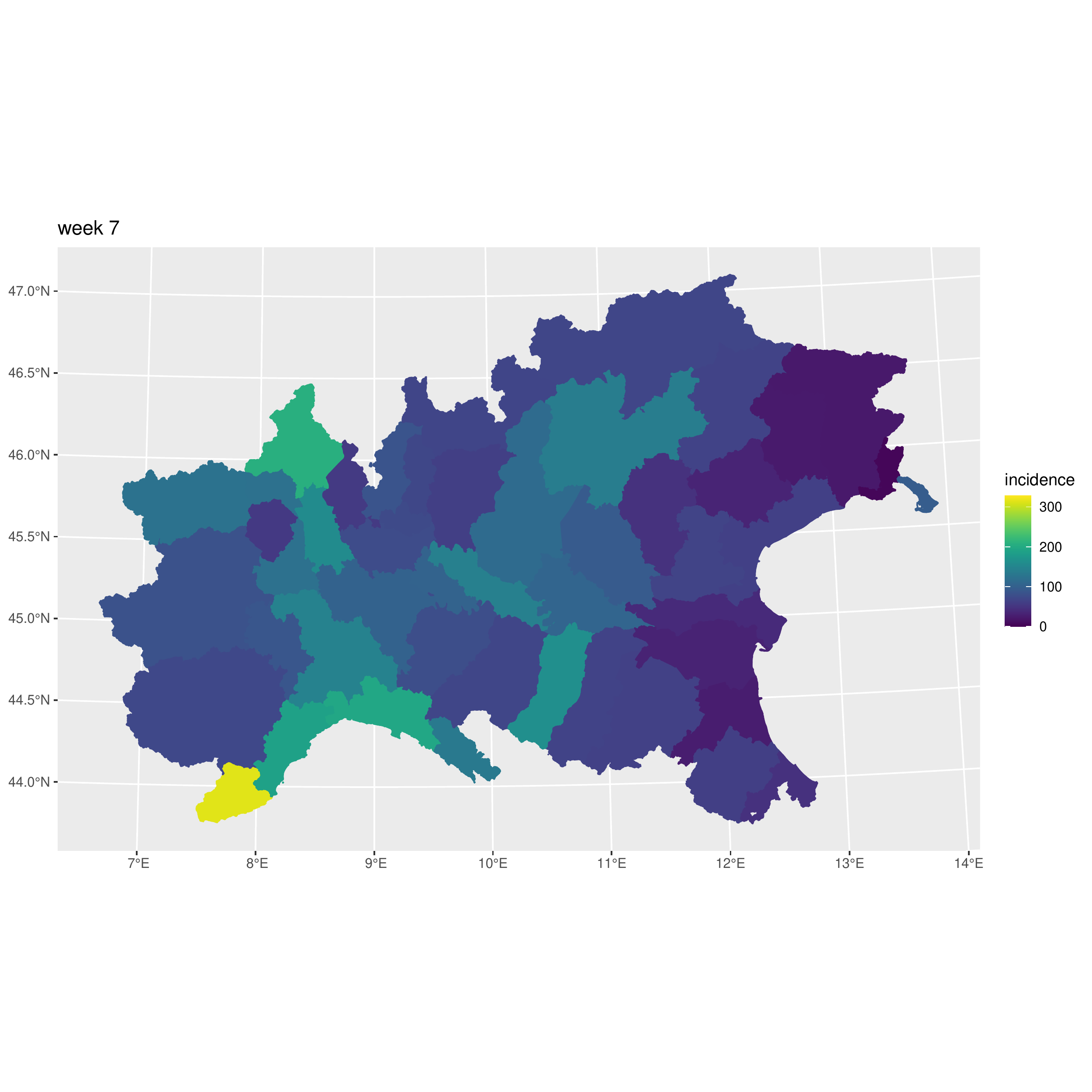}
\includegraphics[width=0.45\textwidth]{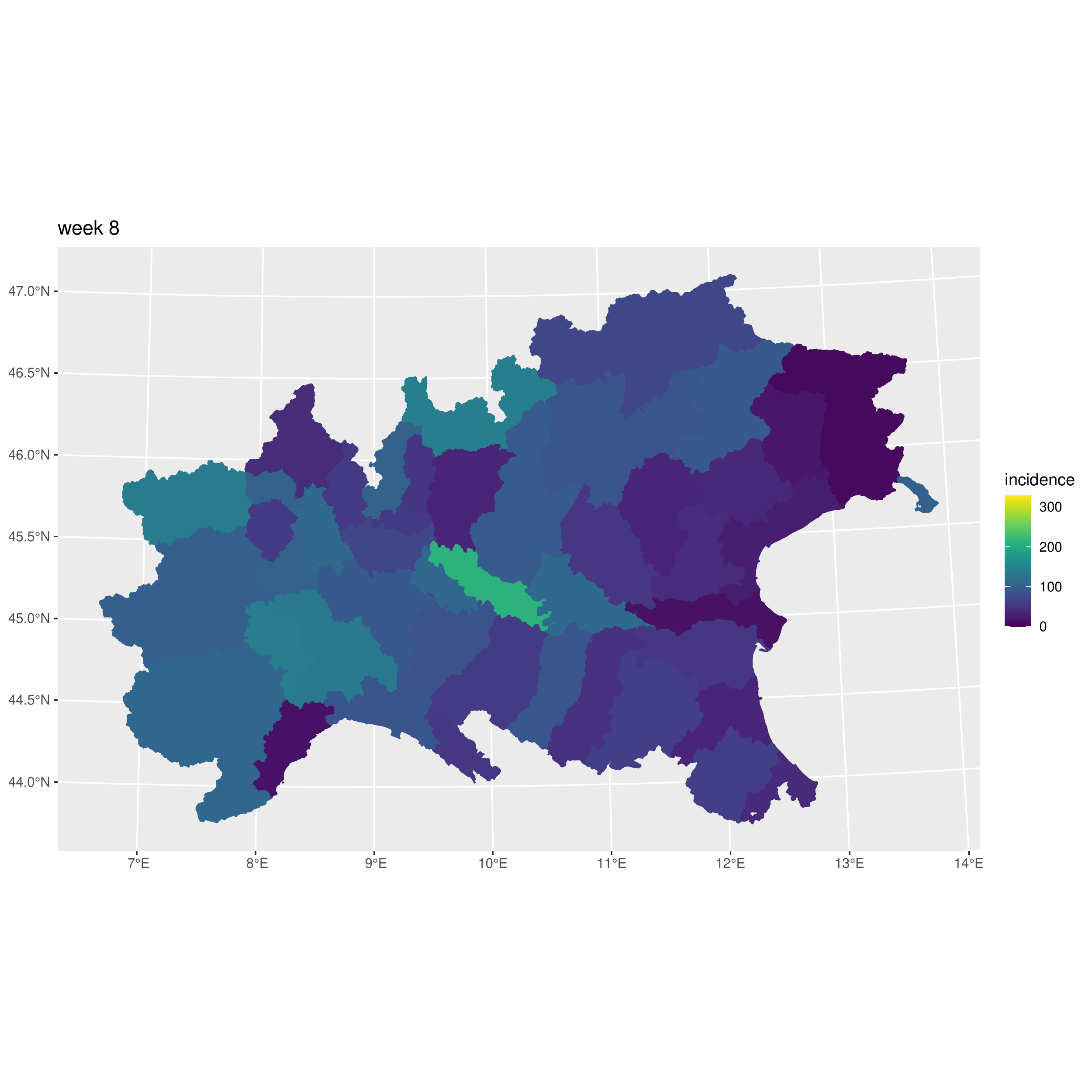}\\
\includegraphics[width=0.45\textwidth]{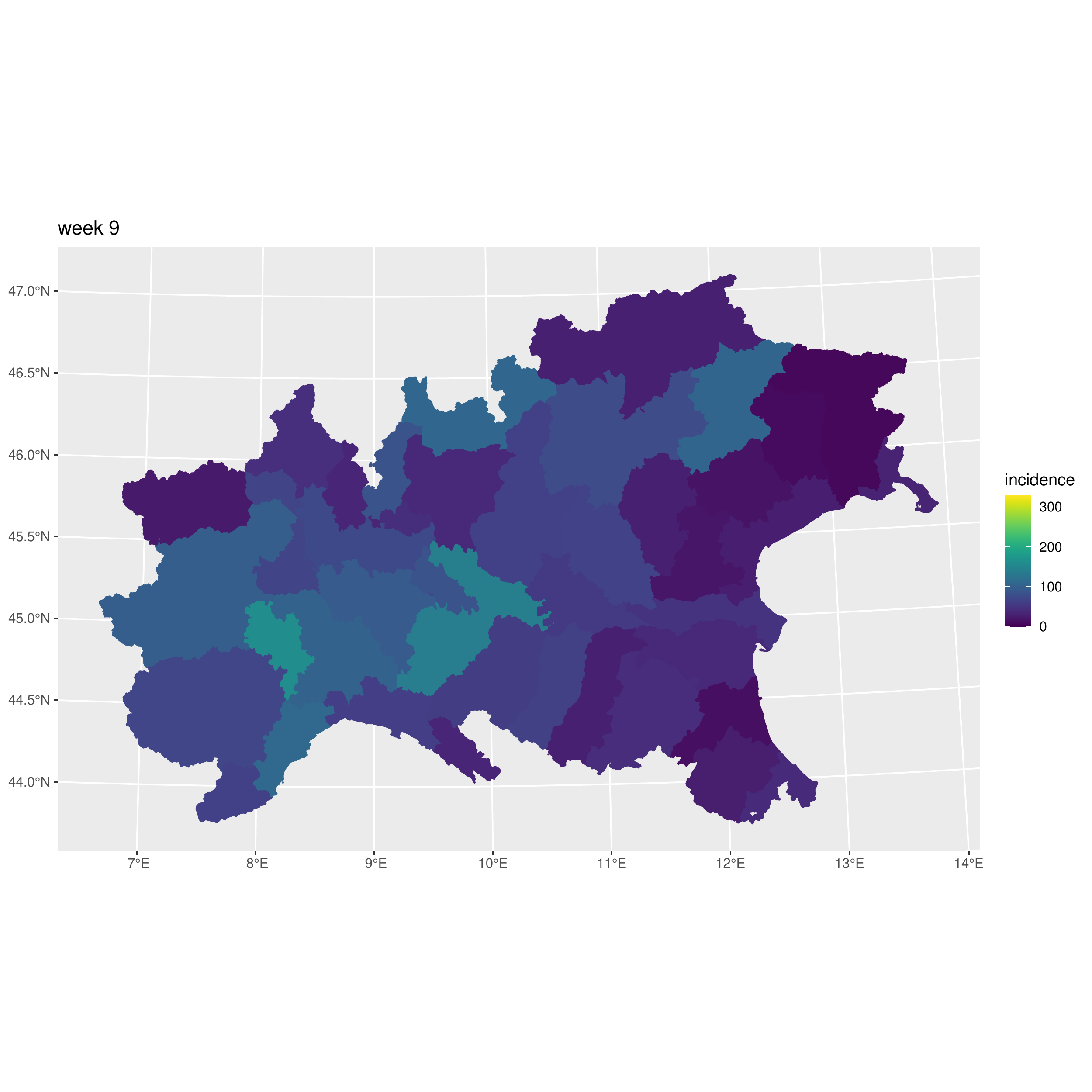}
\includegraphics[width=0.45\textwidth]{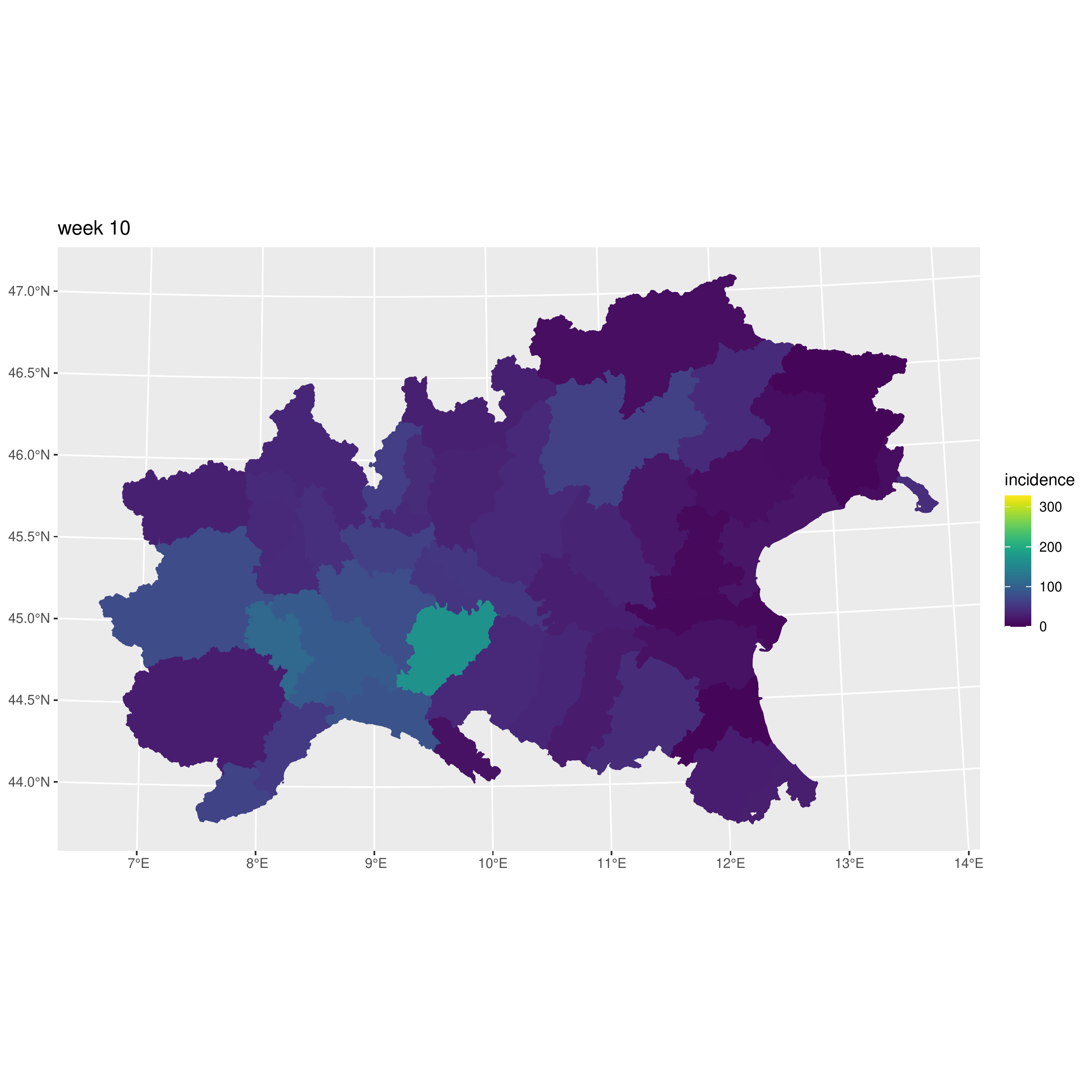}\\
\includegraphics[width=0.45\textwidth]{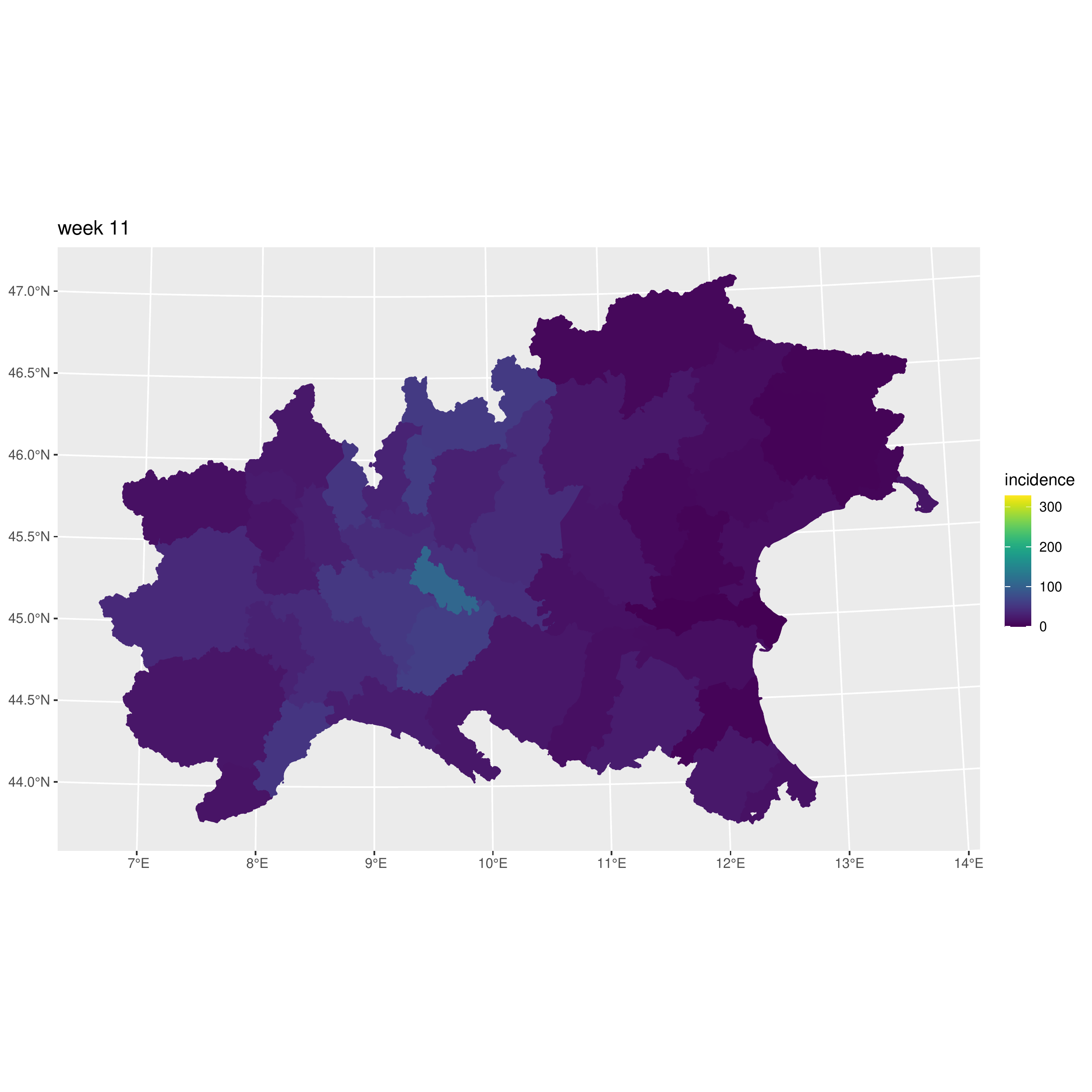}
\includegraphics[width=0.45\textwidth]{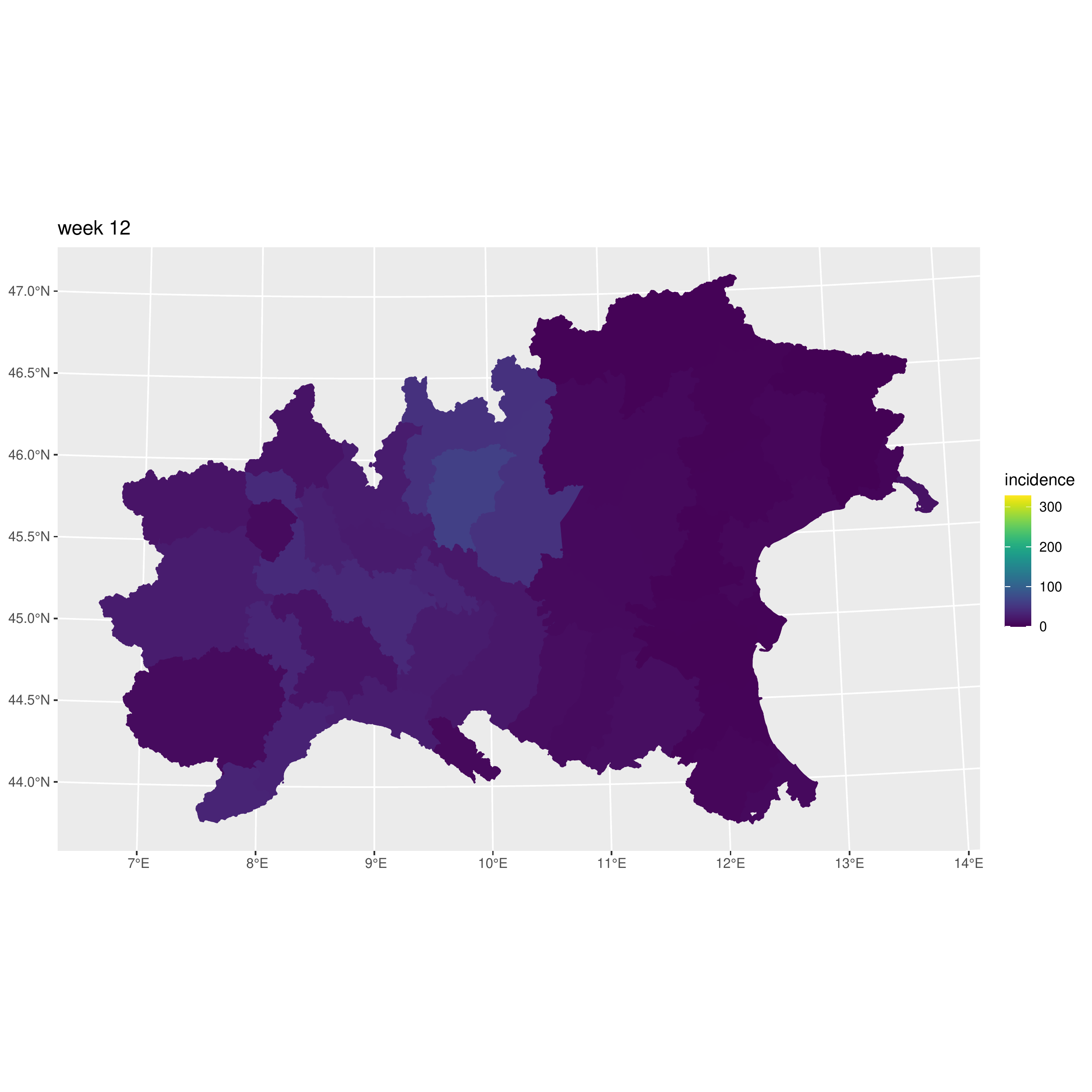}
\caption{Observed weekly cases of Covid-19  per 100000 residents in North Italy during first wave, from week 7 to week 12.}\label{fig:covid_vpplot_space_time}
\end{figure}

\begin{figure}[h]
\includegraphics[width=0.45\textwidth]{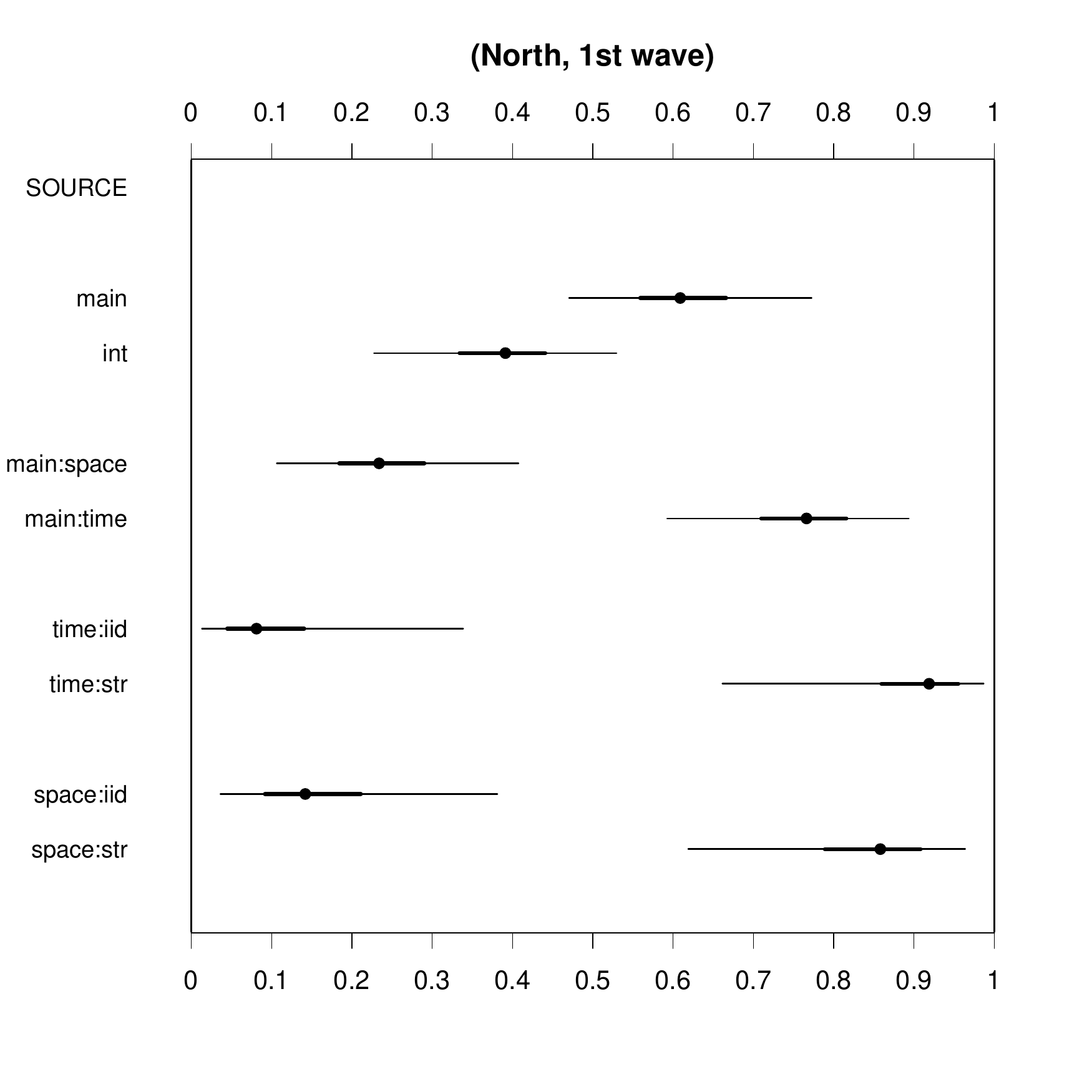}
\includegraphics[width=0.45\textwidth]{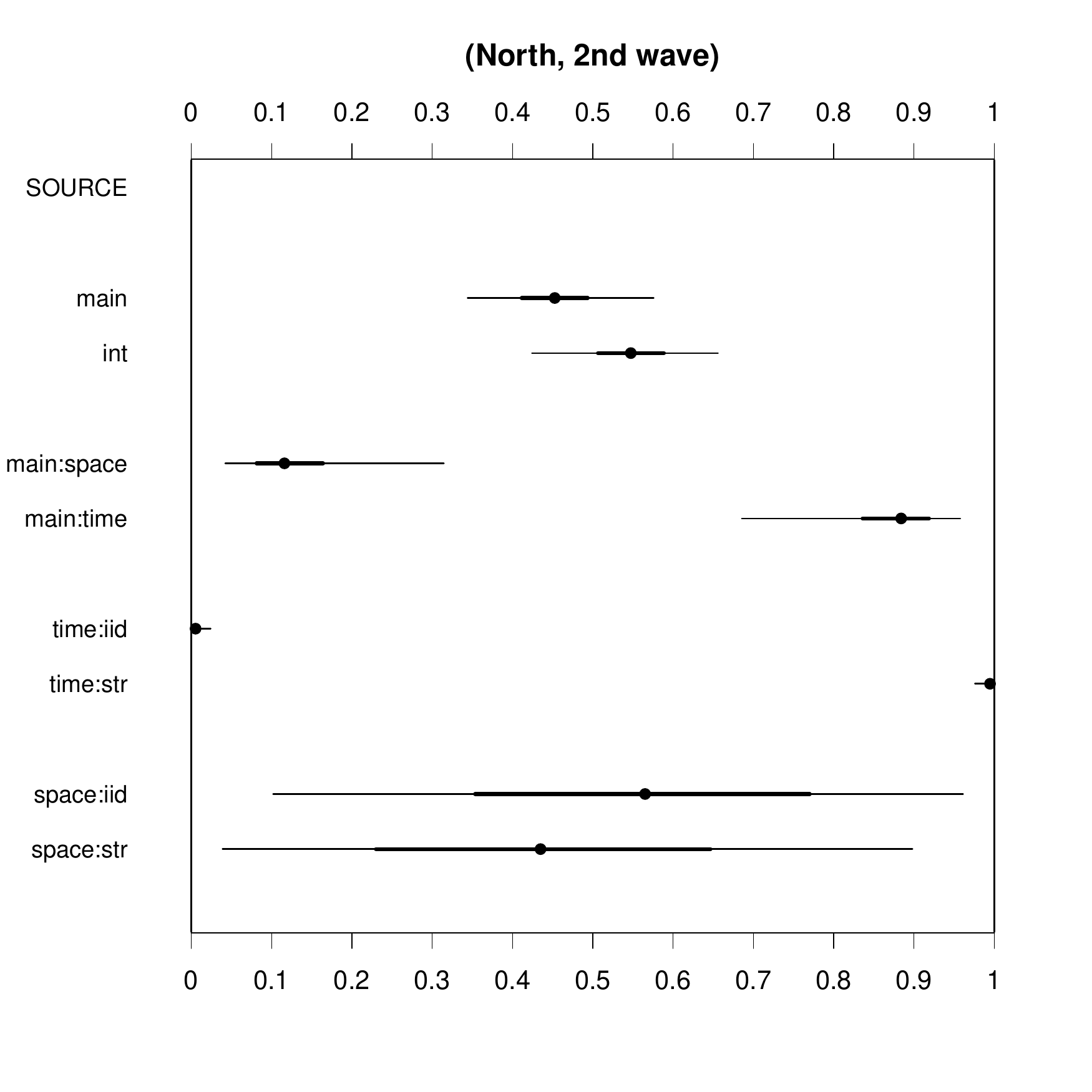}\\
\includegraphics[width=0.45\textwidth]{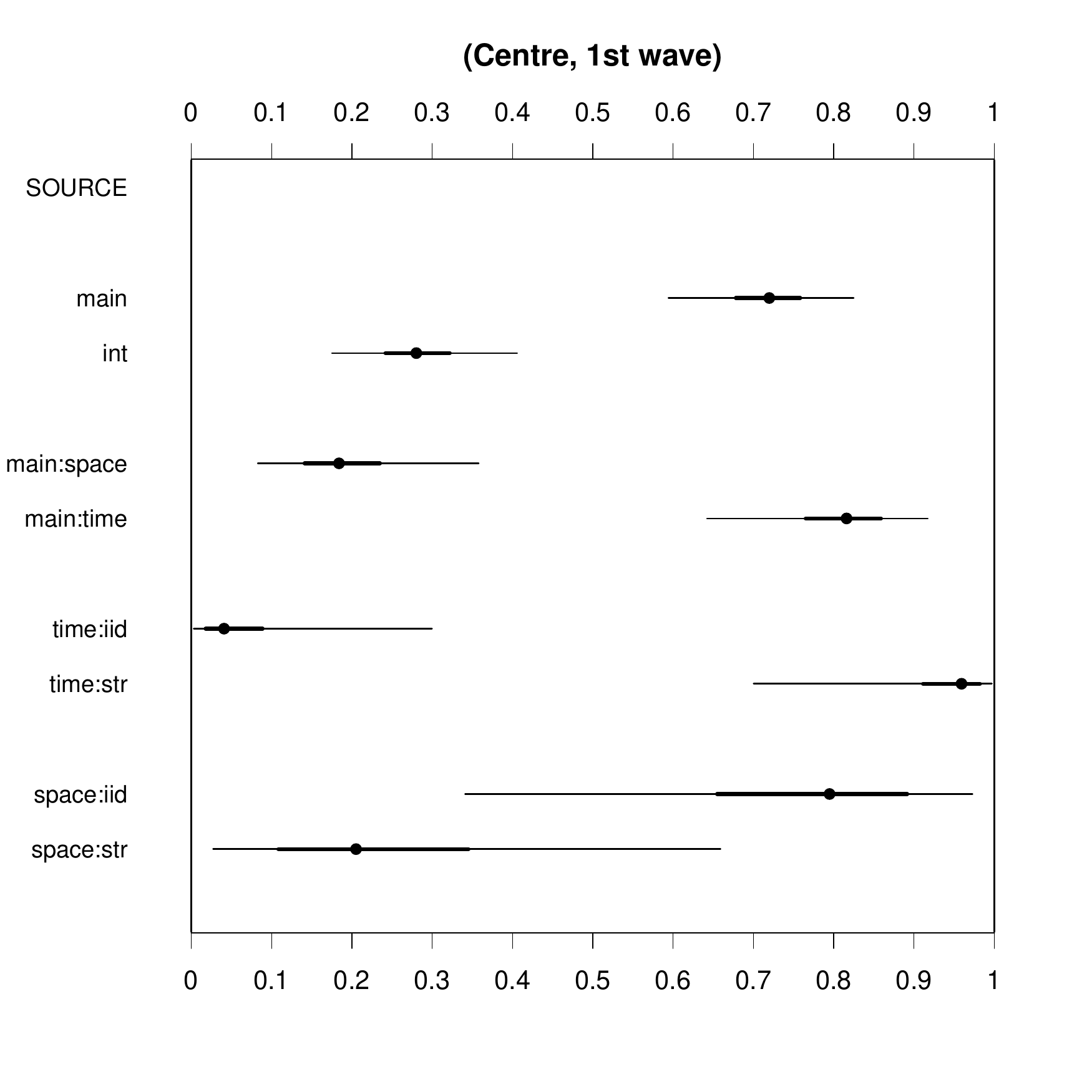}
\includegraphics[width=0.45\textwidth]{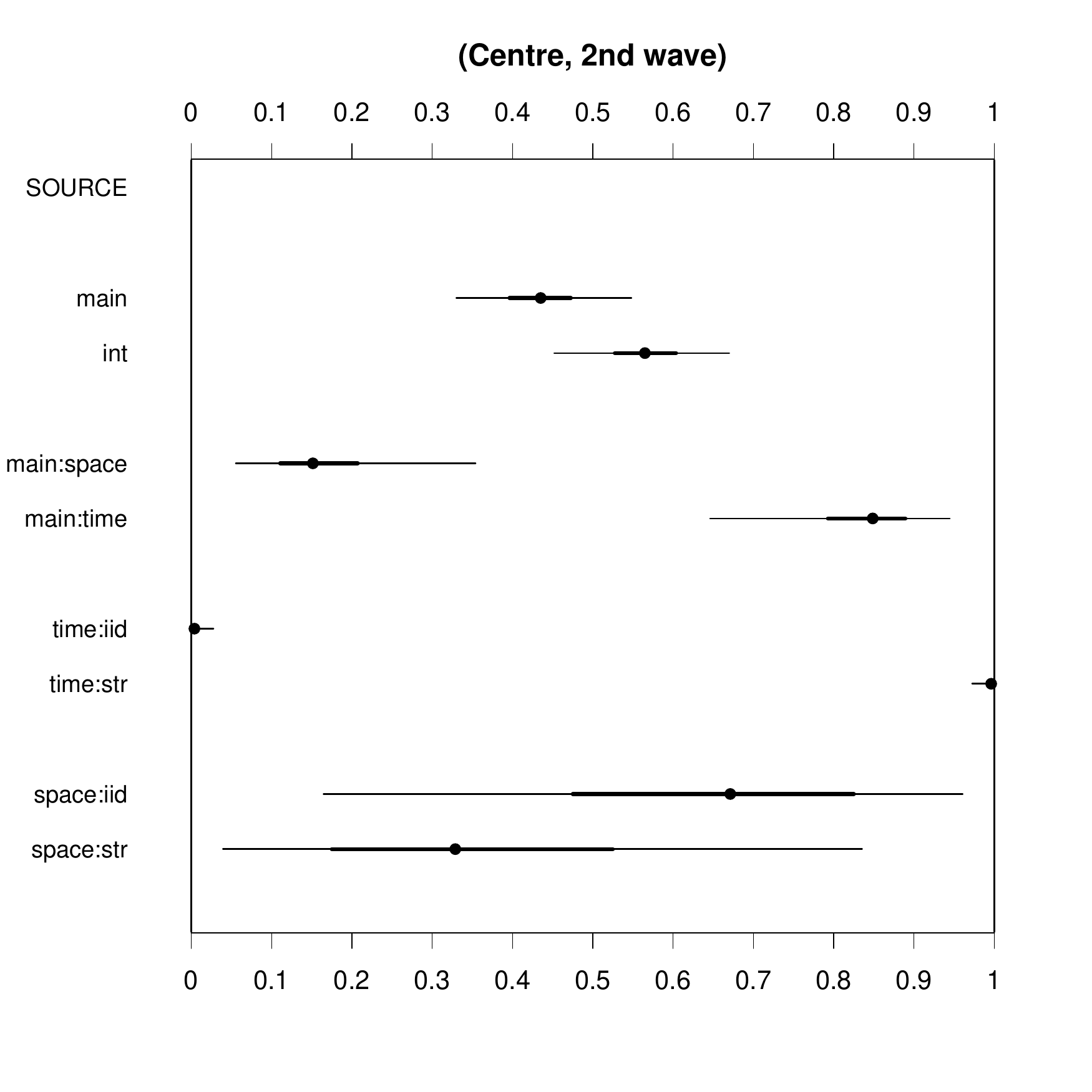}\\
\includegraphics[width=0.45\textwidth]{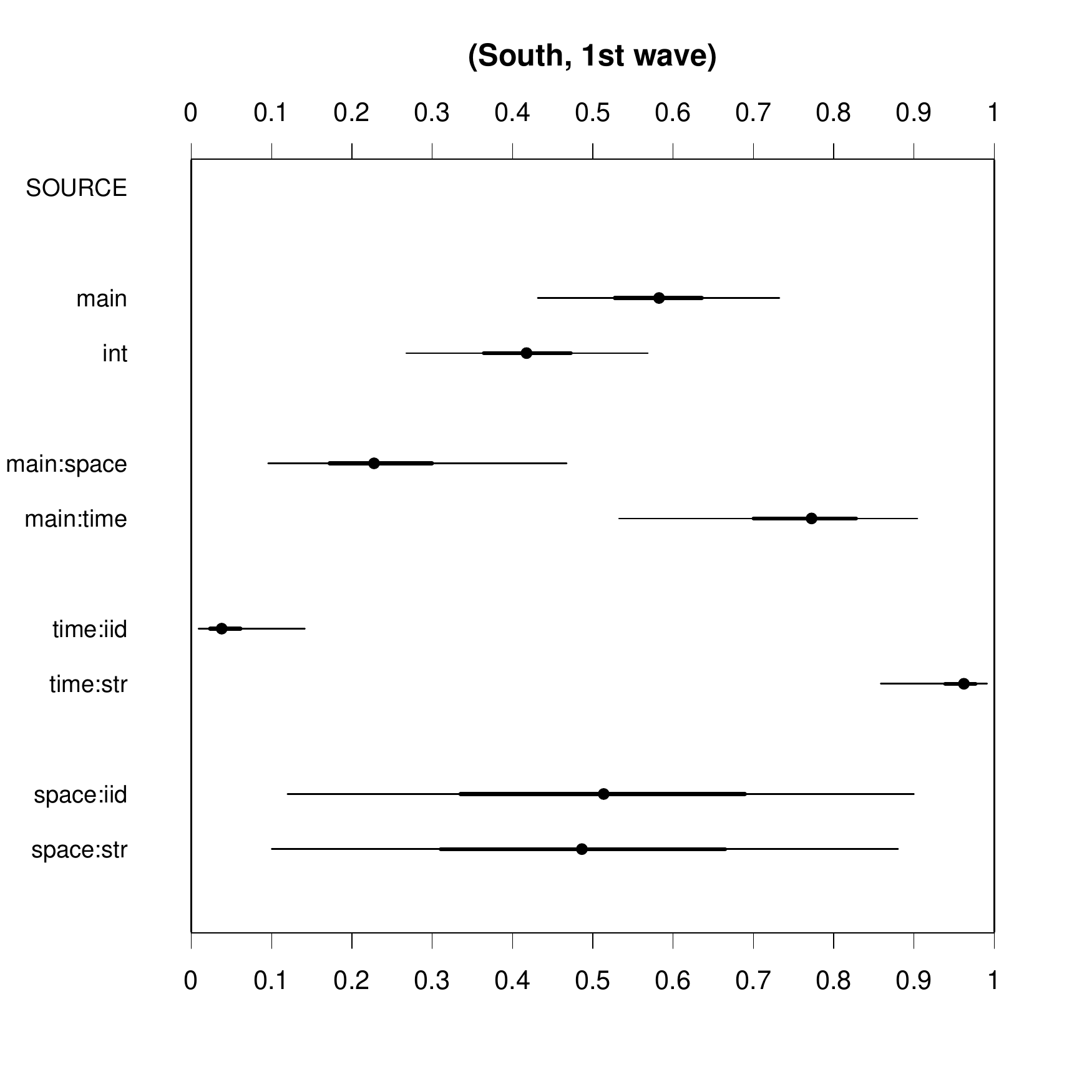}
\includegraphics[width=0.45\textwidth]{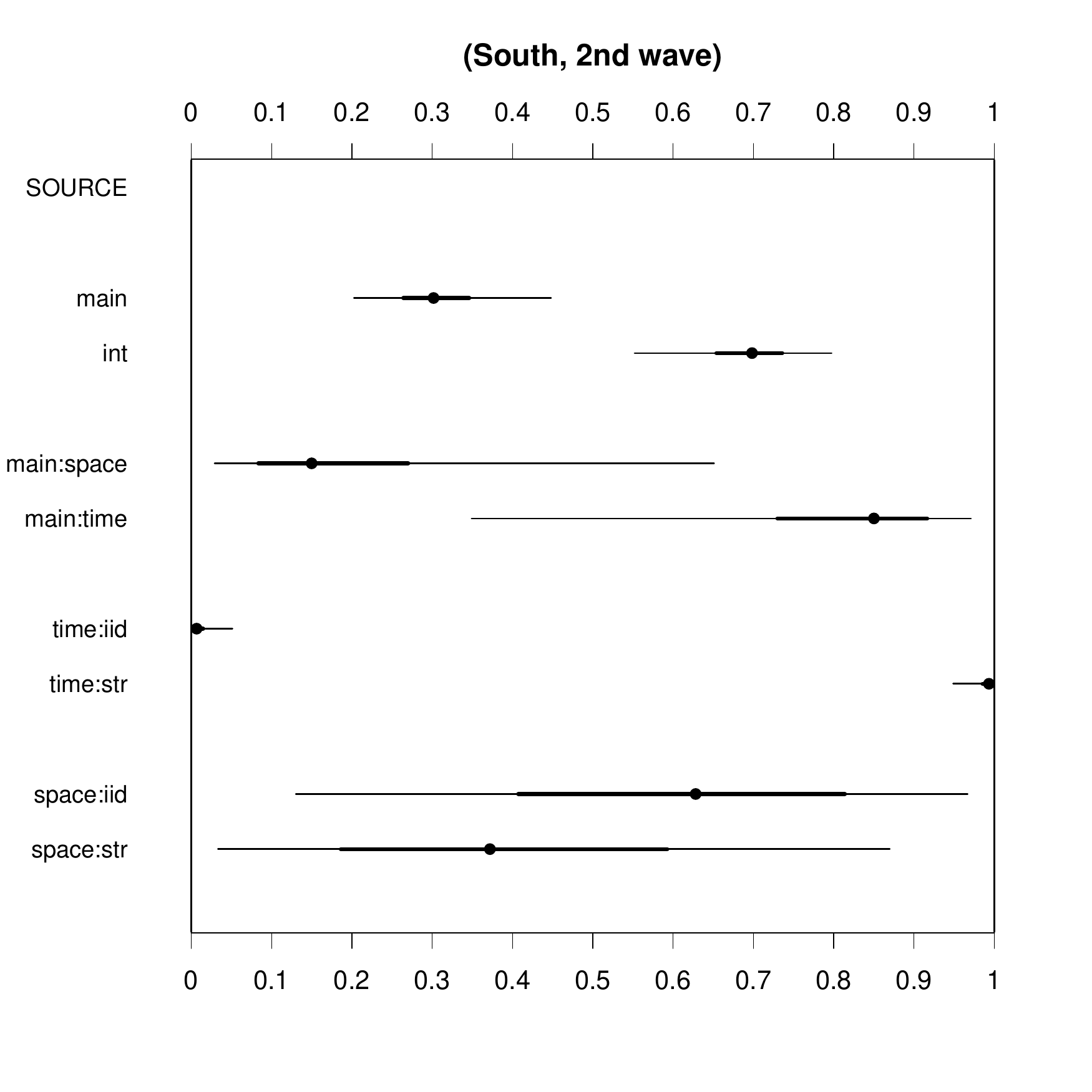}
\caption{Variance partitioning plots for the Covid-19 example; the analysis here refer to different subset of the data for each combination of the factors geographical area, with levels north (N), centre (C) and south (S), and pandemic wave, with levels W1 and W2.}\label{fig:covid_vpplot_space_time}
\end{figure}

\begin{landscape}
\begin{table}[h]
\small\sf\centering
\caption{Variance partitioning table for Ohio lung cancer, comparing the four interaction types.}
\begin{tabular}{llcccc}
 \hline 
{\bf source} & {\bf estimator}& {\bf type I} & {\bf type II} & {\bf type III} & {\bf type IV} \\
  \hline
\texttt{main }   &   $1-\hat\gamma$ &    0.952 (0.913, 0.979) & 0.958 (0.923, 0.981) & 0.973 (0.940, 0.991)& 0.960 (0.924, 0.983)\\
\texttt{int}     &  $\hat\gamma$ &    0.048 (0.021, 0.087) & 0.042 (0.019, 0.077) & 0.027 (0.009, 0.060)& 0.040 (0.017, 0.076)\\
  \hline
\texttt{main:space}& $\hat\phi$&    0.875 (0.765, 0.946) & 0.874 (0.763, 0.943) & 0.878 (0.770, 0.945)& 0.874 (0.758, 0.944)\\
\texttt{main:time}    & $1-\hat\phi$ & 0.125 (0.054, 0.235) &  0.126 (0.057, 0.237) & 0.122 (0.055, 0.230)  &0.126 (0.056, 0.242) \\
  \hline
\texttt{main:time:iid} &$\hat\psi_1$& 0.069 (0.010, 0.229) & 0.058 (0.003, 0.221)& 0.050 (0.002, 0.203) & 0.056 (0.003, 0.214)\\
\texttt{main:time:str} & $1-\hat\psi_1$ & 0.931 (0.771, 0.990) & 0.942 (0.779, 0.997) & 0.950 (0.797, 0.998) &0.944 (0.786, 0.997) \\
  \hline
\texttt{main:space:iid} & $ \hat\psi_2$ & 0.658 (0.273, 0.925) & 0.693 (0.388, 0.917)& 0.706 (0.372, 0.927) & 0.676 (0.361, 0.912)\\
\texttt{main:space:str} &$1-\hat\psi_2$ & 0.342 (0.075, 0.727) & 0.307 (0.083, 0.612)  &0.294 (0.073, 0.628) & 0.324 (0.088, 0.639)\\
 \hline 
\end{tabular}
\label{tab:ohio_anova}
\end{table}
\end{landscape}

\newpage

\section{R code}
\label{appendix:Rcode}
Below the \RINLA code to fit model (6) in our paper, with type 4 interaction, to the Covid-19 dataset in Section 4.2. Note that the model can be estimated using the usual \texttt{inla} call; the R package \texttt{inlaVP} was written to aid the user in setting the interaction type, building the constraints and defining the joint prior. The R package \texttt{inlaVP} is not on CRAN yet, but it is available on github.

\begin{lstlisting}
rm(list=ls())
library(INLA)
# install inlaVP using devtools
library(devtools)
install_github("massimoventrucci/inlaVP")
library(inlaVP) 

## load the data and create interaction index
data(covid_italy)
n1 <- length(unique(covid_italy$id.week))
n2 <- italy_graph$n
dat.tmp <- expand.grid(id.week=1:n1,
                       id.province=1:n2)
dat.tmp$id.int <- 1:(n1*n2)
dat <- merge(covid_italy, dat.tmp,
             by=c("id.week", "id.province"),
             all.x=TRUE)
dat.sort <- dat[order(dat$id.int),] # IMP: sorting the interaction indices is needed

# the graph for Italy is disconnected (3 connected component 'cc'):
# set one separate intercept for each cc of size > 1
intercept <- rep(NA, italy_graph$n)
for(i in seq_along(italy_graph$cc$nodes))
  if (length(italy_graph$cc$nodes[[i]]) > 1) intercept[italy_graph$cc$nodes[[i]]] <- i
intercept <- as.factor(intercept)
dat.sort <- merge(dat.sort, data.frame(id.province=1:italy_graph$n, intercept.cc=intercept))

## inla call
library(INLA)
inla.setOption(num.threads = "1")
# setting 1 core is needed when using joint prior (jp) inside control.expert = list(jp = ...),

# define the interaction model
f.time <- m(dat.sort$id.week, igmrf.type = 'rw1')
f.space <- m(dat.sort$id.province, igmrf.type = 'besag', g=italy_graph)
set.int <- control.interaction(
  m1 = f.time,
  m2 = f.space,
  interaction.type = 4)

# define the joint prior
jp.vp.m2 <- function(theta, theta.desc = NULL) {
  ### the user must specify 'hyper', with the scaling parameters of the PC priors for tau and gamma:
  hyper <- list(prec=list(u=2/0.31, a=0.01),
                gamma=list(u=0.95, a=0.99))
  fun_striid <- function(theta)
  {
    tau <- inlaVP:::theta.to.tau.striid(theta)
    gamma <- inlaVP:::theta.to.gamma.striid(theta)
    phi <- inlaVP:::theta.to.phi.striid(theta)
    psi1 <- inlaVP:::theta.to.psi1.striid(theta)
    psi2 <- inlaVP:::theta.to.psi2.striid(theta)
    return(c(phi,gamma,tau,psi1,psi2))
  }

  if (!is.null(theta.desc)) {
    for(i in seq_along(theta.desc))
      print(paste0("    theta[", i, "]=", theta.desc[i]))
  }
  if (inlaVP:::theta.to.phi.striid(theta) >=0 & inlaVP:::theta.to.phi.striid(theta) <=1 &
      inlaVP:::theta.to.psi1.striid(theta) >=0 & inlaVP:::theta.to.psi1.striid(theta) <=1 &
      inlaVP:::theta.to.psi2.striid(theta) >=0 & inlaVP:::theta.to.psi2.striid(theta) <=1 ){
    lprior <-  INLA:::inla.pc.dprec(prec=inlaVP:::theta.to.tau.striid(theta),
                                    u= hyper$prec$u, alpha=hyper$prec$a, log=TRUE) +
      inlaVP:::pc.gamma(gamma=inlaVP:::theta.to.gamma.striid(theta),
                        lambda=inlaVP:::pcprior.interaction.lambda(
                          u=hyper$gamma$u, alpha=hyper$gamma$a),
                        log=TRUE) +
      log(abs(det(numDeriv:::jacobian(fun_striid, as.numeric(theta), method="Richardson"))))
  } else {
    lprior <- -.Machine$double.xmax
  }
  return(lprior)
}
jpr.vp <- inla.jp.define(jp.vp.m2)

# run inla
res.covid <- inla(y ~  1 + intercept.cc +
                    f(id.time,
                      model = 'rw1',
                      constr = T,
                      scale.model = T) +
                    f(id.space,
                      model = 'besag',
                      graph = italy_graph,
                      adjust.for.con.comp = T,
                      constr = T,
                      # if adjust.for.con.comp = T,
                      # then 'constr = T' interpreted as a sum-to-zero constr on each cc of size > 1
                      scale.model = T) +
                    f(id.int,
                      model = "generic0",
                      Cmatrix = set.int$Rkron,
                      constr = F,
                      extraconstr = set.int$constr) +
                    f(id.time2, model = 'iid') +
                    f(id.space2, model = 'iid'),
                  data = list(y = dat.sort$new_cases,
                              intercept.cc = dat.sort$intercept.cc,
                              id.time = dat.sort$id.week,
                              id.time2 = dat.sort$id.week,
                              id.space = dat.sort$id.province,
                              id.space2 = dat.sort$id.province,
                              id.int = dat.sort$id.int,
                              pop = dat.sort$pop_province),
                  family = 'binomial', Ntrials = pop,
                  control.expert = list(jp = jpr.vp),
                  control.predictor = list(link = 1),
                  control.compute = list(config = TRUE,
                                         dic = TRUE,
                                         waic = TRUE,
                                         cpo = TRUE), 
                  verbose = T)

## VP plot
vp.plot(res.covid, main = paste('Vp plot'))
\end{lstlisting}


\end{document}